\documentclass[prd,twocolumn,showpacs,nofootinbib,amsmath,amssymb,superscriptaddress]{revtex4-1}

\usepackage{graphicx}
\usepackage{dcolumn}
\usepackage{bm}
\usepackage{amsmath,amssymb,bbm,mathrsfs,nicefrac}
 \usepackage{array,multirow}
 \usepackage{float}
\usepackage{tikz}
\usepackage{amsmath}
\usepackage{amssymb}
\usepackage{graphicx,textcomp,float,gensymb,wrapfig, enumitem,comment,dsfont,framed,slashed,appendix,wrapfig,xcolor}
\usepackage{ bbold }
\usepackage{braket}
\usepackage{ mathrsfs }
\usepackage{etoolbox,hyperref,makecell}
\usepackage{feynmp}
\usetikzlibrary{arrows.meta}
\usepackage{empheq}
\usepackage[normalem]{ulem}
\usepackage{scrextend}
\usepackage{slashed}
\usepackage{xspace}
\newcommand{\bea}{\begin{eqnarray}}
\newcommand{\eea}{\end{eqnarray}}

\definecolor{mygreen}{rgb}{0.1, 0.6, 0.1}

\newcommand{\love}{\Lambda}
\newcommand{\loveNotRescaled}{\lambda}
\newcommand{\loveOne}{\Lambda_1}
\newcommand{\loveTwo}{\Lambda_2}
\newcommand{\loveS}{\Lambda_s}
\newcommand{\loveA}{\Lambda_a}
\newcommand{\mans}{\rm MANS\xspace}% Mirror-Matter Admixed Neutron Star
\newcommand{\manss}{\rm MANSs\xspace}% Mirror-Matter Admixed Neutron Stars
\newcommand{\sm}{\textrm{\mbox{\tiny SM}}} 
\newcommand{\dm}{\textrm{\mbox{\tiny DM}}}%MNS
\newcommand{\bigsm}{\textrm{SM}\xspace} 
\newcommand{\bigdm}{\textrm{DM}\xspace}%MNS
\newcommand{\ns}{\rm NS\xspace}%NS
\newcommand{\mns}{\rm MNS\xspace}%MNS
\newcommand{\mnss}{\rm MNSs}%MNSs
\newcommand{\gw}{\rm GW\xspace}%gravitational wave
\newcommand{\Mdm}{M_{\dm}}
\newcommand{\Msm}{M_{\sm}}
\newcommand{\Mmans}{M_{\textrm{\mbox{\tiny MANS}}}}
\newcommand{\Rin}{R_{\textrm{\mbox{in}}}}%inner radius of MANS
\newcommand{\Rout}{R_{\textrm{\mbox{out}}}}% outer radius of MANS
\newcommand{\Mone}{M_1}% mass of lighter star in the binary
\newcommand{\Mtwo}{M_2}% mass of heavier star in the binary
\newcommand{\Mbin}{M_{bin}}% total mass of the binary

\newcommand{\old}[1]{}%{{\color{red} \sout{#1}}}
\newcommand{\new}[1]{{ #1}}%{{\color{blue} #1}}

\usepackage[utf8]{inputenc}

\begin{document}

\title{
Dark Matter or Regular Matter in Neutron Stars? \\ How to tell the difference  from the coalescence of compact objects
}
\author{Maur\'icio Hippert}
\affiliation{Illinois Center for Advanced Studies of the Universe, Department of Physics, University of Illinois at Urbana-Champaign, Urbana, IL 61801, USA}
\author{Emily Dillingham}
\affiliation{Department of Physics, Berea College, 101 Chestnut Street, Berea, KY 40404, USA}
\author{Hung Tan}
\affiliation{Illinois Center for Advanced Studies of the Universe, Department of Physics, University of Illinois at Urbana-Champaign, Urbana, IL 61801, USA}

\author{David Curtin}
\affiliation{Department of Physics, University of Toronto, Toronto, ON M5S 1A7, Canada}

\author{Jacquelyn Noronha-Hostler}
\affiliation{Illinois Center for Advanced Studies of the Universe, Department of Physics, University of Illinois at Urbana-Champaign, Urbana, IL 61801, USA}

\author{Nicol\'as Yunes}
\affiliation{Illinois Center for Advanced Studies of the Universe, Department of Physics, University of Illinois at Urbana-Champaign, Urbana, IL 61801, USA}

\date{\today}

\begin{abstract}
The mirror twin Higgs model is a candidate for (strongly-interacting) complex dark matter, which mirrors SM interactions with heavier quark masses. 
A consequence of this model are mirror neutron stars -- exotic stars made entirely of mirror matter, which are significantly smaller than neutron stars and electromagnetically dark. 
This makes mergers of two mirror neutron stars detectable and distinguishable in gravitational wave observations, but can we observationally distinguish between regular neutron stars and those that may contain some mirror matter?
This is the question we study in this paper, focusing on two possible realizations of mirror matter coupled to standard model matter within a compact object: (i) mirror matter captured by a neutron star and (ii) mirror neutron star-neutron star coalescences.
Regarding (i), we find that (non-rotating) mirror-matter-admixed neutron stars no longer have a single mass-radius sequence, but rather exist in a two-dimensional mass-radius plane.   
Regarding (ii), we find that binary systems with mirror neutron stars would span a much wider range of chirp masses and completely different binary Love relations, allowing merger remnants to be very light black holes.
The implications of this are that gravitational wave observations with advanced LIGO and Virgo, and X-ray observations with NICER, could detect or constrain the existence of mirror matter through searches with wider model and parameter priors.

\end{abstract}

\maketitle

% \tableofcontents

%%%%%%%%%%%%%%%%%%%%%%%%%%%%%%%%%%%%%%%%%%%%%%%%%%%%%%%%%%%%%%
\section{Introduction}

\begin{figure}[htb]
    \includegraphics[clip=true,width=8cm]{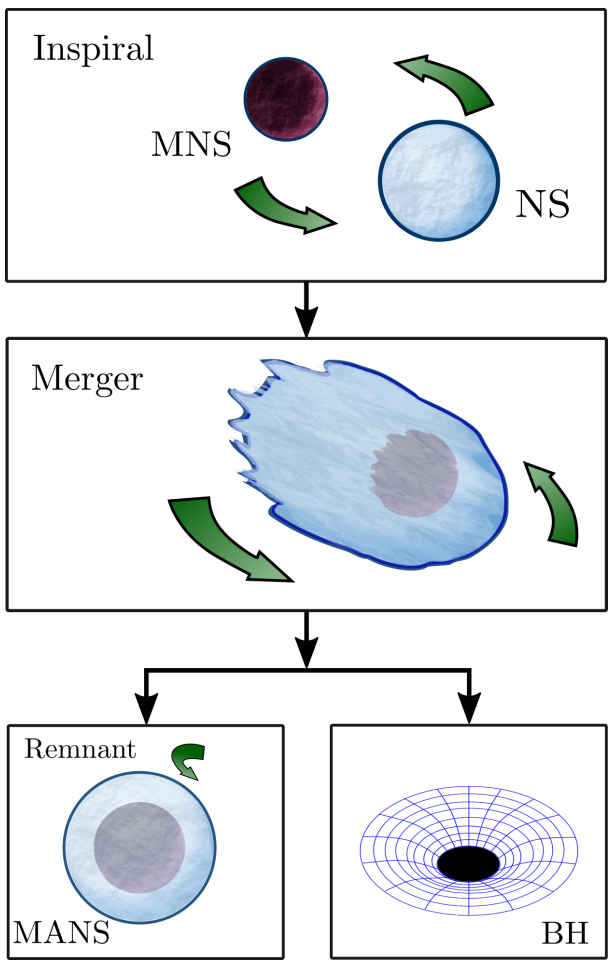}
    \caption{Schematic representation of the coalescence between a NS and a MNS. After inspiralling, radiating energy in GWs (top panel), the two stars collide and merge (middle panel). Depending on the stellar masses, the system either relaxes to an admixed mirror star (bottom left panel) or collapses to a small black hole (lower right panel).}
    \label{fig:Evolution}
\end{figure}

Compact objects present a tantalizing opportunity to study dark matter candidates. Their strong gravitational fields present a unique environment, and their properties are coming under increased scrutiny thanks to gravitational wave (GW) and X-ray observations.
Most dark matter candidates consider one weakly self-interacting species due to simplicity \cite{Lisanti:2016jxe,Lin:2019uvt}. One of the consequences of this assumption is that neutron stars (NSs) can then contain dark matter only through direct (gravitational) capture or through non-gravitational interaction channels, leading to admixed stars (i.e.~those with a mixture of standard model (SM) matter and dark matter) \cite{Li:2012qf,Mukhopadhyay:2015xhs,Baym:2018ljz,Ellis:2018bkr,Dengler:2021qcq,Kain:2021hpk,Sen:2021wev,Jimenez:2021nmr,Miao:2022rqj}.
Admixed NSs can have a small dark matter core 
\cite{Jimenez:2021nmr}, or a dark matter halo that would not affect the visible radius \cite{Miao:2022rqj}; sometimes, such stars can be composed almost entirely of dark matter, leading to extremely tiny dark compact objects (with masses of $\sim 10^{-8}M_{\odot}$)\cite{Horowitz:2019aim}. 
Alternatively, some studies have looked into NSs composed of  hidden sector nucleons from a dark-QCD sector \cite{Maedan:2019mgz},  fundamental asymmetric dark matter fermions \cite{Narain:2006kx,Goldman:2013qla,Tolos:2015qra,Kouvaris:2015rea,Mukhopadhyay:2016dsg,Gresham:2018rqo,Jimenez:2021nmr,Collier:2022cpr,Das:2020ecp,Leung:2022wcf}, or asymmetric bosons \cite{Rutherford:2022xeb,Giangrandi:2022wht}. These studies generally consider simplified interactions of only 1 or 2 particle species. An extensive review of dark matter capture within NSs can be found in~\cite{Deliyergiyev:2019vti}.

In recent years, however, the possibility of complex, strongly self-interacting dark matter candidates has been suggested \cite{Alexander:2016aln,Curtin:2019ngc,Curtin:2020tkm} and the consequence for compact objects explored \cite{Hippert:2021fch,Ryan:2022hku}. Specifically, mirror matter \cite{Blinnikov:1982eh,Blinnikov:1983gh,Khlopov:1989fj,Berezhiani:1995am,Chacko:2005pe} within the mirror Twin Higgs model \cite{Chacko:2005pe, Chacko:2016hvu, Chacko:2018vss,Curtin:2019ngc, Curtin:2020tkm} is a nearly identical copy of the SM in its matter content and gauge interaction, except that the masses of fundamental particles are scaled up by a factor $f/v$, where $f$ and $v$ are vacuum expectation values of SM and mirror-sector Higgs fields, respectively. This model is highly motivated as a solution to the hierarchy problem for $f/v \sim 3-7$. 
The twin top quarks are neutral under all SM gauge interactions, but their interaction with the SM Higgs nevertheless stabilizes its mass and solves the little hierarchy problem without predicting large signals at the LHC~\cite{Burdman:2014zta,Craig:2015pha}. This is in contrast to other solutions to the hierarchy problem like TeV-scale supersymmetry, which predicts large LHC signals due to strong production cross sections for new particles, like stops and gluinos~\cite{ATLAS:2021hza}. 
Additionally, because mirror matter contains multiple species that are strongly interacting, it is possible for mirror matter to clump together to form mirror neutron stars (MNSs) \cite{Hippert:2021fch}. MNSs were found to be very similar to SM NSs, except that they are significantly smaller, with masses $M\sim (0.5-1) M_\odot$ and radii $R\sim (4-8)$ km \cite{Hippert:2021fch}. These MNSs are entirely new hypothesized, electromagnetically dark, compact objects with completely different mass-radius sequences from SM NSs. Importantly, since MNSs are entirely electromagnetically dark GWs from MNS mergers are the best method to detect them. Their distinct tidal deformability and mass range makes them distinguishable from SM NSs or SM stellar-mass black holes.

While mergers between MNSs are very distinct from standard astrophysical gravitational wave signals and represent a spectacular discovery opportunity for new physics, the precise abundance and distribution of MNSs, and hence the rate of their mergers, is almost impossible to predict for a given microphysical model of mirror matter. Depending on the mirror matter distribution in our universe, other astrophysical events involving mirror matter may be much more common. It is therefore vital to consider the full range of possible phenomena involving mirror matter and neutron stars.

What would happen if a NS and a MNS were to coalesce? Schematically, the evolution of such a binary is depicted in Fig.~\ref{fig:Evolution}. First, the two compact objects would spiral around one another.  While their separation is large, they could be effectively treated as point particles in the post-Newtonian approximation~\cite{Blanchet:2013haa}. Because of GW emission, their orbit would shrink and eventually their separation would become small enough that the stars would tidally deform each other. As their separation continues to decrease, the surface of the NS and that of the MNS would cross each other and their stellar interiors would begin to overlap. Unlike mergers of two NSs or two MNSs, this would not be a standard collision because mirror matter and regular matter only interact gravitationally.\footnote{Mirror matter and SM matter do interact via Higgs exchange, but the effect is negligible for astrophysical processes.} After the MNS sinks to the center of the NS, the final remnant would either be a black hole or a stable NS with a mirror matter core, depending on the mass of the progenitor stars.

The development of a quantitative description of the coalescence sketched above requires the combination of various techniques. In the inspiral, post-Newtonian theory can be used to describe the orbital motion of the stars and their tidal deformations. Indeed, the previous study~\cite{Hippert:2021fch} has focused on this inspiral stage and the possibility of using measurements of the tidal deformabilities to distinguish between NS binaries and \mns binaries. 
In this work we extend the inspiral analysis to NS-\mns binaries.

The merger stage can only be described via a two-fluid, numerical relativity simulation in which the fluids only interact gravitationally. However, at this point in time, such a numerical relativity code does not exist. 
Even without such a code, however, one can still study the hypothetical remnant object: a ``mirror-matter admixed NS'' (\mans), 
partially composed of SM matter and mirror matter. There are three possibilities to create \mans~: (i) mirror matter accreting into the cores of a SM NS, 
(ii) SM matter accreting into the cores of a MNS, and 
(iii) MNSs and NSs merging into a stable admixed mirror star remnant. In case (i) [(ii)], one would have very tiny mirror matter [SM] cores that depend on the available mirror matter [SM matter] from their surroundings and the precise nature of the matter-mirror and matter interactions (see e.g.~\cite{Curtin:2020tkm} for a study of Twin Higgs mirror matter accretion in white dwarfs).
In case (iii), most NS and MNS mergers would produce light black holes, but in certain rare cases, stable admixed mirror star remnants would remain with large mirror matter cores. 

In this paper, we study the properties of objects formed in both formation scenarios of \manss~and address their observational prospects.  We essentially ask ourselves the following question: how can we distinguish between SM NSs, MNSs, and \manss from GW observations with advanced LIGO and Virgo, or X-ray observations with NICER? To answer this question, we first construct two-fluid solutions to the Tolman-Oppenheimer-Volkoff (TOV) equations, which describe non-rotating stars (either SM NSs with $f/v = 1$, or MNSs or \manss with $f/v > 1$) in general relativity~\cite{Goldman:2013qla,Kain:2020zjs,Naidu:2021nwh,Gleason:2022eeg}. Our primary findings are the following:
\begin{itemize}
   \item Given an equation of state (EoS), the mass-radius sequence of stable NSs is continuously connected to the mass-radius sequence of stable MNSs through mass-radius sequences of stable \mans. This means that \manss have a two-dimensional (2D) mass-radius \textit{plane}, instead of a one-dimensional mass-radius curve, given a SM EoS. 
   \item \manss can have the ultimate \emph{mass-radius twins}, i.e.~stars with the exact same mass \textit{and} radius, but different DM fraction, and thus, composition. These ultimate twins, however, can be told apart by their tidal deformabilities, $\Lambda$.
    \item 
    The mass-radius plane of stable \manss is nearly independent of the dark sector's $\Lambda_{QCD}$ and 
    extends to much lighter (smaller mass) and much tinier (smaller radius) objects, relative to stable NSs above the Chandresekhar limit.
    \item The relation between the tidal deformability and the mass (or the compactness) of \mans is also two-dimensional. This plane extends to a minimum $\Lambda\sim 10$, but not just for very massive objects, but rather also for light objects with masses as low as $\sim 0.75 M_\odot$. 
    \item The inspiral of a NS and a \mns and that of a MNS and another MNS fills a unique phase space of symmetric mass ratio and chirp mass, unreachable by NS binaries. The symmetric mass ratio can be as low as $0.16$ (corresponding to a mass ratio of $0.3$), while the chirp mass can be as low as $0.5 M_\odot$ (corresponding to a total mass of $1 M_\odot$). 
    \item The remnant of the merger between a NS and a MNS primarily produces very light black holes, with masses between $\sim (1.3,3.7) M_\odot$. Occasionally, such a merger may also produce a stable \mans with a mass of $\Mmans\sim (1.2,1.6)M_\odot$ and radius $R_{\rm{\mans}}\sim (8,9.5)$ km. The merger of a MNS binary can lead to black holes as small as $M\sim 0.8 M_{\odot}$.
    \item If a NS and MNS merge and  both are above their respective Chandrasekhar limits\footnote{The NS Chandrasekhar limit is about $M\sim 1.4 M_{\odot}$. There have already been observations of objects below that limit \cite{Strobel:1999vn,2022NatAs.tmp..224D} with X-Ray observation. From those X-Ray observations, one can estimate the mass and sometimes the radius as well.  However,  time-integrated measures  have been found to be susceptible to systematic error (especially due to assumptions about the atmospheres) such that radii can differ by up to $\sim 50$\% \cite{2013ApJ...764..145C,2012MNRAS.423.1556S}.    }, the admixed core of the remnant \mans~can have masses up to $\Mdm\sim 0.4 M_\odot$ and radii as large as $4.5$ km. Thus, these cores can make up approximately one third of the star's mass and are distributed up to half of their total radius. 
  \item If \mans are formed by other means (e.g. accreting matter), then a much wider 2D mass-radius sequence is possible that could lead to either DM halos or cores.
\end{itemize}

The results summarized above differ from other recent studies on mirror matter admixture on NSs \cite{Sandin:2008db,Ciarcelluti:2010ji,Kain:2021hpk} and binary NS mergers \cite{Emma:2022xjs}. These other studies explored mirror matter candidates that, apart from discrete symmetries, are identical to SM particles. Our work is unique in that we consider mirror Twin Higgs matter, with fundamental masses scaled up by $f/v>1$, and we employ \textit{first-principle results} \cite{Hanhart:2008mx,McNeile:2009mx,Walker-Loud:2008rui,LHPC:2010jcs,Syritsyn:2009mx,Albaladejo:2012te,Horsley:2013ayv}, from  lattice QCD and chiral perturbation theory, to scale the EoS with $f/v$ \cite{Hippert:2021fch}. For $f/v=1$, our model EoS is also tuned to meet current observational constraints on the NS mass-radius relation, while mirror matter is considered in the range $f/v \gtrsim 3$ favoured by LHC constraints~\cite{Burdman:2014zta,Craig:2015pha}. 
This makes our EoS models for the dark matter sector directly relevant for the solution to the Hierarchy Problem as well as highly realistic.

The remainder of this paper presents the details that led to the conclusions summarized above, and it is organized as follows. Section~\ref{sec:micromodel} briefly reviews our microscopic model for the EoS of NS matter developed in~\cite{Hippert:2021fch}, and how we extend it to the mirror sector. Section~\ref{sec:isolated-methods} outlines general methods for the calculation of two-fluid stellar properties and their stability. Section~\ref{sec:isolated-results} presents results regarding the properties of isolated NSs, MNSs and \mans. Section~\ref{sec:inspiral} investigates and discusses the GW imprint of a NS/MNS inspiral. Section~\ref{sec:post-merger} describes the possible remnants of a NS/MNS merger. Section~\ref{sec:detect} summarizes the signatures of mirror from various merger scenarios, and Sec.~\ref{sec:conclusion} concludes and presents an outlook of future work.

%%%%%%%%%%%%%%%%%%%%%%%%%%%%%%%%%%%%%%%%%%%%%%%%%%%%%%%%%%%%%
\section{Microscopic Model}
\label{sec:micromodel}

In this short section, we briefly describe our microscopic model for the EoS of NSs and MNSs~\cite{Hippert:2021fch}. All of the presented results are derived under the simplifying approximations of vanishing temperature and slow rotation, $T\simeq 0$ and $\Omega \simeq 0$.

For the description of matter in the NS core, we employ a relativistic mean-field nuclear model, suplemented by a free gas of electrons and muons, so that charge neutrality and chemical equilibrium under weak interactions are satisfied.
This model features scalar, vector, and vector-isovector interactions and is tuned to reproduce properties of nuclear matter,\footnote{The nuclear physics inputs of our model are the incompressibility, symmetry energy, nucleon Dirac mass and binding energy per nucleon at saturation density. The set of values that are employed here and the corresponding parameters can be found in~\cite{Hippert:2021fch}.} as well as current constraints from NS observations~\cite{Miller:2019cac,Riley:2019yda,Miller:2021qha,Riley:2021pdl,TheLIGOScientific:2017qsa,Abbott:2018exr,Abbott:2018wiz}. At lower densities, we match the model to a Baym-Pethick-Sutherland EoS for the NS crust \cite{Baym:1971pw} via an interpolation procedure. A detailed account of our model can be found in~\cite{Hippert:2021fch} and references therein. Results for NS properties obtained with this model are reviewed in Sec.~\ref{sec:isolated-ns}.

For our description of matter inside a MNS, we adopt the same model, but rescale parameters to account for the different microphysics of the Twin Higgs mirror sector. Our dark-matter model is then fixed by the ratio $f/v$ between the expectation values of the Higgs and Mirror Higgs. We assume this ratio to lie in the range $f/v\simeq (3 - 7)$, as larger values would fail to naturally solve the hierarchy problem in the minimal model~\cite{Burdman:2014zta}, and smaller values are ruled out by Higgs coupling measurements at the LHC~\cite{Aad:2019mbh}.

All the elementary masses in our model ---that is, the ones of leptons and quarks--- are thus rescaled by the same factor $f/v$, defined such that the SM is recovered when $f/v=1$. However, because of the non-perturbative scale $\Lambda_{\textrm{QCD}}$ and the spontaneous breakdown of chiral symmetry of QCD, this translates to non-trivial scalings for the  nuclear-model couplings and the baryon mass with $f/v$. These scalings are extracted from low-energy nuclear-physics phenomenology and first-principle results from chiral perturbation and lattice QCD (see \cite{Hippert:2021fch} for extensive details and suggestions for new calculations from lattice QCD).
In the absence of these inputs, extra couplings are scaled according to dimensional analysis, but for those couplings the details of this procedure have minimal impact on the overall EoS scaling. 
Details of these scalings can be found in~\cite{Hippert:2021fch}. 

%%%%%%%%%%%%%%%%%%%%%%%%%%%%%%%%%%%%%%%%%%%%%%%%%%%%%%%
\section{Stellar Structure Equations}
\label{sec:isolated-methods}

In this section, we review the equations that we use to calculate the structure of MNSs. We also review simple criteria for the stability of these stars and the calculation of their Love number. These results are very well known in the case of ordinary compact stars (see e.g.~\cite{Yagi:2013awa} and references therein), but here we present them with an extra layer of generality \cite{Li:2012qf,Mukhopadhyay:2015xhs,Baym:2018ljz,Ellis:2018bkr,Nelson:2018xtr,Dengler:2021qcq,Kain:2021hpk,Sen:2021wev,Jimenez:2021nmr,Miao:2022rqj}.

Besides the single component stars already discussed in Refs.~\cite{Hippert:2021fch}, we address a third possibility, consisting of two-component isolated stars, made of both SM and mirror matter.
This choice is motivated not only by completeness but by the possibility of such stars existing, as a result of dark-matter admixture in NSs. We refer the reader to Sec.~\ref{sec:isolated-2comp} for details.
In the case of such stars, because interactions between ordinary and dark matter are assumed to be negligible, the structure equations, stability analysis and calculation of Love number must be modified accordingly, as we present below \cite{Dengler:2021qcq}.

%------------------------------------------------
\subsection{Two-Fluid Configurations}

The structure of (mirror) NSs follows from the equilibrium between the push of pressure and the pull of gravity, summarized through the equation of hydrostatic equilibrium. In the case of two-component stars, however, the lack of significant interactions between dark and visible matter prevents their equilibration with one another.\footnote{It is interesting to consider the impact of small possible couplings, like a kinetic mixing between the SM photon and the twin photon, which may result in non-negligible interactions between matter and mirror matter in this case~\cite{Curtin:2020tkm,Curtin:2019ngc,Curtin:2019ngc}, but we leave this for future work.
}. Instead, each fluid component must independently neutralize gravitational forces, resulting in two independent ``equilibrium'' pressures, say, $p_{\textrm{SM}}$ and $p_{\dm}$. Nonetheless, the two fluids interact via gravitation, and the Einstein equations must be solved in the presence of both fluids at the same time \cite{Goldman:2013qla,Li:2012qf}.

Two-component stars are odd creatures, and possess two zero-pressure surfaces, which encompass the two independent fluids. In principle, these fluids can rotate with different angular velocities without violating hydrostatic equilibrium. Nonetheless, after sufficiently long times, tidal effects tend to synchronize their rotations. In the following sections, for the sake simplicity, we assume no differential rotation between the two fluid components.
Under this assumption, the two-fluid system we consider is no different from a single-fluid system, except that there are two stress-energy tensors on the right-hand side of Einstein's field equation.

Assuming two independent fluids amounts then to writing the stress-energy tensor as the linear combination
\begin{equation}
    T^{\mu\nu} = T_{\sm}^{\mu\nu} + T_{\dm}^{\mu\nu}\,,
\end{equation}
where each component is independently conserved, i.e.~
\begin{equation} \label{eq:EPcons}
    {T_{\sm}^{\mu\nu}}_{;\nu} = {T_{\dm}^{\mu\nu}}_{;\nu} = 0
\end{equation}
Also, assuming global thermodynamic equilibrium, the stress-energy tensor for each component can be written as
\begin{equation}\label{eq:ideal}
\begin{split}
    T_{\sm}^{\mu\nu} &= (\epsilon_{\sm}+p_{\sm})u_{\sm}^\mu u_{\sm}^\nu + p_{\sm} g^{\mu\nu} \\
    T_{\dm}^{\mu\nu} &= (\epsilon_{\dm}+p_{\dm})u_{\dm}^\mu u_{\dm}^\nu + p_{\dm} g^{\mu\nu}.
\end{split}
\end{equation}
From the above equations, it is clear that the standard case of a single fluid that leads to a SM-only NS can be recovered by taking $p_{\dm} = 0 = \epsilon_{\dm}$. Thus, from here on, we concentrate on calculations for the more general case in which two fluids are present.

%------------------------------------------------
\subsection{Hydrostatic Equilibrium}
\label{sec:TOV2}

Let us review the hydrostatic equilibrium equations in the case of two independent fluids, relevant for two-component stars with \old{dark-matter} \new{admixed} cores \cite{Goldman:2013qla,Li:2012qf}.  These equations generalize the well-known Tolman-Oppenheimer-Volkoff (TOV) equations for single-fluid stars, which can be recovered as a particular case, by setting to zero the density and pressure of one of the fluids.

We consider a static two-fluid hybrid star in spherical symmetry. The line element squared can therefore be written as $ds^2 = -e^{2 \tau} dt^2+e^{2 \sigma}dr^2+r^2d\Omega^2$, with signature $(-,+,+,+)$, where $\tau$ and $\sigma$ are functions of the (areal) radial coordinate $r$ and $d\Omega^2$ is the line element squared of a two-sphere. The only non-zero component of the four velocity of a static-fluid is the time component. Therefore, $u_{\sm}^\mu = u_{\dm}^\mu = \mathcal{N}(1,0,0,0)$, where $\mathcal{N}$ is a normalization factor that ensures $u_\mu u^\mu = -1$.

With the four velocity known and a few algebraic manipulations, Eqs.~\eqref{eq:EPcons} and the Einstein field equation simplify to a set of TOV-like equations for two fluids:
\begin{subequations}
\begin{align}
    &\frac{dp_i}{dr} = -\left( \epsilon_i+p_i \right)\frac{m+4 \pi r^3\left(p_{\sm}+p_{\dm}\right)}{r\left(r-2m\right)}\label{eq:dp1dr},\\
    &\frac{dm}{dr} = 4 \pi \left( \epsilon_{\sm} +\epsilon_{\dm} \right) r^2 \label{eq:dmdr},\\
    &\frac{dN_i}{dr} = 4 \pi\frac{n_i}{\sqrt{1-{2 m}/{r}}} r^2 \label{eq:dNdr},
\end{align}
\end{subequations}
with $i=\{\bigsm, \bigdm\}$. Here, $p_{\sm,\dm}$ and $\epsilon_{\sm,\dm}$ are the pressure and energy density of the SM and DM fluids, respectively, while $m = [1 - \exp(-\lambda)]/2$ is the enclosed mass function. Notice that Eq.~\eqref{eq:dp1dr} governs the hydrostatic equilibrium of the fluids separately. In Eq.~\eqref{eq:dNdr}, $n_{\sm,\dm}$ and $N_{\sm,\dm}$ are the baryon number density and the total number of baryons and mirror baryons, respectively. 

We follow  \cite{Yagi:2013awa} closely for the standard numerical implementation to solve the two-fluid TOV equation, except for the initial conditions and stopping conditions. For the initial conditions, we need to specify the energy densities of both the SM species and the DM species, $\epsilon_{c}^\sm$ and $\epsilon_{c}^\dm$, at the center of the star. Let us say then that we choose some value of $\epsilon_{c}^\sm$ and $\epsilon_{c}^\dm$ and begin to integrate the equations out from the center. At some radial coordinate, the pressure of one of the fluids (either the SM or the DM fluid) drops to $10^{-8}$ of the corresponding central pressure. That value of the radial coordinate defines the ``radius of the admixed core $\Rin$,'' which contains both fluids. Of course, that is not all of the star, since the density and pressure of the other fluid can still be large at that radius. The integrator then continues, with the energy density and pressure of the first fluid set to zero, until the pressure of the second remaining fluid drops to $10^{-8}$ of the corresponding central pressure. The radial coordinate value at that place defines the outer radius of the entire star $\Rout$ and the integrator stops. The total mass $\Mmans$ of the star is then simply the value of $m$ evaluated at the radius $\Rout$.

The description of this implementation presented above anticipates one of the main results of this paper: stable two-fluid stars have (2D) mass-radius planes instead of (1D) mass-radius curves. Mathematically, this tracks back to the need to specify two initial conditions, $\epsilon_c^{\dm}$ and $\epsilon_c^{\sm}$, to find a stellar solution to the two-fluid TOV equations. This means that one can independently change $\epsilon_c^{\dm}$ while keeping $\epsilon_c^{\sm}$ constant (thus, increasing the number of mirror baryons in the star), or alternatively independently change $\epsilon_c^{\sm}$ while keeping $\epsilon_c^{\dm}$ constant (thus, increasing the number of SM baryons in the star). As we will show in the next section, the sequence of stable stellar configurations is, indeed, represented by a plane in mass and radius. 

%------------------------------------------------
\subsection{Tidal Deformability}
\label{sec:deformability}

Tidal deformation emerges under strong external tidal fields.
If a \mans is in a binary orbit with another compact object, at some point the \mans and the other compact object will be a distance $r_{12}$ appart such that the \mans will become tidally deformed due to the gravitational pull of the other compact object. In this case, the tidal deformability can be calculated within this two fluid approach as well, see
\cite{RafieiKarkevandi:2021hcc,Dengler:2021qcq}.  Just as SM NSs, the two-fluid \mans can be tidally deformed when the distance $r_{12}$ between the stars is small enough, i.e. $\Rout \ll r_{12} \ll R_{\rm ext}$, where $R_{\rm ext}$ is the radius of curvature of the source of the external field (the other compact object) and $\Rout$ is the radius of the entire two-fluid \mans. A multipolar expansion of the metric exterior to the star can be written as~\cite{Hinderer:2007mb}:
\begin{equation}\label{eq:tidal_metric}
\begin{split}
	-\frac{1+g_{tt}}{2}  = & -\frac{\Mmans}{\Rout}- \frac{3 Q_{ij}^{(tid)}}{2 \Rout^3} \left(n^i n^j - \frac{1}{3} \delta^{ij} \right) \\
	& + {\cal{O}}\left(\frac{1}{\Rout^4}\right) 
	+ \frac{1}{2} {\cal{E}}_{ij} \Rout^2 n^i n^j + {\cal{O}}\left(\Rout^3\right)\,,
\end{split}
\end{equation}
where ${\cal{E}}_{ij}$ is the (quadrupole) tidal tensor field, 
$Q^{(tid)}_{ij}$ is the corresponding tidally-induced and traceless quadrupole moment tensor, and $n^i = x^i/\Rout$ is a field-point unit vector. The tidal deformability $\lambda$ can be defined by $Q^{(tid)}_{ij} = -\loveNotRescaled \, {\cal{E}}_{ij}$. The dimensionless tidal deformability can then be written as
\begin{equation}\label{eq:Lambda}
    \love = \loveNotRescaled / \Mmans^5,
\end{equation}
where $\Mmans$ is the mass of the entire two-fluid star obtained from the two-fluid TOV equations. 

The quantity \(\loveNotRescaled\) can be calculated by solving the Einstein field equations interior to the star at first order in the tidal perturbations. The version of this equation for a single fluid star is presented e.g.~in~\cite{Yagi:2013awa}. To obtain the corresponding equation for a two-fluid star, one may naively replace $\epsilon \to \epsilon_{\sm} + \epsilon_{\dm}$, $p \to p_{\sm} + p_{\dm}$, and $m \to m_{\sm} + m_{\dm}$ in the relevant equations, where $m_i$ can be obtained using the equation $dm_i/dr = 4 \pi \epsilon_i r^2$ with $i \in \{\sm,\dm\}$. However, because both fluids exert gravitational attraction on each other, one must employ Eq.~\eqref{eq:dp1dr} to write $d\epsilon_i/dr = (d\epsilon_i/dp_i)(dp_i/dr)$,  instead of applying a naive replacement in the corresponding term for one-fluid stars.  

In the end, the Einstein field equations at first order in the tidal perturbation become
\begin{equation}
    r \frac{dy_*}{dr} + y_*^2 + F(r)y_* +r^2 Q(r) = 0 \label{eq:yEq}
\end{equation}
with
\begin{equation}
    F(r) = \frac{r + 4 \pi r^3 (p_\sm+p_\dm - \epsilon_\sm-\epsilon_\dm)}{r - 2m_\sm-2m_\dm}\label{eq:F}
\end{equation}
and
\begin{equation}
\begin{split}
    Q(r)&= \frac{4 \pi r}{r - 2 m_\sm - 2 m_\dm}
    \bigg[5(\epsilon_\sm+\epsilon_\dm) + 9 (p_\sm+p_\dm) + \\
    & \left. \frac{p_\sm +\epsilon_\sm}{{c_s}_\sm^2}+\frac{p_\dm + \epsilon_\dm}{{c_s}_\dm^2}  - \frac{3}{2 \pi r^2}\right] + \\
    &- \left( 2\frac{m_\sm+m_\dm + 4 (p_\sm+p_\dm) \pi r^3}{r (r-2 m_\sm-2 m_\dm )}\right)^2,\label{eq:Q}
\end{split}
\end{equation}
which agrees with~\cite{Dengler:2021qcq}. Here the quantity $y_*$ is defined by $y_* \equiv r h_2'(r)/h_2(r)$, where $h_2$ is related to time-time component of the metric by $g_{tt} = -e^{2\tau}(1+h_2 Y_{2m}(\theta,\phi))$, with $Y_{2m}$ the $\ell=2$ spherical harmonic. Notice that the quantity $h_2$ represents a first-order-in-perturbation term due to tidal effects. The quantity ${c_s}_i^2$ in Eq.~\eqref{eq:Q} is the speed of sound squared of the $i_{th}$ fluid\footnote{Note that the $Q(r)$ term depends on $1/c_{s1}^2$ and $1/c_{s2}^2$, which can diverge during a first-order phase transition. In this radial regime, however, the pressure and density drops violently to keep the ratio finite.}, which is defined as ${c_s}_i^2 \equiv dp_i/d\epsilon_i$. After solving the interior equation, Eq.~\eqref{eq:yEq}, one can match the interior solution to the exterior metric of Eq.~\eqref{eq:tidal_metric} to find the ratio of coefficients $Q^{tid}_{ij}$ and ${\cal{E}}_{ij}$, and thus, to obtain \(\love\). Doing so, one finds
\begin{equation}
\begin{split}\label{eq:tidalLove}
	\love &= \frac{16}{15} \left\{\left(1-2 {C}\right)^2\left[2-y+2 {C}\left(y-1\right)\right]\right\} \left\{ \right.  \\
	&2 {C}\left[6-3y+3 {C}(5y-8)\right]
	  \\
	 &+ 4 {C}^3 \left[ 13-11y+  {C}(3y-2) + 2 {C}^2(1+y) \right] \\ 
	 & \left. +3(1-2 {C})^2\left[2-y+2 {C}(y-1)\right] \ln\left(1-2 {C}\right) 
	 \right\}^{-1}\,,
\end{split}    
\end{equation}
where $C = \Mmans/\Rout$ is the gravitational compactness of the entire star and $y = y_*(\Rout)$.

%------------------------------------------------
\subsection{Stability Analysis}
\label{sec:stability}

The stability of stellar configurations may be determined by analyzing the spectrum of radial density oscillations.  
In a complete analysis, one adds a harmonic perturbation of the form $\xi(r)e^{-i \omega t}$ 
to the metric fields and linearizes the Einstein field equations with respect to the perturbation. The result of this procedure is a Sturm-Liouville problem, whose solution determines the frequency eigenvalues $\{\omega_i\}$. In this approach, unstable radial modes manifest as exponentially increasing solutions, with  $\operatorname{Im} \omega_i > 0$ \cite{Comer:1999rs, Leung:2012vea, Kain:2020zjs}. 
The equation of radial pulsation for a single-fluid compact star was first derived by  Chandrasekhar\cite{Chandrasekhar:1964zz}. 
Here, we resort to a simpler, less rigorous, stability criterion, following \cite{Henriques:1990xg,Nyhan:2022pda}. 
An analogous criterion, for single-fluid stars, is explained in detail by Weinberg~\cite{Weinberg:1972kfs}. 

Assuming invariance under time reversal, each eigenmode $\xi_i(r)$ has a pair of fundamental frequencies $\omega_i^{\pm}=\pm\sqrt{\omega_i^2}$, where $\omega_i \in \mathds{R}$ is an eigenvalue.  
The onset of instability then corresponds to the point at which the lowest-lying eigenvalue $\omega_0^2$ flips sign and becomes negative. 
Precisely at this point, $\omega_0^{\pm}=0$, and the field perturbations are simply $\xi_0(r)$, which is a static configuration, and therefore, in equilibrium. Because equilibrium configurations are uniquely specified by central densities, these modes must correspond to shifts $\epsilon_{c}^i \to \epsilon_{c}^i + \delta \epsilon_{c}^i$.  Moreover, if we approach $\omega_0^2\to 0^+$ from above, these configurations will be joined by arbitrarily slow oscillations, which must leave the total (mirror) baryon number  $N_i$ unchanged. 

All of this implies that at the onset of unstable radial modes, the total \new{mirror baryon and SM} baryon numbers must be \emph{stationary} under variations of $\epsilon_{c}^i$:\old{\footnote{The TOV equations on Eqs.~\eqref{eq:dp1dr} and \eqref{eq:dmdr}  correspond to the condition that $\delta M=0$ when $\delta N_i=0$, so that the total mass is also stationary  \cite{Goldman:2013qla}.}} 
\begin{equation}\label{eq:stationary}
\arraycolsep=2pt\def\arraystretch{1.5}
\left(
\begin{array}{c}
     \delta N_\sm \\
     \delta N_\dm
\end{array}
\right) = 
\left(
\begin{array}{cc}
 {\partial N_\sm}/{\partial\epsilon_{c}^\sm} & {\partial N_\sm}/{\partial\epsilon_{c}^\dm} \\
 {\partial N_\dm}/{\partial\epsilon_{c}^\sm} & {\partial N_\dm}/{\partial\epsilon_{c}^\dm}  
 \end{array}
\right)
\left(
\begin{array}{c}
     \delta\epsilon_{c}^\sm \\
     \delta\epsilon_{c}^\dm
\end{array}
\right)
= 0.
\end{equation}
The existence of nontrivial solutions $\delta\epsilon_{c}^i\neq 0$ demands that the matrix in Eq.~\eqref{eq:stationary} have zero determinant:
\begin{equation}\label{eq:stabcondition}
  \frac{\partial N_\sm}{\partial\epsilon_{c}^\sm} \frac{\partial N_\dm}{\partial\epsilon_{c}^\dm} 
  - \frac{\partial N_\sm}{\partial\epsilon_{c}^\dm}  \frac{\partial N_\dm}{\partial\epsilon_{c}^\sm}=0.
\end{equation}
This is the \old{stability criteria}\new{criteria for the onset of radial instability} for two-fluid stars.

\new{
One might wonder if Eqs.~\eqref{eq:stationary} and \eqref{eq:stabcondition} are consistent with the well-known condition $\partial M/\partial \epsilon_c=0$ in the case of a single fluid. 
For single-fluid stars, Eq.~\eqref{eq:stationary} will only have nontrivial solutions if $\partial N/\partial\epsilon_c=0$. Under the assumption of uniform entropy per baryon, the stellar mass is stationary under any transformation 
$\epsilon(r)\to \epsilon(r) +\delta \epsilon(r)$ that leaves the total baryon number $N$ unchanged  \cite{Weinberg:1972kfs}. This holds if and only if the TOV equation is satisfied, and thus, exclusively for equilibrium configurations. In light of this result, demanding that $\partial N/\partial\epsilon_c =0$ is equivalent to imposing that $\partial M/\partial \epsilon_c=0$, which is the widely used criteria for the onset of instability \cite{Weinberg:1972kfs}. 

The result mentioned above has recently been generalized to the case of two fluid stars \cite{Goldman:2013qla}. 
Given that the entropy per baryon in the SM fluid and the entropy per mirror baryon in the DM fluid are kept uniform, Eqs.~\eqref{eq:dp1dr}-\eqref{eq:dNdr} are satisfied if and only if any transformation $\epsilon_{i}(r)\to \epsilon_{i}(r) +\delta \epsilon_{i}(r)$ satisfying $\delta N_i = 0$, with $i=\{\bigsm,\bigdm\}$, also leaves the mass unchanged, i.e. $\delta M=0$. Equation \eqref{eq:stationary} is precisely the condition that $N_\sm$ and $N_\dm$ are stationary, under which case $M$ is stationary as well. That is, at the onset of instability, $\delta N_\sm=0$, $\delta N_\dm =0$ and $\delta M=0$ under variations of the central energy densities, $(\epsilon_{c}^\sm,\epsilon_{c}^\dm)\to (\epsilon_{c}^\sm+\delta \epsilon_{c}^\sm,\epsilon_{c}^\dm+\delta \epsilon_{c}^\dm)$. 

The physical intuition behind this simpler stability criterion is also left unchanged. 
Suppose a single-fluid star is taken out of equilibrium, with $\epsilon_c \to \epsilon_c + \delta \epsilon_c$, while its total baryon number $N = N(\epsilon_c)$ is kept fixed.  The star is then at a baryon-number difference $\Delta N = N(\epsilon_c) - N(\epsilon_c+\delta \epsilon_c)$ away from equilibrium. Assuming $\partial M/\partial N>0$, if $\Delta N >0$, there is a mass excess, indicating that the gravitational pull is too strong and the star will contract. If $\Delta N < 0$, gravitational attraction is not sufficiently strong and the star expands. In the spirit of Le Chatelier's principle, equilibrium is stable if an increase in density $\delta\epsilon_c>0$ leads to a restoring expansion, with $\Delta N\approx - (\partial N/\partial\epsilon_c)_{\textrm{eq}} \delta\epsilon_c <0$, where the derivative is taken in hydrostatic equilibrium. Therefore, a necessary condition for stable equilibrium is that $\delta N/\delta \epsilon_c>0$, or equivalently $\delta M/\delta \epsilon_c>0$ \cite{glendenningbook}.

For a two-fluid star, the two central densities and two conserved charges are mutually intertwined, making the physical picture less clear. However, an analogous heuristic argument can be found by diagonalizing the matrix $\partial N_i/\partial \epsilon_{c\, j}$, with  $\{i,j\}=\{ \bigdm,\bigsm\}$, in Eq.~\eqref{eq:stationary}. By doing so, one obtains two independent  sets of variables, $(\epsilon_{c}^A, N_A)$ and $(\epsilon_{c}^B, N_B)$
corresponding to eigenvalues $\kappa_A$ and $\kappa_B$, such that:
\begin{equation}\label{eq:stationarydiag}
\arraycolsep=2pt\def\arraystretch{1.5}
\left(
\begin{array}{c}
     \delta N_A \\
     \delta N_B
\end{array}
\right) = 
\left(
\begin{array}{cc}
 \kappa_A & 0 \\
 0 & \kappa_B 
 \end{array}
\right)
\left(
\begin{array}{c}
     \delta\epsilon_{c}^A \\
     \delta\epsilon_{c}^B
\end{array}
\right).
\end{equation}
Because $N_A$ and $N_B$ are linear combinations of $N_\sm$ and $N_\dm$, they are also conserved and are kept fixed as the star is perturbed. Changes to $\epsilon_c^A$ and $\epsilon_c^B$ are then performed independently to find that equilibrium configurations can be stable only if both eigenvalues are positive
\begin{equation}\label{eq:stabilitydiag}
 \kappa_A >0, \qquad \kappa_B > 0.
\end{equation}
This generalizes the widely used stability condition $\partial M/\partial\epsilon_c >0$ to multi-fluid stars. 

Finally, we observe that Eq.~\eqref{eq:stabcondition} provides one condition for two independent variables $\epsilon_c^\sm$ and $\epsilon_c^\dm$. Therefore,
as the mass-radius relation for two-fluid stars is an area, the boundary of the stable region is a set of curves. This is in contrast to the case of a single fluid, for which the mass-radius relation is a curve and stable regions are delimited by points. 
}

\old{Clearly, for a single fluid, Eq.~\eqref{eq:stabcondition} reduces to $\partial N/\partial \epsilon_c = 0$, which, in view of the TOV equations, is equivalent to the commonly used condition $\partial M/\partial \epsilon_c = 0$ }

Henceforth, we will determine stability by \old{checking}\new{using} the condition in Eq.~\eqref{eq:stabilitydiag}, which is in agreement with previous stability studies of two-fluid dark-matter admixed stars \cite{Comer:1999rs, Leung:2012vea, Kain:2020zjs}. 
A rigorous stability analysis of two-fluid stars, using e.g.~our EoS, is left to future work.

%%%%%%%%%%%%%%%%%%%%%%%%%%%%%%%%%%%%%%%%%%%%%%%%%%%%
\section{Properties of Individual Stars}
\label{sec:isolated-results}

Before we study coalescences involving NSs, MNSs and MANSs, we must first address the properties of each of these stars in isolation.
Masses and radii follow directly from the EoSs discussed in Sec.~\ref{sec:micromodel} and the equations of structure (including the two-fluid Tolman-Oppenheimer-Volkoff (TOV) equation), which were reviewed in Sec.~\ref{sec:TOV2}. The calculation of tidal deformabilities, can be computed as described in Sec.~\ref{sec:deformability}.
Results for isolated MNSs and NSs were already presented in \cite{Hippert:2021fch}, but we present them again here for completeness. We also present results for a new type of object (\mans ), which we remind the reader are compact ojbects composed of an admixture of mirror matter ($f/v>1$) and SM matter ($f/v=1$).

\begin{figure}[tb]
    \includegraphics[clip=true,width=8cm]{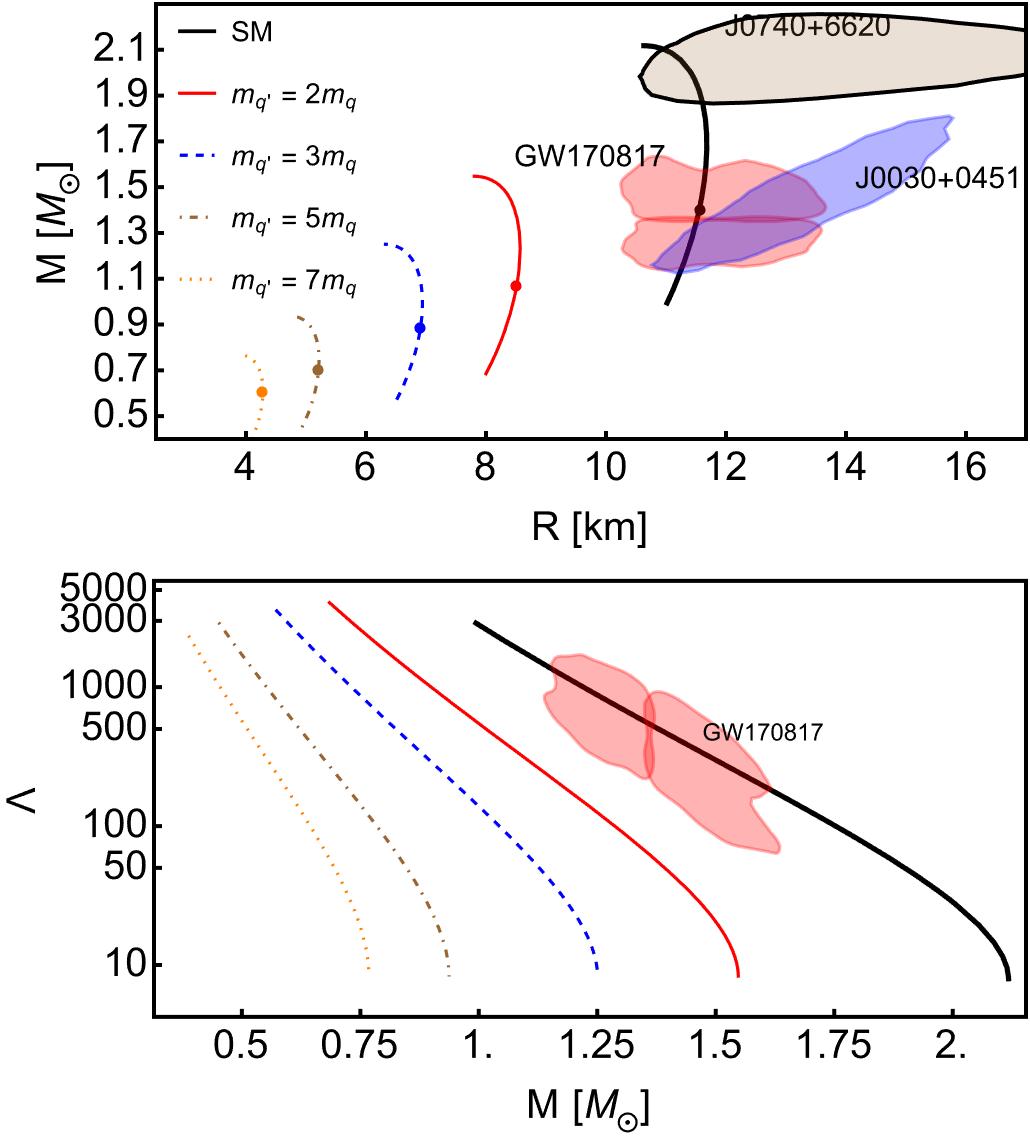}
    \caption{Upper panel: Mass-radius relations for NSs and MNSs with quark masses scaled  by $m_{q'}/m_q = f/v$. Lower panel: tidal deformability-mass for NSs and MNSs with quark masses scaled  by $m_{q'}/m_q = f/v$.
    The blue and beige shaded regions correspond to $2\sigma$ confidence regions, using X-ray observations of pulsars J0030+0451 and J0740+6620, respectively by NICER~\cite{Miller:2019cac,Riley:2019yda,Miller:2021qha,Riley:2021pdl}. Red shaded regions represent $2\sigma$ confidence intervals from GW event GW170817 observed by advanced LIGO/Virgo~\cite{TheLIGOScientific:2017qsa,Abbott:2018exr,Abbott:2018wiz}. 
    }
    \label{fig:MRplotfront}
\end{figure}

%------------------------------------------------
\subsection{Neutron Stars and Mirror Neutron Stars}
\label{sec:isolated-ns}

The NS EoS discussed in Sec.~\ref{sec:micromodel} \cite{Hippert:2021fch} can be inserted into Eqs.~\eqref{eq:dp1dr}, \eqref{eq:dmdr} and~\eqref{eq:yEq}, with $p_{\dm}=\epsilon_{\dm}=0$, to calculate the mass, radius and tidal deformabilities of SM NSs. The resulting mass-radius and tidal deformability-mass relation are shown in Fig.~\ref{fig:MRplotfront} and are consistent with the current observational constraints \cite{Hippert:2021fch}.
The latest NICER measurement of pulsar J0740+6620 \cite{Riley:2019yda,Miller:2019cac} was not used to constrain model parameters in Ref.~\cite{Hippert:2021fch}, but yet, the mass radius curve is consistent with these observations. The aim of the present study is \textit{not} to advocate for a particular model of NS matter. Instead, we simply choose an EoS that is in agreement with known astrophysical observations, and which we can extend to the mirror sector to draw new predictions. An extensive discussion of uncertainties from the SM EoS and mass-radius relation can be found in Ref.~\cite{Hippert:2021fch}, where we show that our predictions for MNS matter are robust against these uncertainties, as well as model details.

\begin{figure*}[tb]
    \centering
    \begin{tabular}{c c}
    \includegraphics[width=8cm]{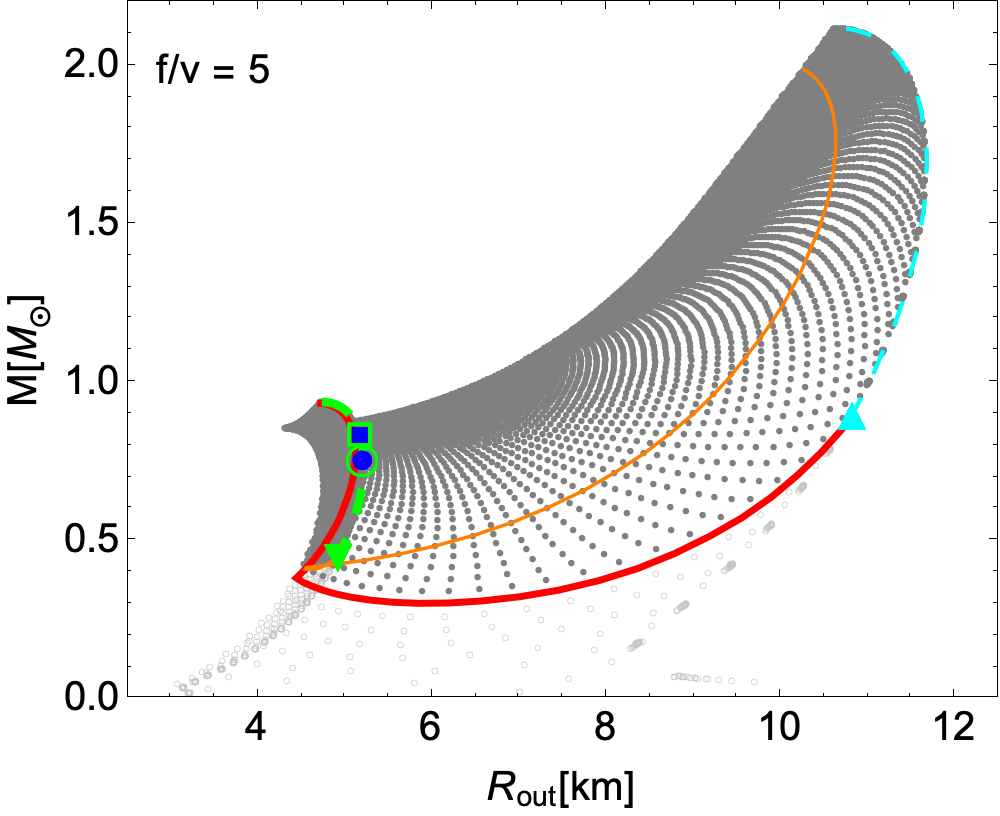} 
    \includegraphics[width=8cm]{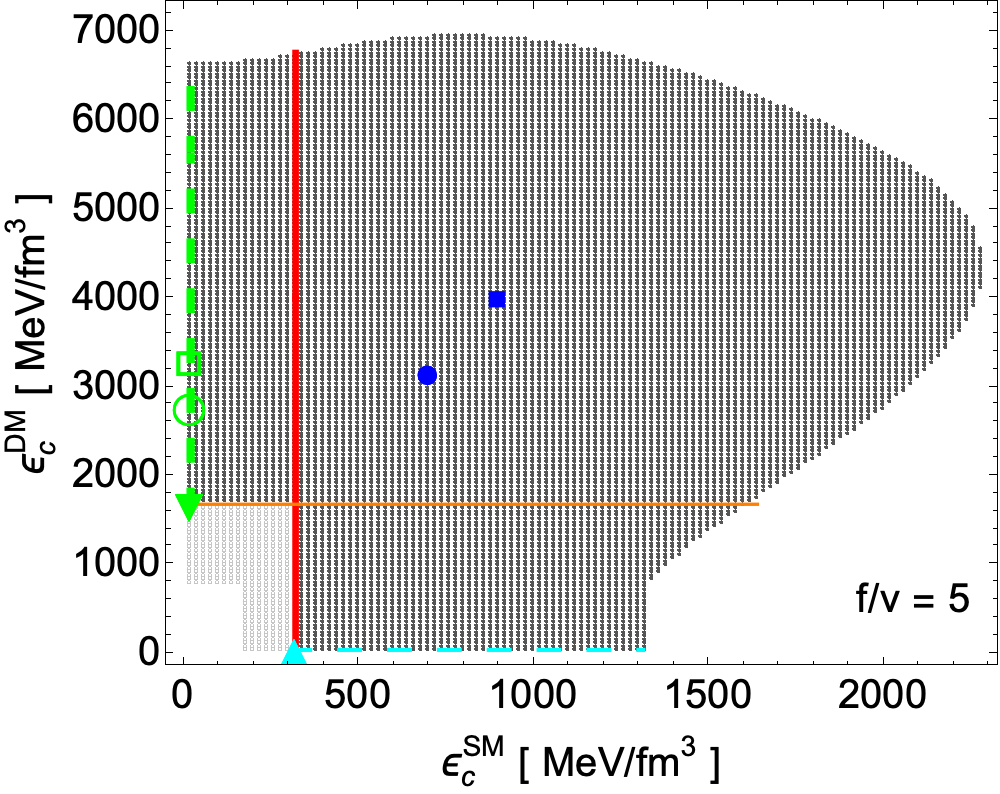} 
    \end{tabular}
    \caption{
    Mass-radius plane of \mans (left panel) for different choices of central SM and mirror-matter energy densities (right panel). The SM NS and the MNS mass-radius sequences are shown with a cyan and a green dashed line respectively. The cyan up triangle and the green down triangle correspond to energy densities at which a SM NS and a MNS would have masses of $64\%$ their Chandrasekhar limits respectively. Increasing the mirror matter and the SM central densities respectively then leads to the red and orange boundaries. 
    Observe that the mapping from $(\epsilon_c^\mathrm{SM}, \epsilon_c^\mathrm{DM})$  to $(R_\mathrm{out}, M)$ is not bijective at low radii, allowing for the possibility of ultimate twins.
    Two examples of ultimate twins are shown with filled blue and open green circles and squares. 
    }
    \label{fig:route_phaseSpace}
\end{figure*}

Similarly, the MNS EoS of Sec.~\ref{sec:micromodel} can be inserted into the same equations, but this time with $p_{\sm}=\epsilon_{\sm}=0$ and specific choices of $f/v$, to calculate the mass, radius and tidal deformabilities of MNSs. 
The resulting mass-radius and $\Lambda$-mass sequences of MNSs are also shown in Fig.~\ref{fig:MRplotfront}, where Chandrasekhar-mass configurations are marked with a filled circle in the mass-radius panel \cite{Hippert:2021fch}, and we calculate the sequences down to $64\% M_{Ch}$. Figure~\ref{fig:MRplotfront} allows us to compare predictions for NSs and MNSs. Observe that, in the phenomenologically relevant range $f/v \simeq 3 - 7$, the mass-radius and $\Lambda$-mass sequences for MNS are shifted to lower masses and smaller radii, and to smaller $\Lambda$ and smaller masses. This is perhaps most clearly seen for the Chandrasekhar-mass configurations, which go from the well-known $\sim 1.4 M_\odot$ value for SM matter to sub-solar masses for MNS. Moreover, the $\Lambda$-mass sequences for NSs and MNSs are very clearly separated, indicating they could be distinguished by their GW signatures \cite{Hippert:2021fch}.

%------------------------------------------------
\subsection{Mirror-Matter-Admixed Neutron Stars}
\label{sec:isolated-2comp}

Besides isolated NSs and MNSs, our model enables us to make predictions for a third class of objects, not addressed in \cite{Hippert:2021fch}: \(\mans\), compact objects akin to NSs and MNSs, but containing \textit{both} SM ($f/v=1$) and mirror matter ($f/v>1$) in their interior.
We will here consider the structure of \mans with an arbitrary fraction of mirror matter, which our two-fluid set up is tailored-made to handle. In particular, we set $p_\sm(\epsilon_\sm)$ to the SM EoS and $p_\dm(\epsilon_\dm)$ to the mirror matter EoS with a choice of $f/v$ in the 2-fluid structure equations [Eqs.~\eqref{eq:dp1dr} and \eqref{eq:dmdr}] to solve for \mans. We vary the fraction of mirror matter to SM through the choice of SM and mirror matter central energy densities. By then varying $f/v$ and these central densities, we are able to study how the astrophysical properties of \(\mans\) are affected by the Higgs VEV in the mirror sector.

Before proceeding, let us comment on how the analysis we carry out here differs from previous related work. Previous studies explored admixed stars with mirror matter, but they set $f/v=1$ i.e. they assumed that the SM EoS was identical to the mirror matter EoS \cite{Sandin:2008db,Ciarcelluti:2010ji,Kain:2021hpk}. Our study is quite different because we use first-principle lattice QCD results to scale the SM EoS with $f/v$, thus using a consistent EoS for both sectors. Most previous work fixed the fraction of dark matter to SM because the specific dark matter capture model studied predicted a certain fraction. The details of capture, however, depend sensitively on the precise nature of the DM self-interactions, the interactions between DM and SM matter, and unknown astrophysical details, such as the precise distribution of mirror matter in our galaxy. Lifting this restriction, the phase space of solutions of admixed stars is clearly two-dimensional, and the properties of \mans vary depending on the choice of fraction, as we will see below. 

%------------------------------------------------
\subsubsection{Full mass-radius region}
\label{subsubsection:full-mass}

The full mass-radius plane of \(\mans\) is shown in the left panel of Fig.~\ref{fig:route_phaseSpace} for $f/v = 5$ and various choices of central densities shown in the right panel of this figure. Pure SM NSs are shown with a dashed cyan line, while pure MNSs are shown with a green dashed line. Everything in between is a \mans, where we now clearly see that what used to be a one-dimensional mass-radius sequence for a NS or a MNS is now a two-dimensional mass-radius plane. In this figure, we fixed $f/v=5$ as a representative example, but we have checked that we find qualitatively similar results for other values of $f/v$.

How low of a mass can \mans have? SM NSs have a Chandrasekhar limit of $\sim 1.4 M_\odot$, 
but lower NS mass measurements have been claimed using X-ray observations~\cite{Strobel:1999vn, 2022NatAs.tmp..224D}.\footnote{These time-integrated measures have been found to be susceptible to systematic error (especially due to assumptions about the atmospheres) such that radii can differ by up to $\sim 50$\% \cite{2013ApJ...764..145C,2012MNRAS.423.1556S}, but we still adopt this measurement as an interesting benchmark.} 
One of the lowest SM NS observed is approximately $0.9 M_\odot$ (i.e.~$\sim 64\%$ of the Chandrasekhar limit), and this is shown with a cyan up triangle in Fig.~\ref{fig:route_phaseSpace}. 
Fixing the SM central density to this value and then increasing the central density of mirror matter leads to the solid red line. For illustration, let us now assume that the smallest MNS possible is also $64\%$ of the Chandrasekhar mass of that MNS sequence, shown with a green down triangle in Fig.~\ref{fig:route_phaseSpace}. Fixing the mirror matter central density, $\epsilon^\dm_c $, to $64\%$ of its Chandrasekhar mass and increasing the SM central density, $\epsilon^\sm_c $, leads to the orange line in the figure. The region in the $(\epsilon_c^\sm,\epsilon_c^\dm)$ below the red line and the orange line is shown with a light gray color in both panels.  

One of the most interesting features of the two-dimensional mass-radius plane is the possibility of \textit{ultimate twins}: stars with the same mass \textit{and} radius but different dark matter fraction. This occurs because the mapping from $(\epsilon_c^\mathrm{SM}, \epsilon_c^\mathrm{DM})$  to $(R_\mathrm{out}, M)$ is not bijective at low radii. 
Examples of such twins are shown with a filled blue circle and square, and an open green circle and square. As is clear from the right panel, these \mans have different ratios of mirror matter to SM matter central densities; the green cases are actually MNSs, while the blue cases are \mans{s}. Although not shown in the figure, there are actually an infinite number of ultimate twins, as one varies the central densities, with the pattern shown in the figure.

\begin{figure}[htb]
    \centering
    \includegraphics[width=\linewidth]{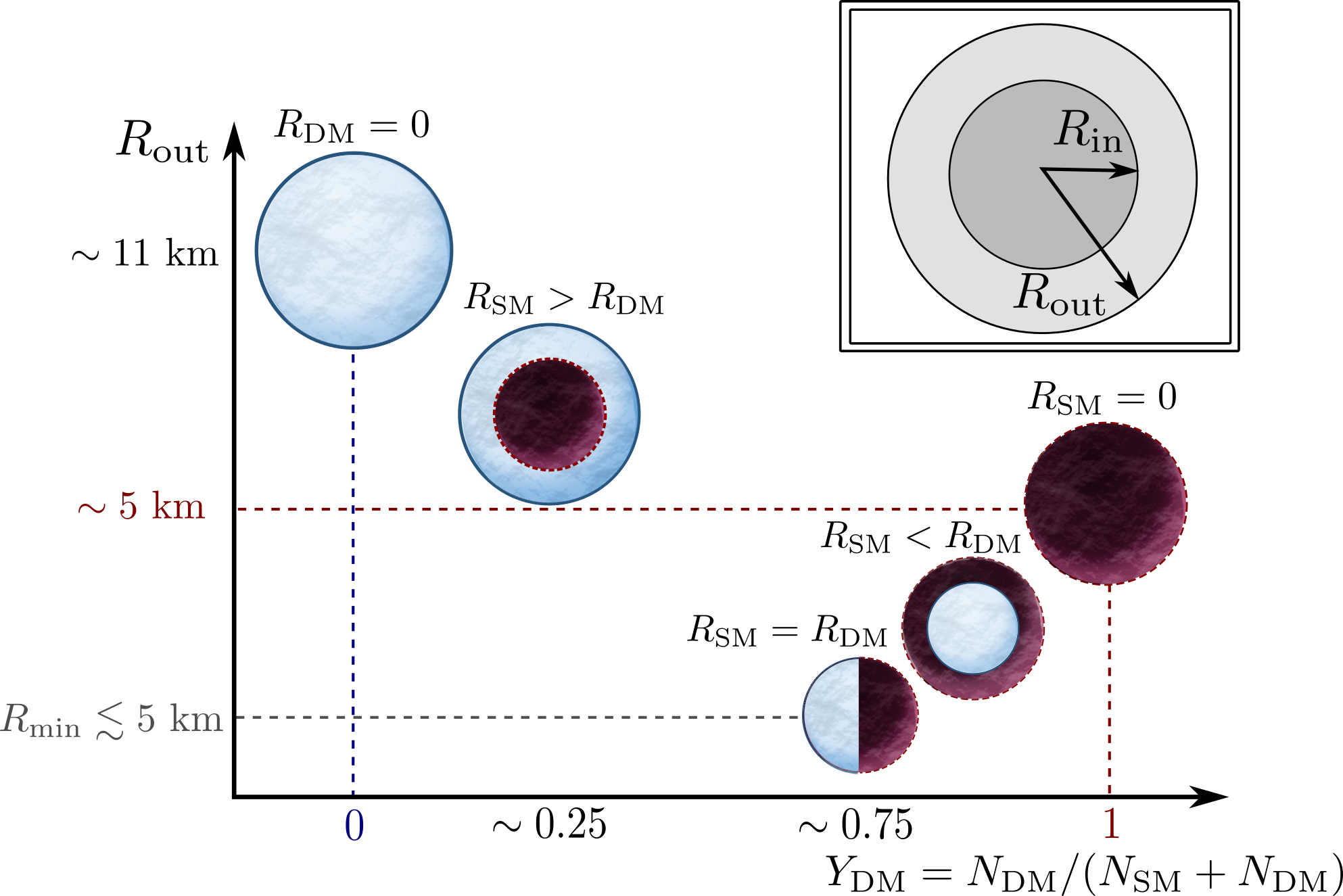} 
  \caption{ Typical outer radius $\Rout$ for stellar configurations with different mirror baryon fractions $Y_\dm \equiv N_\dm/(N_\sm+N_\dm)$. The inset illustrates the definition of the inner and outer radii, $\Rin$ and $\Rout$.
  The minimum value for the outer radius  corresponds to a configuration where both fluids occupy the entirety of the star, that is, when $\Rin = \Rout$, which happens at $Y_\dm \sim 0.75$. For $0< Y_\dm \lesssim 0.75$, we find NSs develop a DM core, $Y_\dm \gtrsim 0.75$, we find a DM halo instead.   
    }
    \label{fig:Radii}
\end{figure}
Let us pause at this point and discuss an unexpected feature of the results presented above: the allowed region of \(\manss\) is not bounded from the left by the \(\mns\) sequence (thick long-dashed green curves in Fig.~\ref{fig:route_phaseSpace}). Consider then what happens to \mans as we start with a SM NS and we begin to add mirror matter while keeping the total mass fixed, as shown schematically in the cartoon of Fig.~\ref{fig:Radii}. As one increases the DM fraction $Y_\dm\equiv N_\dm/(N_\sm+N_\dm)$ by increasing the mirror matter energy density, one can identify the following stages:
\begin{itemize}
    \item \textit{SM NSs}. Initially, $Y_\dm = 0$ and the star is purely made out of SM matter
    \item \textit{MANS with a DM core and a SM halo}. As we increase $\epsilon_c^\dm$ and thus $Y_\dm$, the mirror matter settles at the center of the star (because of the extra concentration of gravitational mass in the admixed region) and the star now has two radii: 
    one which contains some of the SM matter and all of the mirror matter ($R_{\rm in} = R_{\dm}$) 
    and 
    one that contains all of the SM and mirror matter in the star ($R_{\rm out} = R_{\sm}$). 
    The spherical region inside the inner radius, $r < R_{\rm in}$, will be called the core, and for these \mans{s}, it is where all of the mirror matter resides. 
    The shell region outside the inner radius but inside the outer radius, $R_{\rm out} > r >R_{\rm in}$, will be called the halo, and for these \mans, it is composed entirely of SM matter. 
    \item \textit{MANS with equal DM and SM matter}. As we increase $\epsilon_c^\dm$ and $Y_\dm$ further, the radius that contains the mirror matter grows, while the radius that contains the SM matter shrinks. Eventually, one reaches a critical value $Y_\dm=Y_\dm^*$ at which $R_{\rm in} = R_{\dm} = R_{\sm} = R_{\rm out}$ and the entire star is occupied by both fluids. This is when the \mans sequence reaches its minimum radius. 
    \item \textit{MANS with a SM core and a DM halo}. Increasing $\epsilon_c^\dm$ and $Y_\dm$ even further, the radius that contains SM matter is now inside the radius that contains the DM, so $R_{\sm} < R_{\dm}$, and therefore, the \mans now has a SM core ($R_{\rm in} = R_{\sm}$) and a DM halo ($R_{\rm out} = R_{\dm}$). This is the flip case to the \mans with a DM core and a SM halo. 
    \item \textit{MNSs}. Eventually, $\epsilon_c^\dm$ is increased enough that $Y_\dm$ tends to unity and the \mans becomes a MNS. Since the radius that contains the DM grows as we increase $\epsilon_c^\dm$, $R_{\rm out} = R_{\dm}$ while $R_{\rm in} = R_{\sm} =0$. In particular, note that the radius of the MNS is here \textit{larger} than the radius of the \mans with equal DM and SM matter.  
\end{itemize}
From this analysis, one can clearly see that the stellar radius $\Rout$ decreases for $Y_\dm<Y_\dm^*$ and it increases for $Y_\dm>Y_\dm^*$, until at $Y_\dm=Y_\dm^*$ the \mans has the minimum allowed value for $\Rout$. This critical value is the point at which the star is entirely filled with both fluids, so that the maximum possible concentration of mass is achieved. The sequence of stars with $Y_\dm = Y_\dm^*(M)$ therefore delimits the mass-radius region from the left in Fig.~\ref{fig:route_phaseSpace}. 

\begin{figure}[htb]
    \centering
    \begin{tabular}{c c c}
    \includegraphics[width=7cm]{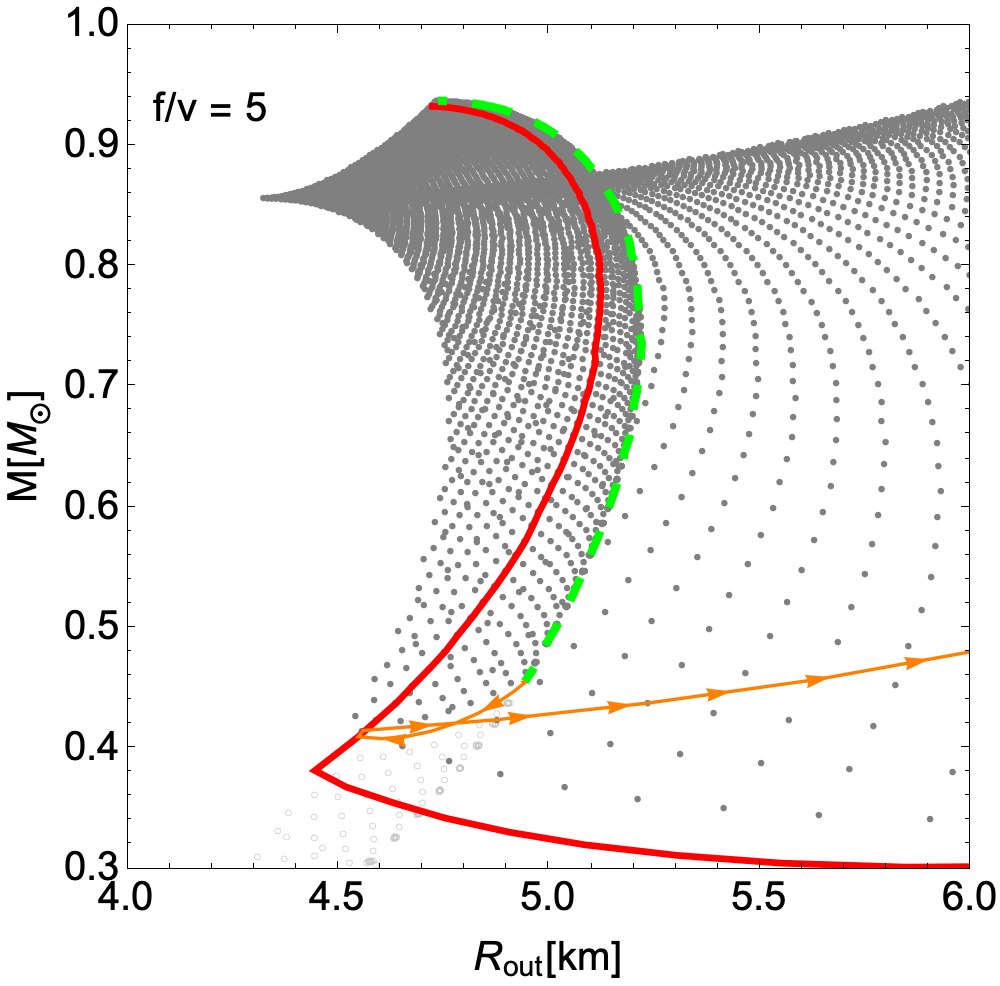}
    \\
    \includegraphics[width=7cm]{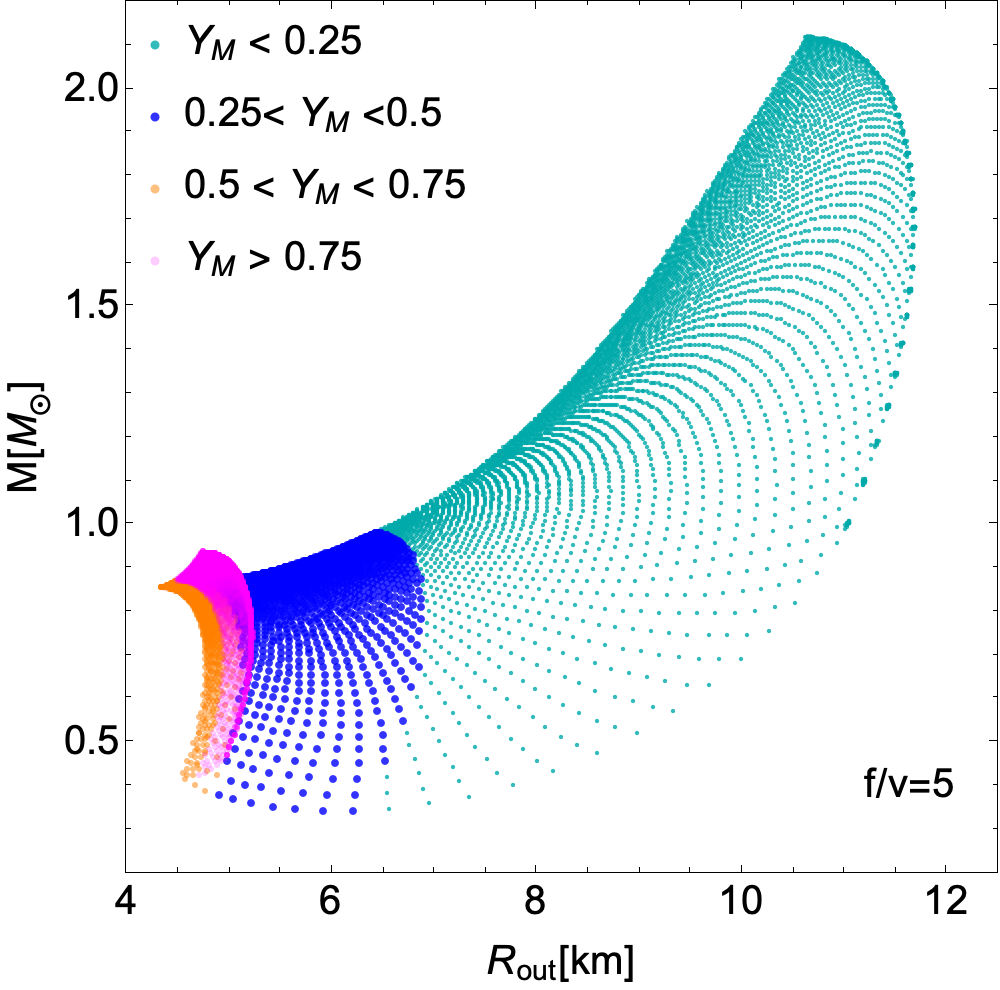}
    \end{tabular}
    \caption{Mass-radius regions, zoomed to the region inside which ultimate twins exist (top) and color-coded to indicate stars with different DM fraction (bottom), fixing $f/v = 5$. The arrows in the top panel indicate the direction in which the SM central energy density increases. Observe that that there is a region (between the left-most boundary and the MNS sequence) inside which different $Y_\dm$ leads to stars with the same mass and radius, which we have defined in this paper as ultimate twins.   
    }
    \label{fig:MR-fullregion-contour}
\end{figure}

A consequence of the results presented above is that the mirror-baryon fraction $Y_\dm$ is indeed zero at the right most boundary (at the SM NS sequence), but it is not unity at the left most boundary, and therefore, $Y_\dm$ must change \textit{non-monotonically} as the \mans radius changes near the \mns sequence. Because the radius is non-monotonic in $Y_\dm$, the region to the left of the \(\mns\) sequence (to the left of the thick long-dashed green curve in Fig.~\ref{fig:route_phaseSpace}) is composed of two overlapping areas with different values of $Y_\dm$. Let us investigate how these overlapping areas are connected . Since $Y_{\dm}$ is one-to-one in $\epsilon_c^{DM}$, one can focus on any constant $\epsilon_c^{\dm}$ curve, such as the orange curve in Fig.~\ref{fig:route_phaseSpace}. Starting at the intersection of this sequence with the MNS sequence (i.e.~the intersection of the thick dashed green curve and the orange curve, which also corresponds to the right-most star in Fig.~\ref{fig:Radii})), as $\epsilon_c^{\sm}$ increases, the mass-radius point moves left along the orange curve, as shown in the top panel of Fig.~\ref{fig:MR-fullregion-contour}, which is just a zoomed version of the left panel of Fig.~\ref{fig:route_phaseSpace}. After hitting the left boundary, where $\Rout = \Rin$ (the central star in Fig.~\ref{fig:Radii}), the mass-radius sequence turns around and begins to move right. When this happens, the region left to the pure \(\mns\) sequence becomes double-valued in $Y_{\dm}$. One can see this more clearly in the bottom panel of Fig.~\ref{fig:MR-fullregion-contour}, which shows the mass-radius plane again, with points of different $Y_\dm$ color-coded. This region between the MNS sequence and the left-most boundary is where the ultimate mass-radius twins live.

%------------------------------------------------
\subsubsection{\manss through Accretion of Dark Matter into Neutron Stars}

\begin{figure*}[htb]
    \centering
    \begin{tabular}{c c}
    \includegraphics[width=8cm]{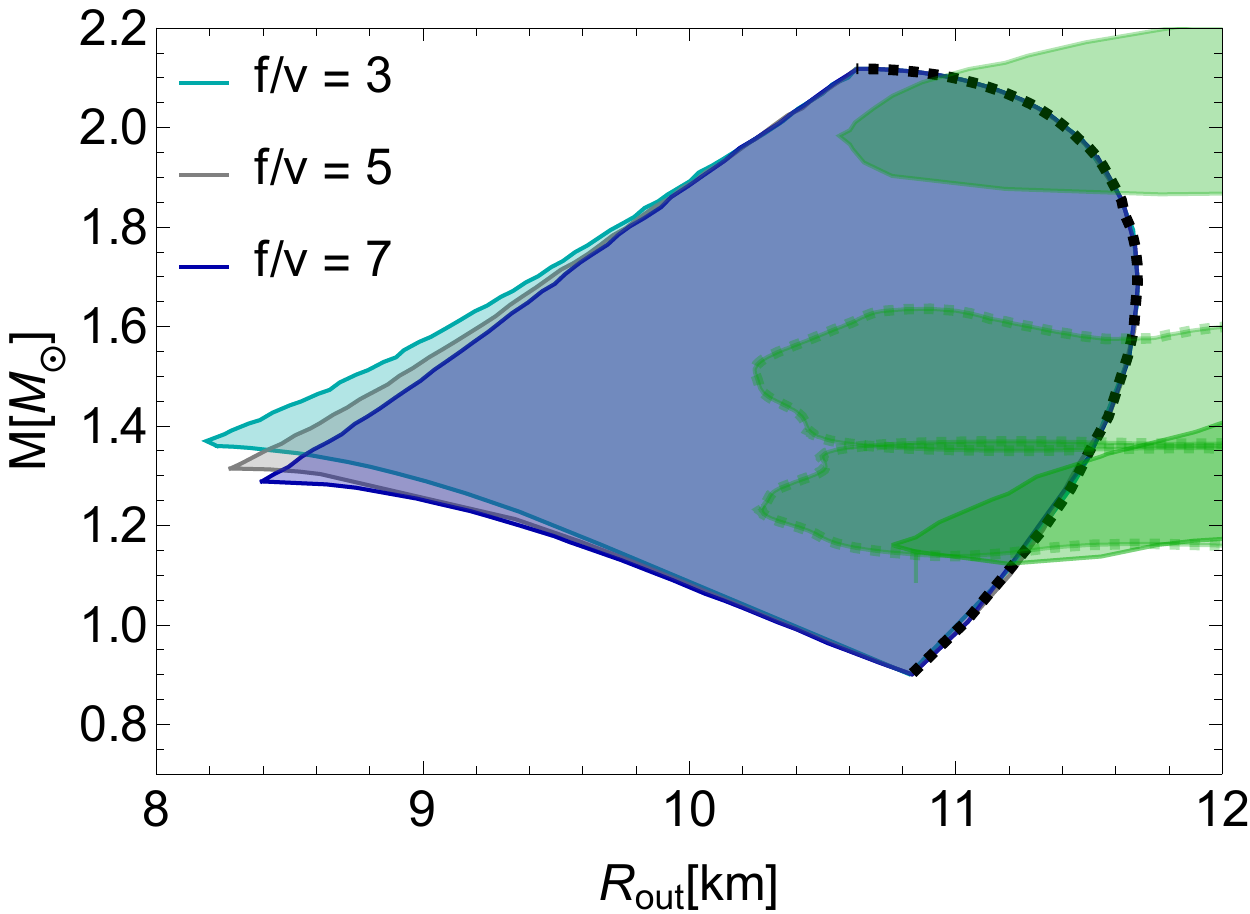}
    \includegraphics[width=8cm]{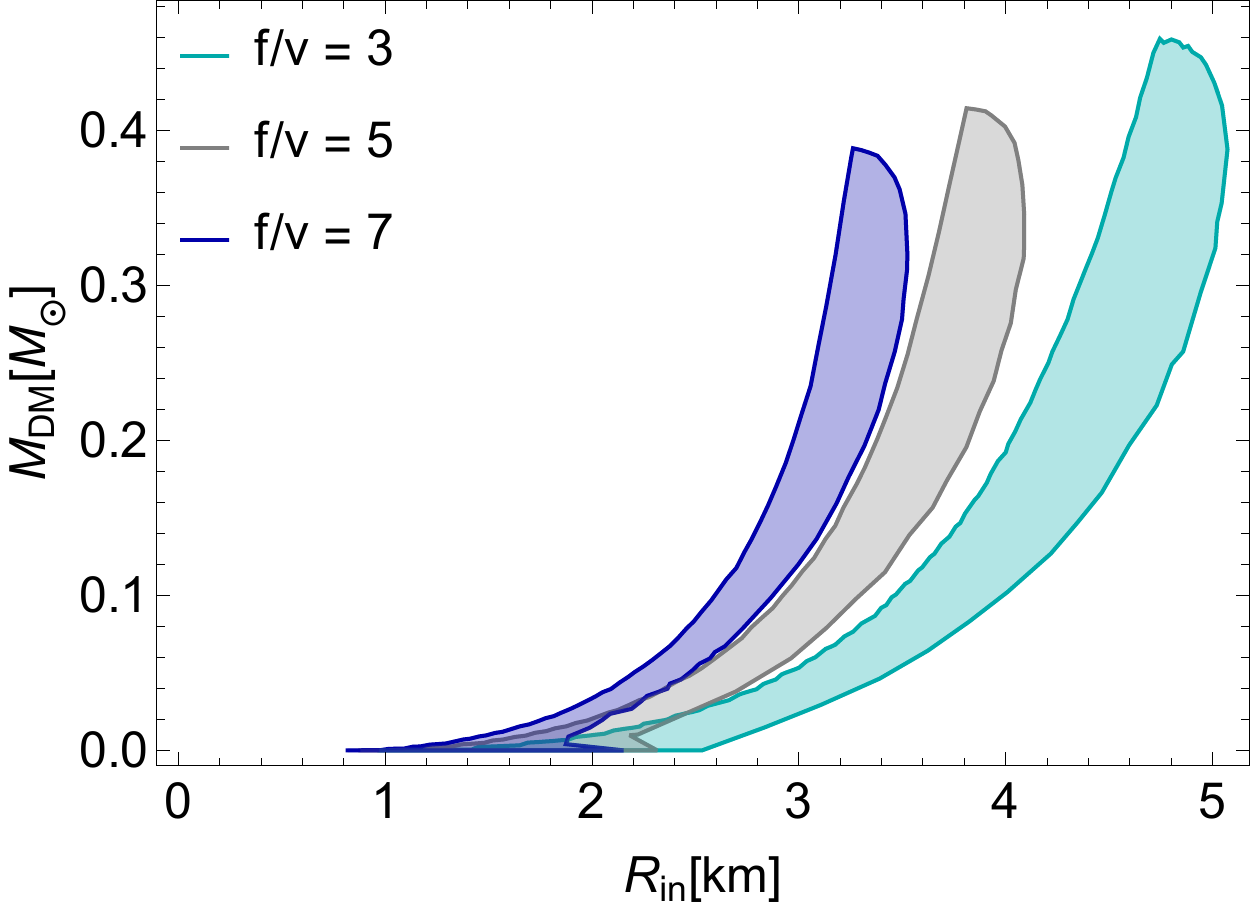} 
    \end{tabular}
    \caption{Left panel: Mass-radius region for \manss that contain a DM core (instead of a DM halo). The black dashed line is the isolated NS mass-radius sequence. Green shaded regions are constraints from LIGO and NICER \cite{Miller:2019cac,Riley:2019yda,Miller:2021qha,Riley:2021pdl,TheLIGOScientific:2017qsa,Abbott:2018exr,Abbott:2018wiz}. The overlapping shaded regions connected to the isolated NS sequence are the mass-radius regions for \manss with a DM halo for different values of $f/v$. Right panel: DM core mass as a function of its radius $R_{in}$ for different values of $f/v$. Observe that the mass-radius planes are mostly insensitive to the value of $f/v$, although the relation between the mass and radius of the DM core is not.  
    }
    \label{fig:MR-MANS}
\end{figure*}

Having explored the entire mass-radius plane that connects NSs and MNSs, we now explore predictions for \manss formed from accretion of dark matter into SM NS. Since SM NSs cannot have masses smaller than some percentage (64\% for this paper) of their respective Chandrasekhar masses ($0.9 M_\odot$ for this paper), the mass of the \mans after dark matter accretion is also constrained from below. Such \mans have a DM core instead of a DM halo, with $R_{\rm in} = R_{\dm} < R_{\sm} = R_{\rm out}$.

The left panel of Fig.~\ref{fig:MR-MANS} shows the mass-radius plane for these objects for various values of $f/v$. Observe that the mass-radius region for these \mans depends very weakly on $f/v$, and thus, on the mirror quark mass $m_q'=(f/v)\, m_q$. The right panel of Fig.~\ref{fig:MR-MANS} shows the mirror-matter mass $M_{\dm}$ as a function of the inner radius $\Rin$. While the inner radius $\Rin$ can change by a factor of $\sim 2$ with varying $f/v$, the maximum value of $M_{\dm}$ changes at most by $\sim 15\%$. 

Why is the mass-radius plane approximately universal with $f/v$ but the $M_\dm$-$\Rin$ plane is not? Let us consider the mass-radius plane first. Near the SM NS sequence, for small $M_\dm$, the DM core is very small (compare to the total radius of the star). The influence of such a core is then to only increase the total mass (by adding DM mass), while keeping the size of the inner radius small (as one can see on the right panel). As the $M_\dm$ increases further, the size of the inner core also increases and the $f/v$ universality begins to be lost because different $f/v$ lead to different sizes of DM cores (as seen again on the right panel). 

\begin{figure}[htb]
    \centering
    \includegraphics[clip=true,width=8cm]{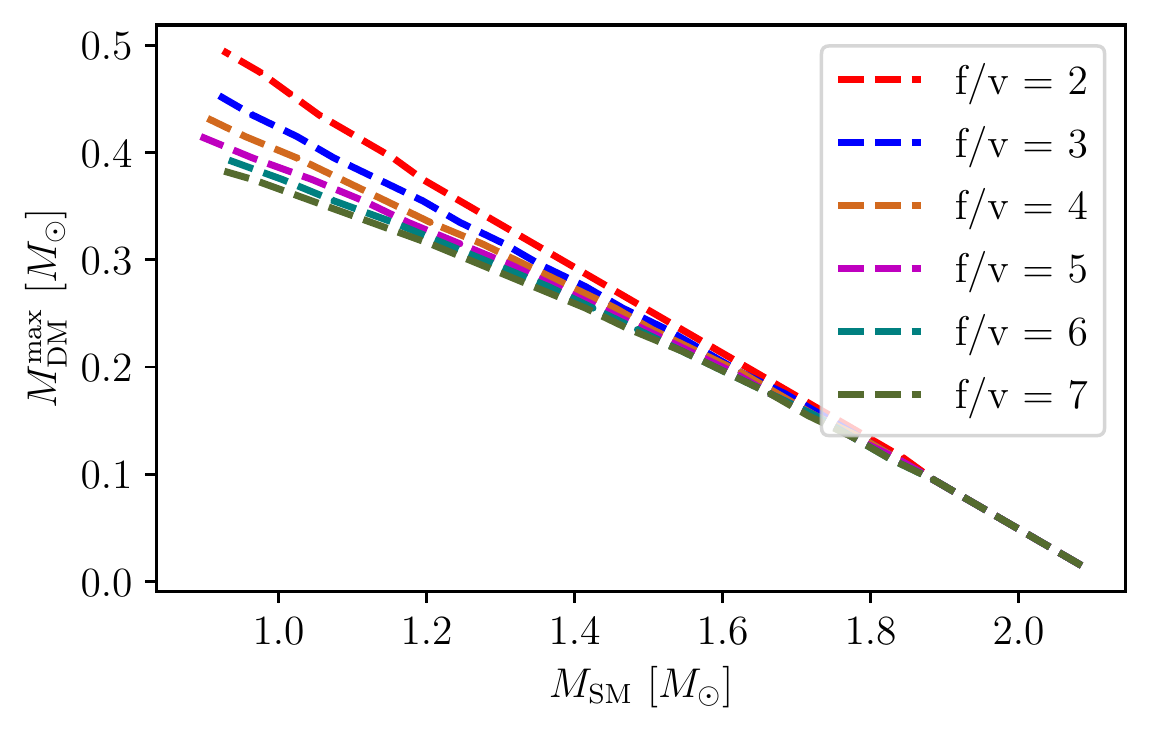}
    \caption{Maximum amount of mirror matter that can be added to a SM NS before it becomes unstable, as a function of its total SM mass.  Different values of f/v are considered for the mirror matter contribution.
    }
    \label{fig:MMvNS}
\end{figure}

A more detailed view of the maximum DM mass, $M_{\dm}^{\textrm{max}}$, as a function of the initial \bigsm mass $\Msm$, is presented in Fig.~\ref{fig:MMvNS}. Indeed, one observes that the curves for different $f/v$ lie very close to each other for small $M_{\dm}^{\textrm{max}}$, partly explaining the behavior observed in Fig.~\ref{fig:MR-MANS}. The dependence of $\Mdm^{\textrm{max}}$ on $f/v$ becomes weaker for decreasing $\Mdm$ and for increasing $f/v$, as $\Rin$ becomes larger. 

\begin{figure}[htb]
    \centering
    \includegraphics[clip=true,width=8cm]{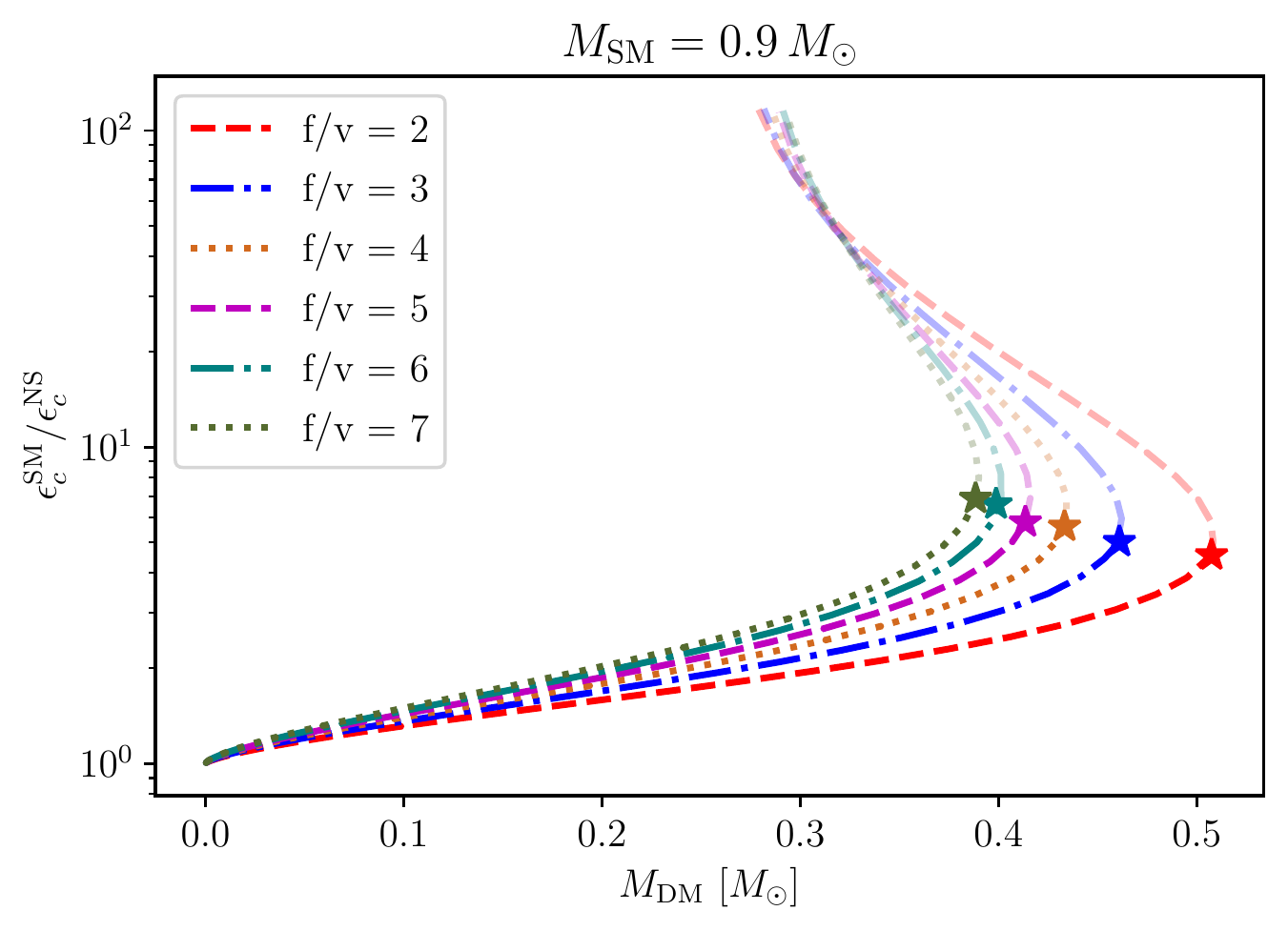}
    \caption{Fractional change in the central energy density of SM matter $\epsilon_c^\sm$, normalized to its value in the absence of dark matter $\epsilon_c^{\textrm{\tiny NS}}$, as a function of the admixed DM mass $M_{\dm}$. Here, we increase $M_\dm$ at a fixed SM mass of $M_{\textrm{SM}}=0.9\, M_\odot$. The point at which the star becomes radially unstable is marked by a star, with faint lines corresponding to unstable configurations. Different colors correspond to different values of $f/v$. 
    }
    \label{fig:epsvMM}
\end{figure}

One might na\"ively expect that adding DM to a NS would correspond to increasing $\epsilon_c^\dm$ at a fixed $\epsilon_c^\sm$, but this, however, is not the case. As \bigdm is added to a \ns, the increase in stellar mass leads to an increase in gravitational pull, which leads to an increase in the SM central density $\epsilon_c^\sm$. In fact, this change in $\epsilon_c^\sm$ can be used to investigate the effect of DM-admixture on the matter distribution and hydrostatic balance of the SM fluid. Results for the fractional change of $\epsilon_c^\sm$ as a function of $M_\dm$ are shown in   Fig.~\ref{fig:epsvMM}, where we fix SM mass at $M_\sm =  0.9 M_\odot$. The point at which the star becomes radially unstable is marked by a star and coincides with the maximum of $M_{\dm}$, as one would expect. Observe that the SM central density can increase by a factor of $\sim 5-7$ due to the admixture of DM. Surprisingly, this change tends to be smaller for smaller values of $f/v$, even though $M_{\dm}^{\textrm{max}}$ tends to be larger. 

%------------------------------------------------
\subsection{Compactness and Tidal deformability of \manss}

\begin{figure}[htb]
    \centering
    \includegraphics[clip=true,width=7cm]{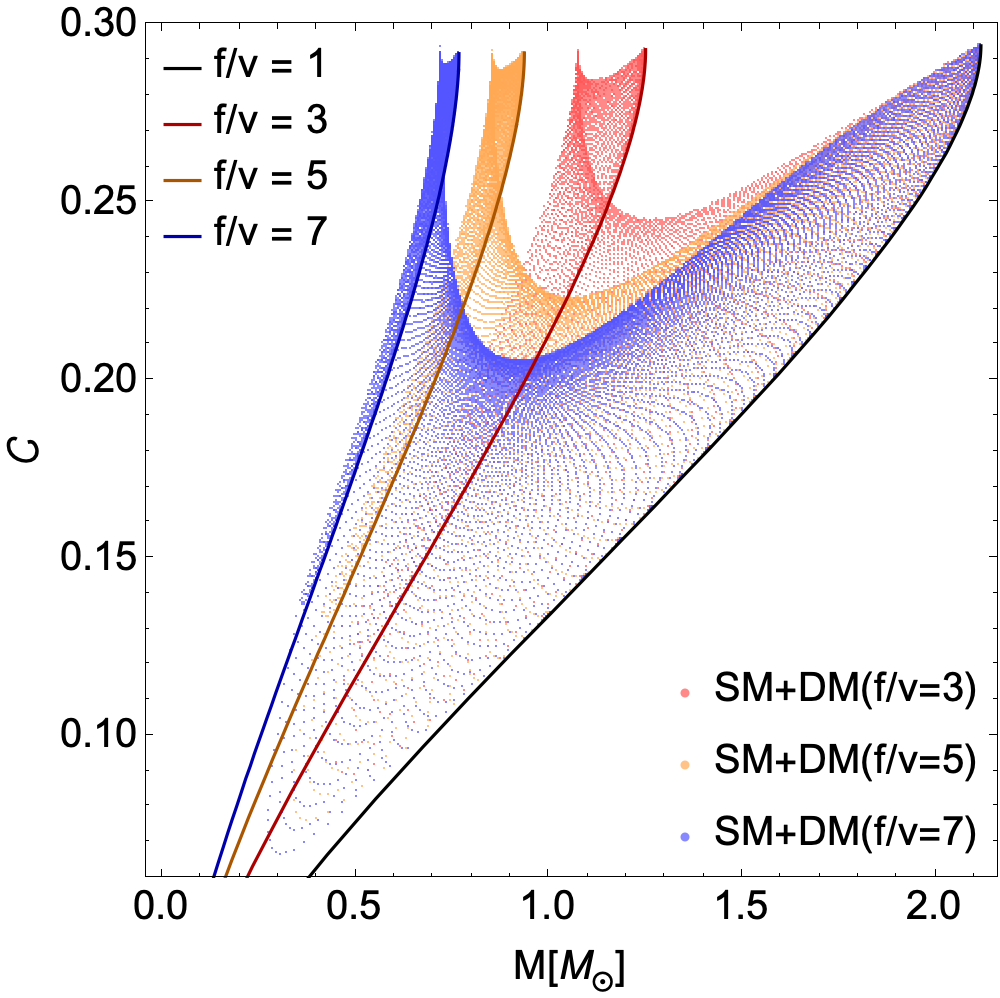} 
    \includegraphics[clip=true,width=7cm]{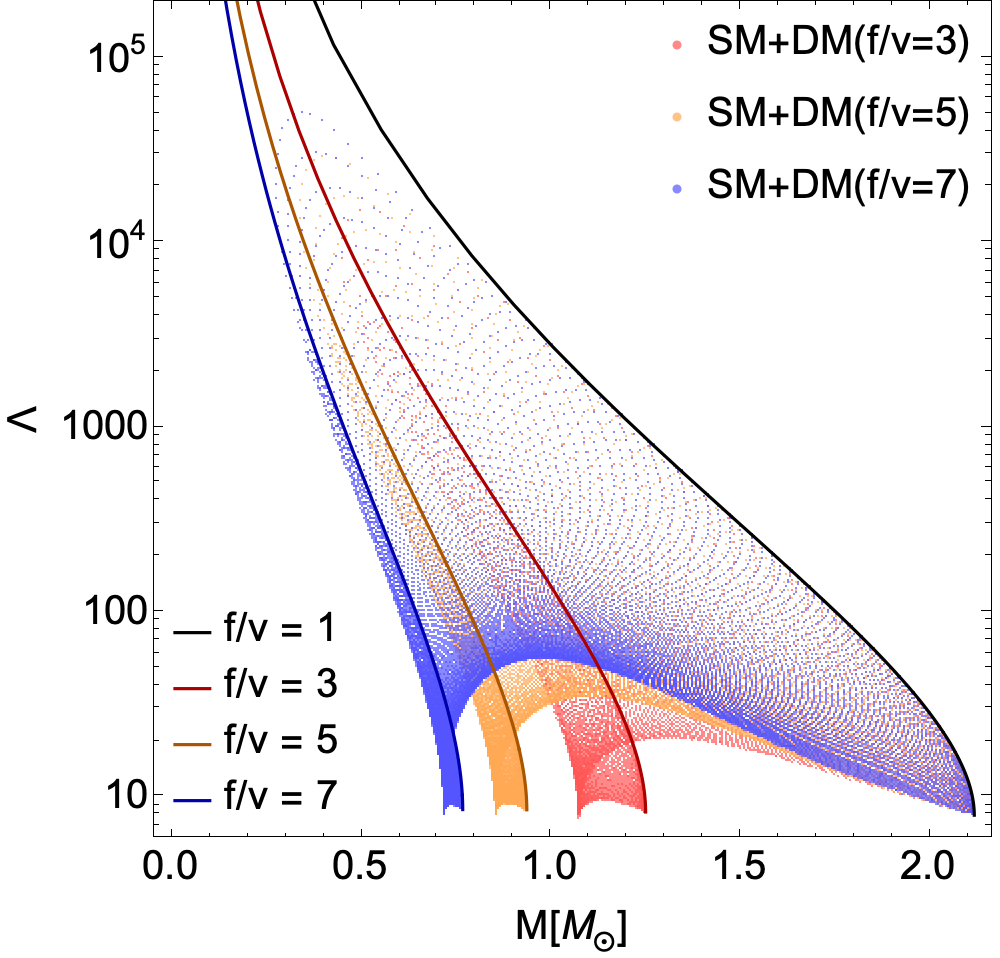}
    \caption{
    Compactness of \manss as a function of total mass (top) and tidal deformability as a function of total mass (bottom) for different values of $f/v$ (shown with different colors). The black solid line corresponds to SM NSs, while the colored solid lines correspond to MNSs with different values of $f/v$. Observe that the compactness of \manss is at most $0.3$, which is comparable to that of SM NSs. Such high-compactness stars also correspond to the maximum mass configurations, which occurs at lower masses for \manss than for SM NSs. Similarly, the tidal deformability of \manss can be extremely low, even at low masses.
    }
    \label{fig:compactness+lambda}
\end{figure}

As we have seen above, \mans{s} can have rather small radii, so one may expect their compactness $\mathcal{C} = M/R_{out}$ to also be large and their tidal deformability to be small. The top panel of Fig.~\ref{fig:compactness+lambda} shows the compactness as a function of the total mass $M$ for \manss with different values of $f/v$. Observe that, even with the addition of \bigdm, \mans~never reach a compactness above $C=0.3$, which is comparable to the maximum compactness of SM NSs. However, for MNSs and DM-rich \mans, this large value of compactness is achieved at much lighter masses. As the mass of a star increases, the star acquires a higher compactness, and therefore, it should become harder to deform and thus it possess a smaller tidal deformability (i.e.~a smaller $\Lambda$). The relation between $\Lambda$ and the total mass $M$ is shown on the middle panel of Fig.~\ref{fig:compactness+lambda}, which corroborates this expectation.

\begin{figure*}[htb]
    \centering
    \begin{tabular}{c c}
    
    \includegraphics[width=8cm]{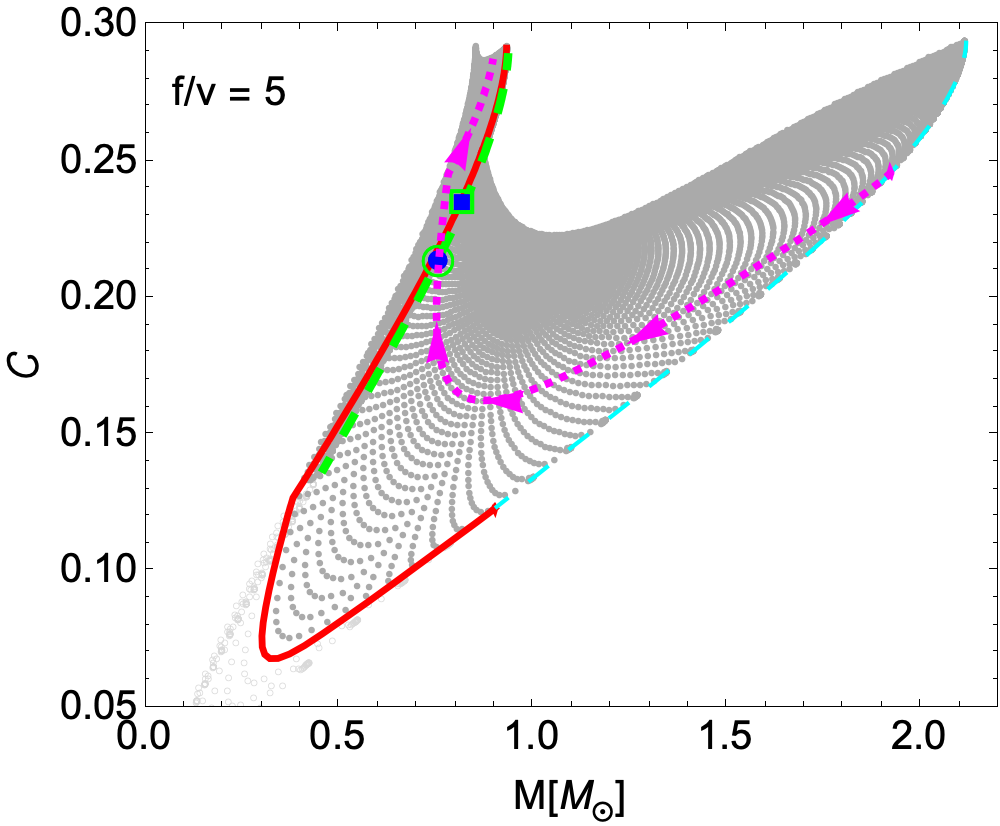} 
    \includegraphics[width=8cm]{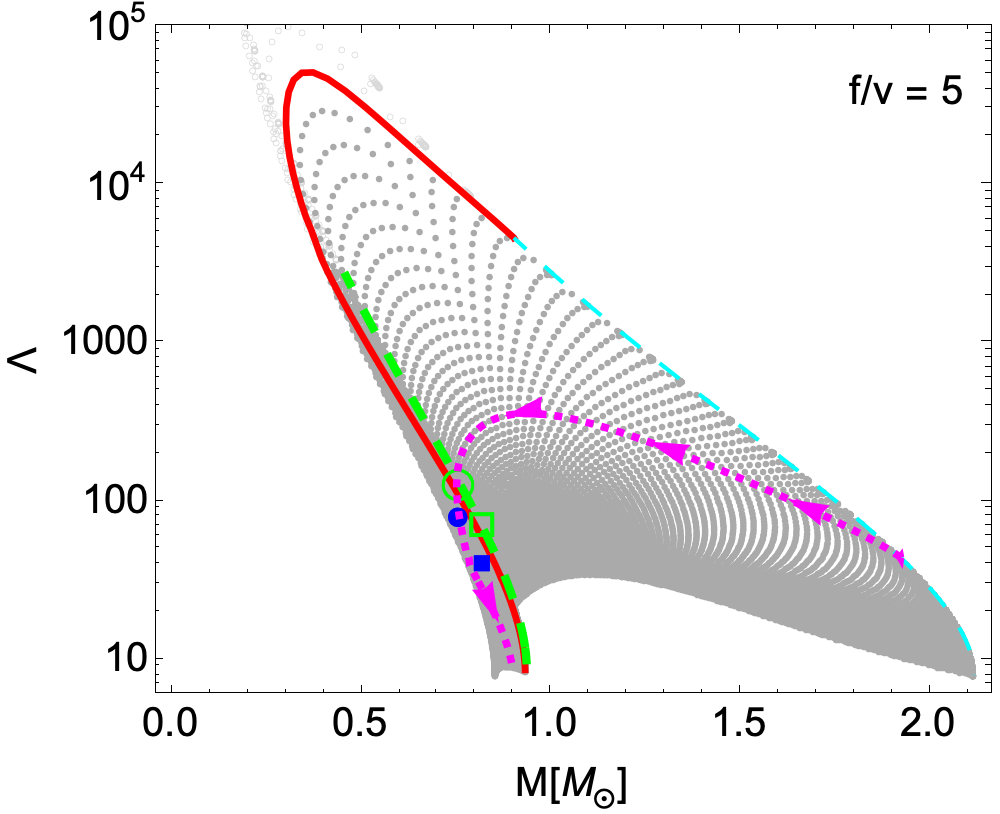}
    \end{tabular}
    \caption{Compactness versus total mass (left) and tidal deformability versus total mass (right) for the $f/v=5$ case. The color coding of curves is the same as in Fig.~\ref{fig:route_phaseSpace}, except that here we also introduce a new sequence (colored magenta), with arrows indicating the direction in which $\epsilon^\dm_c$ increases, which we discuss in the main text. Observe that the behavior of the compactness and the tidal deformability is not monotonic in the total mass. Observe also that ultimate twins have the same mass, radius and compactness, but different tidal deformabilities. }
    \label{fig:route_C_Lambd}
\end{figure*}

The tidal deformability of \manss, however, is not monotonic in the mass of the star, contrary to the intuition presented above. Figure~\ref{fig:route_C_Lambd} shows the compactness as a function of mass (left) and the tidal deformability as a function of mass (right) for the $f/v=5$ case. For concreteness, let us focus on a particular stellar sequence, depicted through the dotted magenta line. When the mirror-matter central density is zero ($\epsilon_c^\dm=0$), the \mans is simply a SM NS, shown at the point where the magenta line connects to the cyan line. As one increases $\epsilon_c^\dm$, the sequence moves to the left (as shown by the arrows on the magenta line), terminating in the ultimate twins shown with the green empty circle and blue filled circle. Observe that for $M \gtrsim 0.85 M_\odot$, as $\epsilon_c^\dm$ increases, the mass and compactness decrease, while the tidal deformability increases. The compactness decreases because, although the radius decreases, it does so slower than the mass. The tidal deformability increases because, as the star becomes lighter, it also becomes easier to deform, as expected. 

This behavior, however, changes drastically when $\epsilon_c^\dm$ has increased enough that  $M \lesssim 0.85 M_\odot$. Now, as the mass continues to decrease, the compactness \textit{increases}, while the tidal deformability \textit{decreases}. The compactness increases because the radius begins to decrease faster than the mass. In turn, the tidal deformability decreases because, as the compactness increases, the star becomes more difficult to deform. We conclude from this then that the tidal deformability is indeed tied to how easy or hard it is to deform a star, but the latter is connected to the compactness of the star, and not just its total mass, when one changes $\epsilon_\dm^c$ while fixing $\epsilon_\sm^c$.

Another interesting feature of Fig.~\ref{fig:route_C_Lambd} refers to the ultimate twins, which we recall are depicted with a green empty circle and a green filled square. As explained earlier, ultimate twins are \manss with the same mass and radius, but different internal DM composition and, thus, different $\epsilon_\dm^c$. As such, ultimate twins must also have the same compactness, which is corroborated by the left panel of Fig.~\ref{fig:route_C_Lambd}. The tidal deformabilities of these twins, however, is \textit{not} the same, as shown in the right panel of this figure. Indeed, the twin with a combination of mirror matter and SM matter (the blue filled square and filled circle in Figs.~\ref{fig:route_C_Lambd} and~\ref{fig:route_phaseSpace}) has a lower tidal deformability than the twin that only possesses mirror matter (the green empty circle and empty square in Figs.~\ref{fig:route_C_Lambd} and~\ref{fig:route_phaseSpace}). 

Why is this? The reason that for a two-fluid star the tidal deformability does not just scale with the compactness can be gleaned from the definition of $\Lambda$ in Eq.~\eqref{eq:yEq} and~\eqref{eq:Q}. As one can see from these equations, $\Lambda$ is a function of both the compactness of the star \textit{and} the variable $y_*$, evaluated at the surface of the star. From the differential equation that $y_*$ must satisfy, one can schematically argue that $y_*(R_{\rm out})$ will depend on the pressure and energy density inside the star, the averaged value of which can be related to the compactness of a star in a single-fluid model. For a two-fluid model, however, the average pressure and density do not just scale with the compactness. In fact, for a two-fluid model, the averaged value of the pressure is much larger for a two-fluid star than for a single-fluid star. A larger amount of interior pressure translates into a star that is more difficult to deform, and thus one with a lower tidal deformability. This is indeed what we find with ultimate twins, which always have one member that is very near the MNS sequence (essentially a single fluid star with $\epsilon^\sm_c \approx 0$) and one member in the \mans plane (a two fluid star with $\epsilon^\sm_c \neq 0 \neq \epsilon^\dm_c$). Therefore, the \mans member of the twin has a larger averaged pressure, is more difficult to deform, and thus has a smaller $\Lambda$.

The different tidal deformabilities suggests the possibility that GW observations of the inspiral of stellar-mass compact objects could be used to detect \mans{s}. We have already seen that the $\Lambda$-$M$ relation of \mans{s} lies on a plane instead of a line. Therefore, a set of measurements of $\Lambda$, each with sufficient accuracy, could allow us to reconstruct the $\Lambda$-$M$ relation and determine its dimensionality. To further determine whether there are ultimate twins in this plane, one would have to measure the mass, radius and tidal deformability of the same star, which may be achievable if one can infer not just $\Lambda$, which is the quadrupolar (electric-type) tidal deformability, but also the octopolar one. This could be achievable with third-generation detectors. 

\begin{figure}[htb]
    \centering
    \begin{tabular}{c c}
    \includegraphics[clip=true,width=7cm]{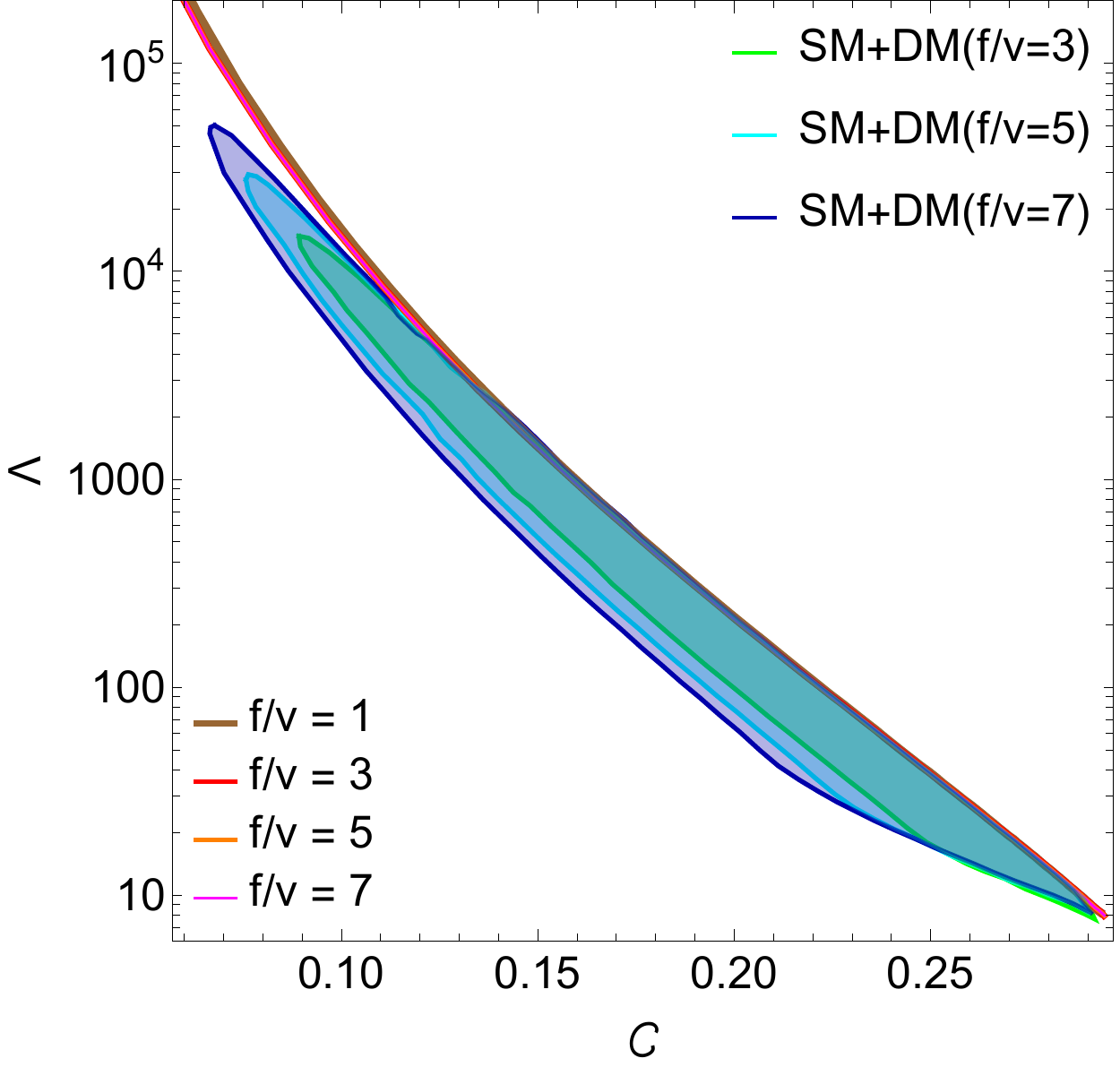}
    \end{tabular}
    \caption{Tidal deformability as a function of compactness for different values of $f/v$. Observe that while SM NSs and MNSs (solid black and color lines) all share an approximately insensitive $\Lambda$-$C$ relation, but \mans do not. Instead, the $\Lambda$-$C$ relation for \mans is not a one-dimensional curve, but rather a two-dimensional place, which can differ significantly from the relation for SM NSs and MNSs. }
    \label{fig:route_C_Lambd_FV}
\end{figure}

Inferences on the compactness, and thus the radius, of \mans through the $\Lambda$-C relation, however, must be done with great care. This is because,  although SM NSs share a nearly EoS insensitive relation between the tidal deformability and the compactness, this universality is lost in \mans, as shown in Fig.~\ref{fig:route_C_Lambd_FV}. Indeed, observe how SM NS sequences and MNS sequences all share a nearly identical $\Lambda$-$C$ relation, \manss do not, and in fact, they possess a two-dimensional $\Lambda$-$C$ relation. Therefore, the measurement of the tidal deformability of \manss cannot be associated with a single compactness (and thus a single radius). 

\begin{figure*}[htb]
    \centering
    \includegraphics[width=\linewidth]{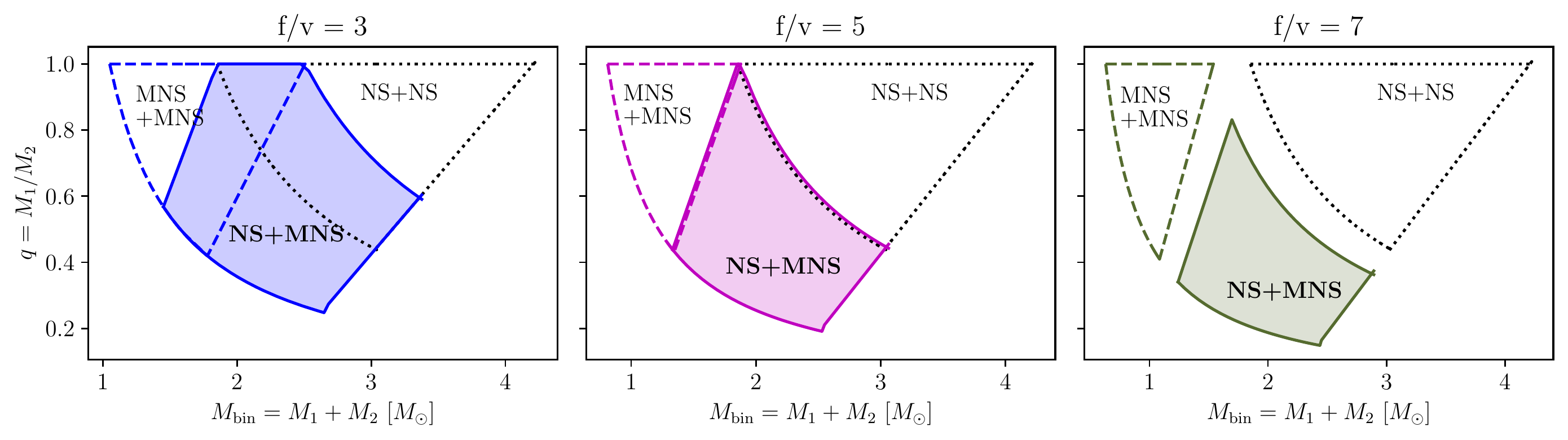}
    \caption{Mass ratio $q$ vs. total mass $M$ of binary systems for three different values of $f/v$. The shaded region enclosed by solid lines corresponds to binaries composed of one NS and one MNSs, while the region delimited by dashed lines corresponds to binaries of two MNSs. For reference, we also show the region corresponding to SM NS binaries, delimited by the black dotted lines.
    }
    \label{fig:q_vs_mtot}
\end{figure*}

%%%%%%%%%%%%%%%%%%%%%%%%%%%%%%%%%%%%%%%%%%%%%%%%%
\section{Inspiralling mixed binaries}
\label{sec:inspiral}

In this section, we present predictions for the observable signatures of inspiraling binaries of NSs and MNSs. We present results for the chirp mass and mass ratio between these stars and find that MNS-MNS and NS-NS inspirals produce unique GW signatures. 

%------------------------------------------------
\subsection{Mass Ratio Properties}
\label{sec:NS-MNS-insp}

Having determined the structure of isolated NSs, MNSs and \manss, we now investigate binary systems composed of NSs and MNS, specifically  NS-\(\mns\), and \(\mns\)-\(\mns\) binaries. 

The properties of these binaries are limited by the formation mechanisms for these objects. For instance, there are no known mechanisms to create NSs below a certain threshold $M^{(\sm)}_{\textrm{min}}$. 
To limit ourselves to viable NS and MNS masses, we implement minimum mass thresholds according to the estimated constraint $M\gtrsim M^{(\textrm{SM})}_{\textrm{min}}\simeq 0.9-1\,M_\odot$ for NSs \cite{Strobel:1999vn}. 
For definiteness, and to avoid the risk of being overly restrictive, we take  $M^{(\textrm{SM})}_{\textrm{min}} = 0.9\,M_\odot$.
To extrapolate this estimate to the mirror sector, we fix the ratio between $M_{\textrm{min}}$ and the Chandrasekhar mass of \mns{s} at $M^{(\textrm{SM})}_{\textrm{min}}/M_{\textrm{Ch.}} = 0.64$. 
In practice, this yields a scaling of $M_{\textrm{min}}$ with the (mirror) baryon mass $m_B'$:
\begin{equation}
\frac{M^{(\dm)}_{\textrm{min}}}{M^{(\textrm{SM})}_{\textrm{min}}} =\frac{M^{(\dm)}_{\textrm{Ch.}}}{M^{(SM)}_{\textrm{Ch.}}} = \left(\frac{m_B}{m_B'}\right)^2\,,
\label{eq:massthreshold}
\end{equation}
where $m_B$ is the SM baryon mass and $m_B'/m_B$ as a function of $f/v$ is taken from Ref.~\cite{Hippert:2021fch}. These are the same restrictions we placed on the minimum mass of \mans in Sec.~\ref{subsubsection:full-mass}. 

With this in mind, let us now look at the allowed mass ratio $q\equiv\Mone / \Mtwo$ and  total mass $\Mbin = \Mone + \Mtwo$ regions for NS-NS, NS-MNS and MNS-MNS binaries, using the convention $\Mone\leq \Mtwo$. Results are shown in Fig.~\ref{fig:q_vs_mtot}, where the shaded areas correspond to the predicted region for NS-MNS inspirals. The areas corresponding to NS-NS and MNS-MNS binaries are enclosed by dotted and dashed lines, respectively. Each panel corresponds to a different value of $f/v = 3, 5$ and $7$.
For all values of $f/v$, we find significant regions in the $q$-$\Mbin$ space that are exclusive of NS-MNS and MNS-MNS inspiral and mergers, indicating that these systems can be effectively distinguished from the binary masses alone. For $f/v\gtrsim 5$, we find no overlap between the three different regions. For example, one could have a $(0.5,0.5)M_\odot$ binary MNS inspiral, which would lead to a mass ratio of 1 and a total mass of $1 M_\odot$, which is not possible for a binary NS inspiral. 

\begin{figure*}[htb]
    \centering
    \includegraphics[width=\linewidth]{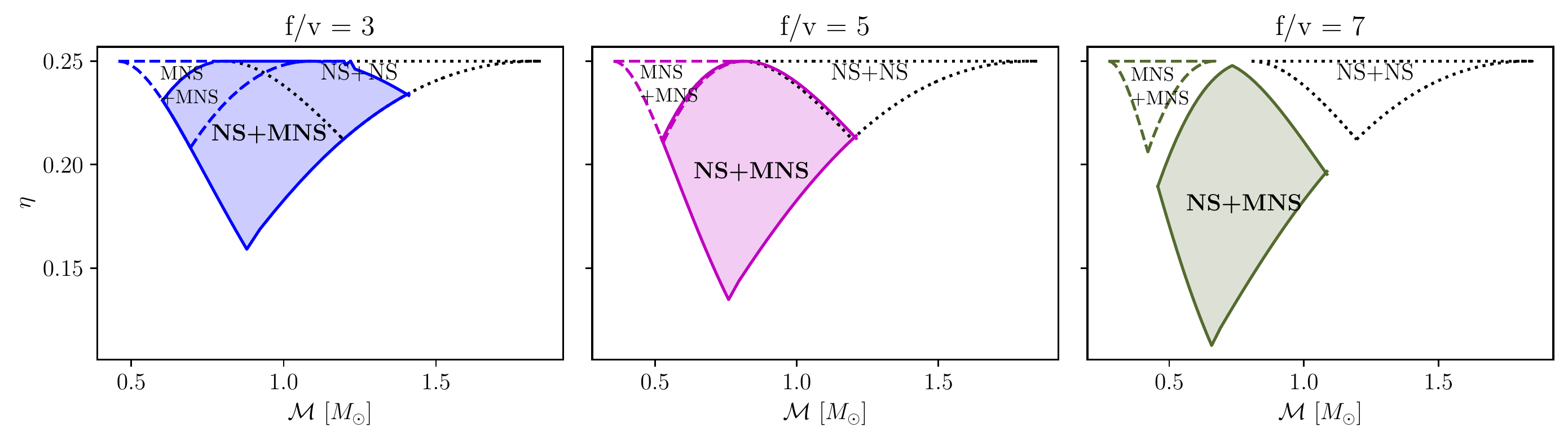}
    \caption{Symmetric mass ratio $\eta$ vs. chirp mass $\mathcal{M}$ of binary systems for three different values of $f/v$. The shaded region enclosed by solid lines corresponds to binaries composed of one NS and one MNSs, while the region delimited by dashed lines corresponds to binaries of two MNSs. For reference, we also show the region corresponding to SM NS binaries, delimited by the black dotted lines.}
    \label{fig:eta_vs_chirp}
\end{figure*}

%------------------------------------------------
\subsection{Gravitational Wave Signatures}

In the event of a NS-\mns~binary coalescence, \gw~radiation from the inspiral phase would provide the best candidate for signatures of its exotic nature. In a coalescence, the evolution of the orbital and, therefore, \gw, frequency is determined by the chirp mass $\mathcal{M} \equiv \eta^{3/5} (M_1+M_2)$ of the binary, where $\eta \equiv M_1 M_2/(M_1+M_2)^2$ is the symmetric mass ratio. The chirp mass and symmetric mass ratio regions relevant to NS-NS, NS-\mns~and \mns-\mns~inspirals are shown in Fig.~\ref{fig:eta_vs_chirp}. The low values of chirp mass for \mns-\mns~and NS-\mns~binaries indicate that the inspiral phase would be responsible for most of the detectable GW radiation emitted by such binaries. The merger of such compact objects would be outside the sensitivity band of second-generation ground-based detectors.

The inspiral of binaries containing \mns{s} would also be less loud than those composed of SM NSs. The chirp mass of the binary system determines the magnitude of the GW signal, since the latter scales with $\sim \mathcal{M}^{5/6}/D_L$, where $D_L$ is the luminosity distance to the source. As can be seen from the figure, the chirp mass of a NS-\mns~binary is $\mathcal{M}\sim 0.9\,M_\odot$ for $f/v=3$, and $\mathcal{M}\sim 0.7\,M_\odot$ for $f/v=7$, which is smaller than $\mathcal{M}\sim 1.2 M_\odot$ for NS-NS binaries. This suggests that GW signals from NS-\mns~inspirals should be $\sim 60 - 80\%$ that of NS-NS binaries, and therefore, although weaker, they should still be detectable. Such a detection would then allow us to distinguish a NS-MNS binary from a NS-NS binary just from the chirp mass measurement. 

GW emission in the late inspiral phase is also characterized by the tidal deformabilities of the inspiraling stars. The tidal deformabilities of the two stars, \(\loveOne\) and \(\loveTwo\), can be calculated by solving Eq.~\eqref{eq:yEq} with a given EoS \cite{Flanagan:2007ix}. Using current GW detectors, only a certain combination of the individual tidal deformabilities  \(\loveOne\) and \(\loveTwo\), the so-called ``chirp deformability''\cite{Favata:2013rwa,Yagi:2013baa}, can be measured. This implies that without any additional information, the individual tidal deformabilities are degenerate and cannot be solved for from a GW measurement. One way to break this degeneracy is to use EoS insensitive relations, including the binary Love relation~\cite{Yagi:2015pkc}, or some functional form for the EoS~\cite{LIGOScientific:2018cki}.  However, both approaches assume that the merging compact objects have a single EoS that can describe both stars.  In the case of an inspiraling NS-MNS, this is a poor assumption because each compact object has a different EoS, therefore completely breaking the universality of the binary Love relation.

\begin{figure}
    \centering
    \includegraphics[width=0.9\columnwidth]{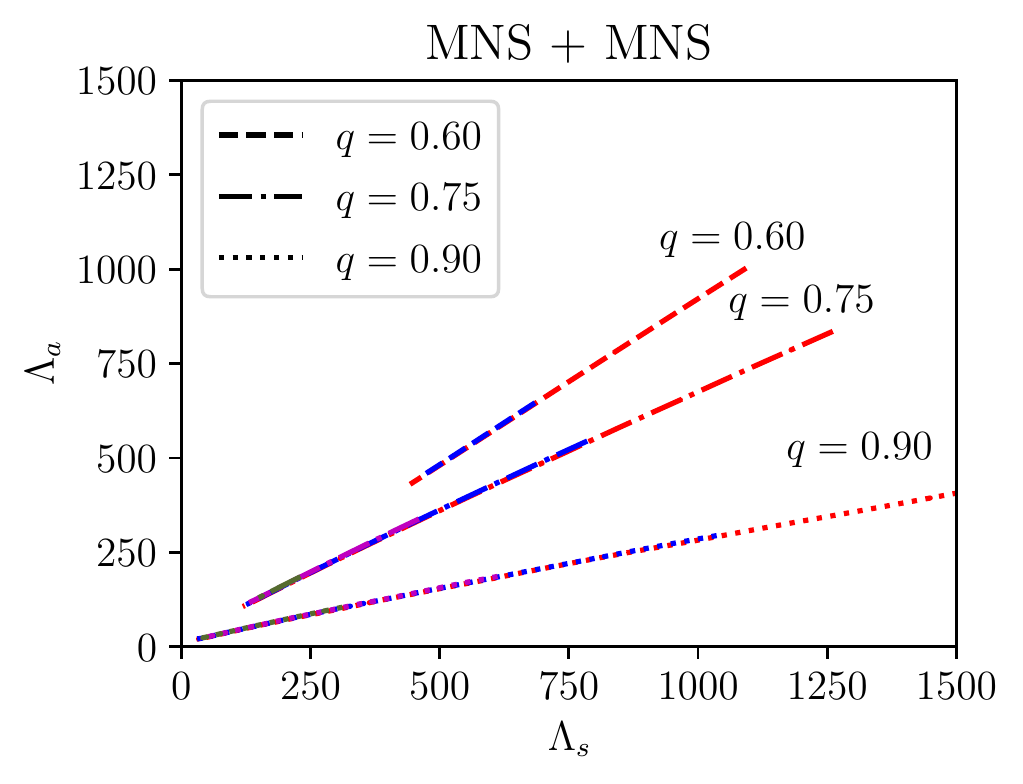} 
    \includegraphics[width=0.9\columnwidth]{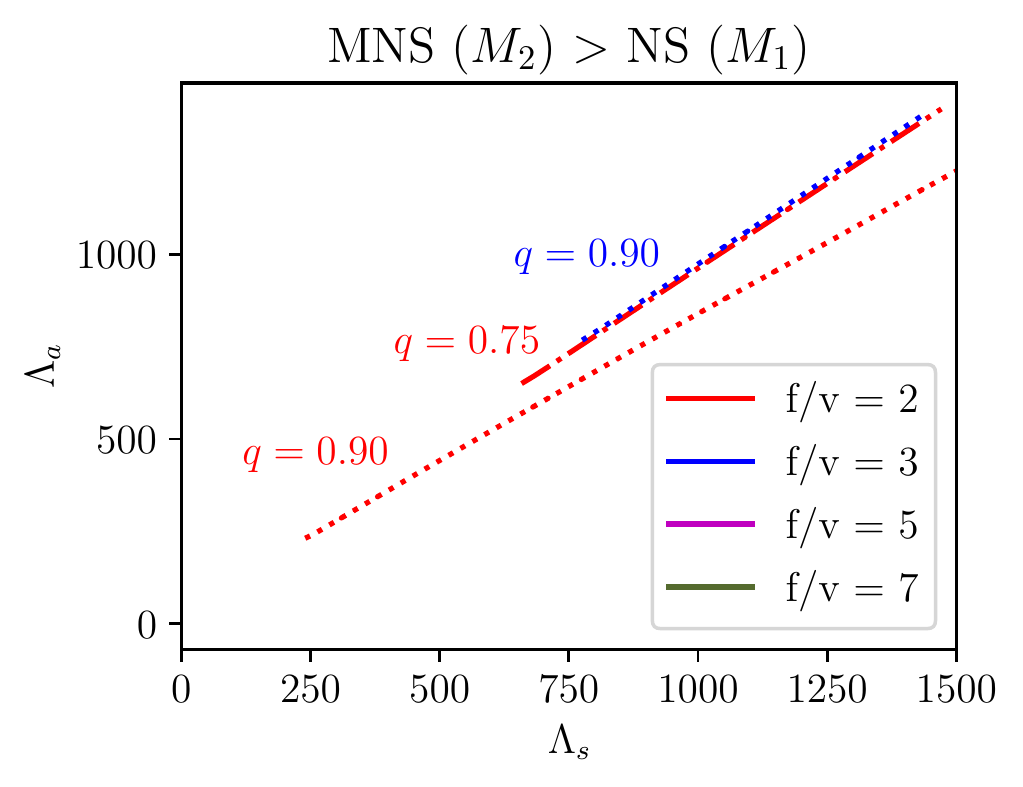}
    \includegraphics[width=0.95\columnwidth]{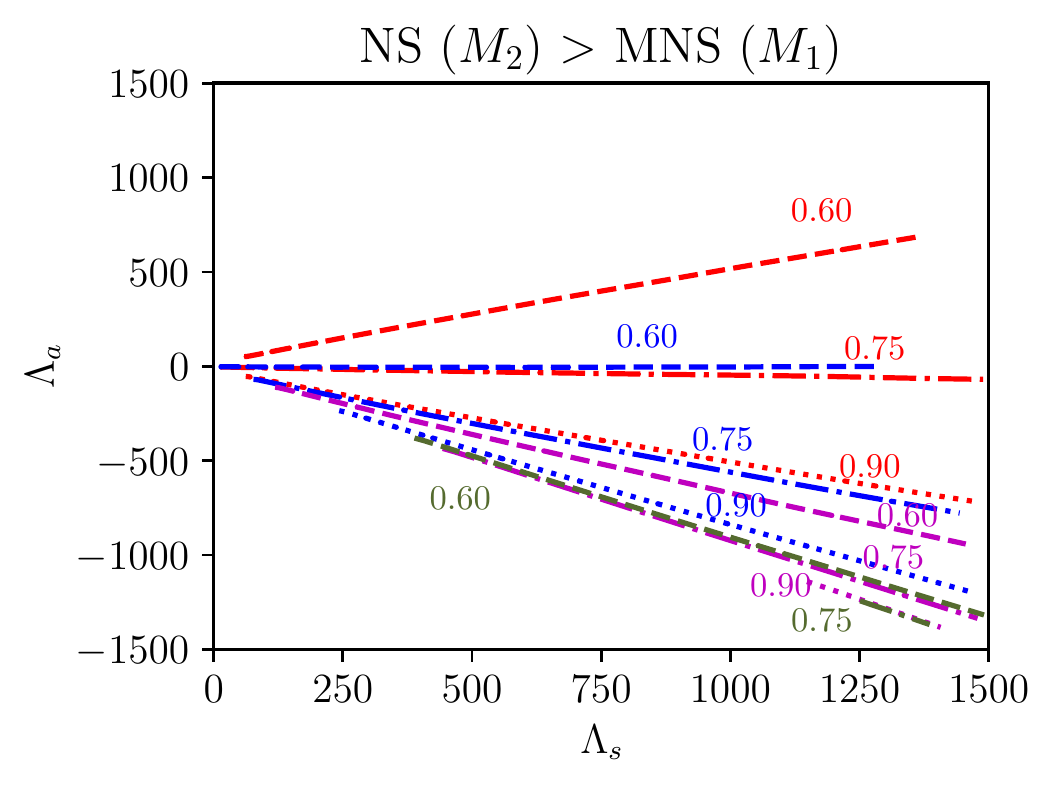}
    \caption{Binary Love relations for binaries of two MNSs (top), a MNS with a heavier NS companion (middle), and a MNS with a lighter NS companion (bottom). The asymmetric binary Love number $\loveA\equiv (\loveOne-\loveTwo)/2$ is shown as a function of the symmetric Love number $\loveS\equiv (\loveOne+\loveTwo)/2$. Different colors correspond to different values of $f/v$, while the distinct line styles represent different mass ratios  $q\equiv M_1/M_2$. Observe that, in the top panel, curves for fixed values of $q$, but different values of $f/v$, lie on top of each other, and therefore present EoS-insensitivity. In the middle panel, two of the curves are very close to each other, but they correspond to different mass ratios. In the bottom panel, the EoS-insensitivity is completely lost. 
    }
    \label{fig:binary-love}
\end{figure}

To explore the potential breakage of the binary Love relations, we plot the relations for \mns-\mns and \ns-\mns binaries in Fig.~\ref{fig:binary-love}. The top panel shows the binary Love relations of \mns-\mns binary systems with both stars with the same $f/v$. One can clearly see that the EoS insensitivity is preserved for different values of $f/v$, keeping $q$ fixed, which is unsurprising for \mns-\mns mergers because they both come from the same EOS. In that panel, NS-NS binaries are shown in black, and they lie on top of the other curves. The middle panel shows the binary Love relations of a lighter NS with a heavier MNS. We cannot calculate the relations for lower mass ratios and high $f/v$ because, from Fig.~\ref{fig:MRplotfront}, MNSs become lighter for higher $f/v$, and for some $q$ and $f/v$, there are no MNS that are heavy enough. In this case, we see that the EoS insensitivity is lost\footnote{The $f/v=2$ and $q=0.9$ curve is almost on top of the $f/v=3$ and $q=0.60$ curve, by pure coincidence.}, with the $f/v=2$ curve lying far from the $f/v = 3$ curve for fixed $q=0.90$. The bottom panel shows the binary Love relations for a higher NS with a lighter MNS. Because \mns{s} tend to be lighter than \ns, this is the most common scenario of a \mns-\ns binary. Observe, once more, that the EoS insensitivity is completely lost because, for the same mass ratio $q$ (dashed curves for example), there is no overlap for curves with different values of $f/v$. Observe also that, since \mns{s} are typically less deformable than \ns{s} (due to their higher compactness), the asymmetric Love number $\Lambda_a$ can be negative. This occurs for both $f/v>4$ and $q>0.60$, and tends toward $\Lambda_a = -\Lambda_s$ as $f/v$ and $q$ increase, making $\Lambda_1$ much smaller than $\Lambda_2$.

Before continuing with a discussion of the post-merger phase, let us end with a warning. Many GW searches today use a prior on the chirp mass and symmetric mass ratio that may exclude MNS stars all together. Indeed, only a fraction of the allowed $\eta$-$\mathcal{M}$ region overlaps with the region expected to be occupied by NS-NS binaries. Because of this, GW events from binary systems including one or two MNSs run the risk of being overlooked in a template-based coherent analysis. To detect such events, the priors on the chirp mass and the symmetric mass ratio should be updated to include the relevant mass ranges (i.e. much lower masses). Alternatively, one could rely on wavelet methods or excess power methods, both of which do not rely on templates, to detect such events, although such approaches may be less efficient than template-based searches. 

%%%%%%%%%%%%%%%%%%%%%%%%%%%%%%%%%%%%%%%%%%%%%%%%%
\section{Post-Merger Remnant}
\label{sec:post-merger}

In this section, we discuss the possible remnants of a MNS-NS collision. In particular, we present results for possible stable \mans remnants. We also comment on the formation of black holes when a stable stellar configuration is not achieved.

\begin{figure*}[htb]
    \centering
    \includegraphics[width=8cm]{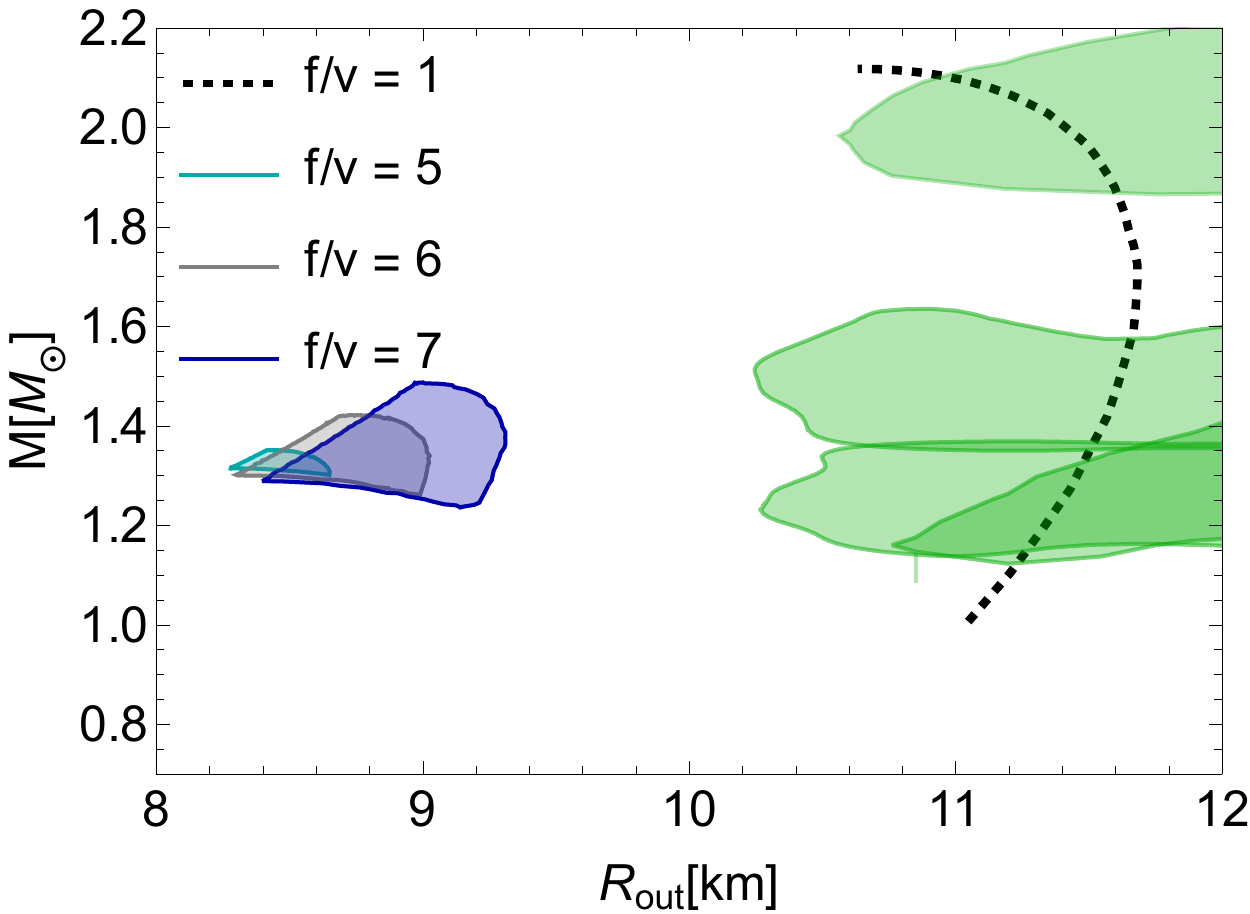}     
    \includegraphics[width=8cm]{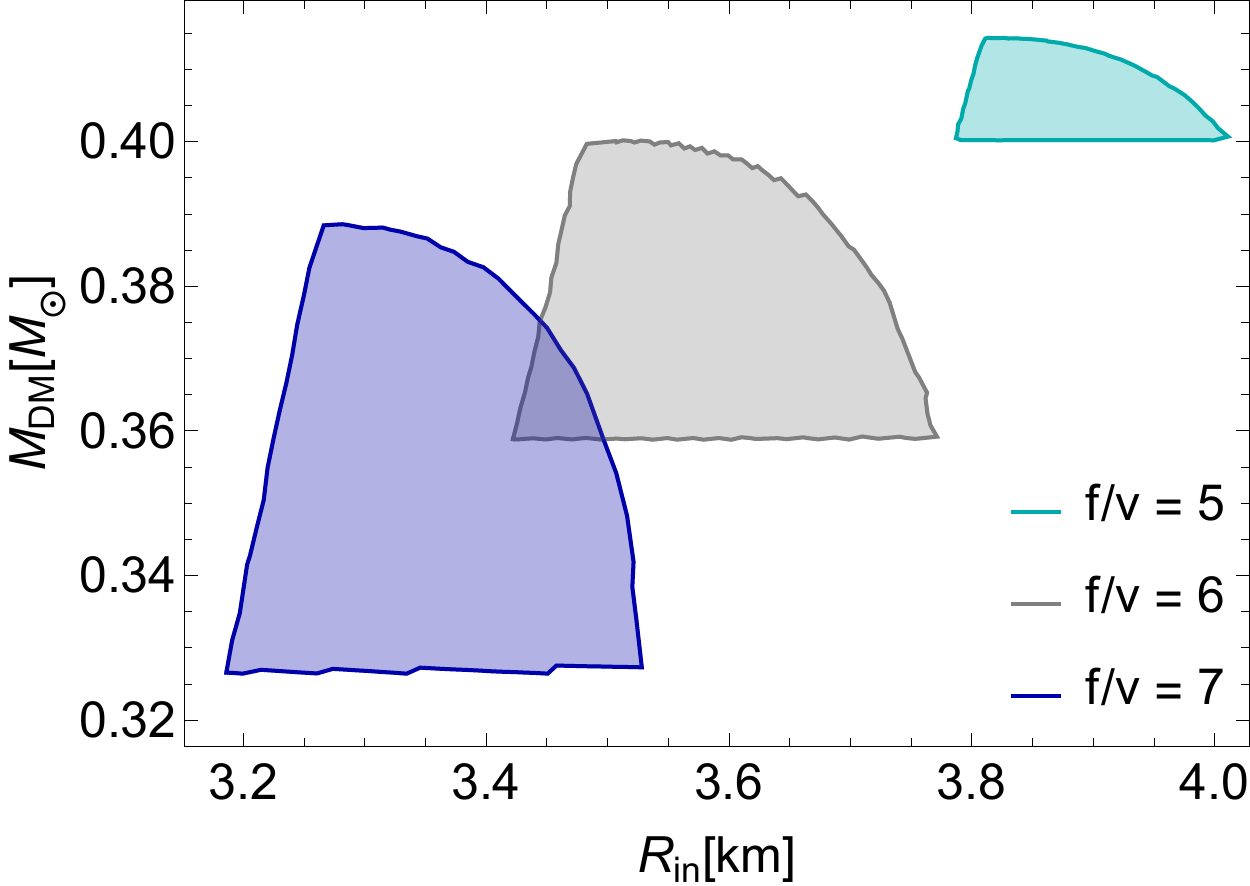}
    \caption{Left panel: Mass-radius region for stable remnants of a NS-MNS merger.  The black dashed line is the isolated NS mass-radius sequence. Green-shaded regions are constraints from LIGO and NICER \cite{Miller:2019cac,Riley:2019yda,Miller:2021qha,Riley:2021pdl,TheLIGOScientific:2017qsa,Abbott:2018exr,Abbott:2018wiz}. The overlapping shaded regions are the allowable mass-radius region for stable remnants, for different values of $f/v$. Right panel: The shaded regions correspond to the allowed dark-matter mass and radius for the mirror-matter core of the stable remnants of NS-MNS mergers. Stable remnants are not expected for $f/v\lesssim 4$.
    }
    \label{fig:cutoff}
\end{figure*}

\begin{figure}[htb]
    \centering
    \includegraphics[clip=true,width=8cm]{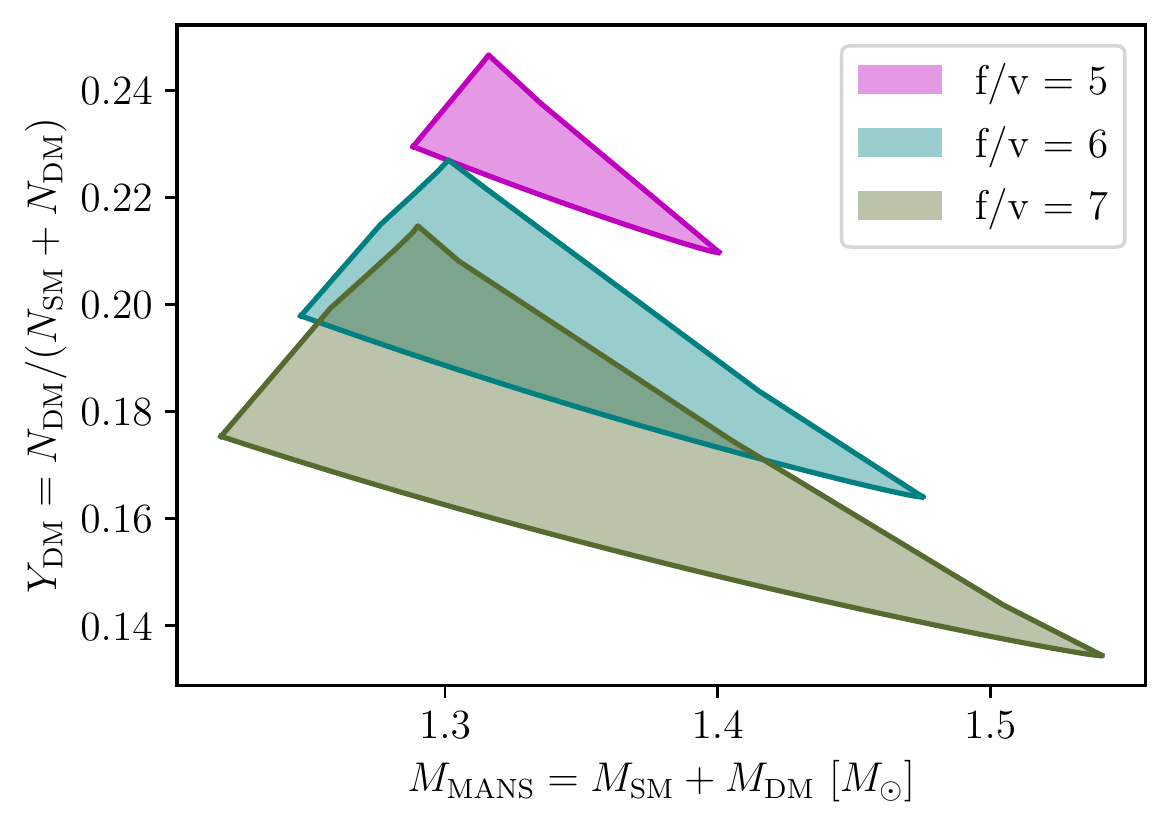}
    \caption{Mirror matter fraction $Y_\dm$ versus total mass $\Mmans$ for stable remnants of NS-MNS coalescences. Allowed regions for different $f/v$ are represented as shaded areas of different color. Stable remnants are not expected for $f/v\lesssim 4$.
    }
    \label{fig:remY-vs-M}
\end{figure}

%------------------------------------------------
\subsection{NS-MNS Collision}

A NS-MNS binary merger would be a unique event. 
Because there are no interactions via QCD, QED, or weak interactions that exist between mirror matter and the SM, one cannot directly change the SM EoS of NSs to add in the contribution of mirror matter.  In other words, collisions of NSs and \mnss~ would not be like NS-NS collisions at all because they would be like ``ghosts passing through each other.'' Rather, they are likely to continue to inspiral well passed the regime where their surfaces are overlapping. Once inside each other, the nested stars are likely to keep on spinning at different rotational angular frequencies, even as differential rotation within each fluid subsides,  because the only mechanism for angular-momentum transfer between the two fluids in the remnant is gravitational, including tidal effects.
While the observational signature of NS-MNS mergers can be interesting due to the above differences to NS-NS mergers, 
conclusions would have to rely on input from computer simulations, which are not yet available. We thus leave the discussion of the merger phase of NS-MNS binaries for future work.

%------------------------------------------------
\subsection{\mans~Remnants}
\label{sec:2comp-remnant}

After a sufficiently long time, transient effects should die out and a  NS-MNS merger would lead to a black hole or to a \mans, like the ones discussed in Sec.~\ref{sec:isolated-2comp}. 
In both cases, the resulting configurations are restricted by the relevant mass thresholds for each of the colliding stars. Let us then consider these end-states and discuss whether they would lead to interesting observational signatures. 

 Let us once again consider the viable star requirement ($M_{\rm NS} \geq 0.9 \,M_\odot$)  but this time we will also apply this constraint to \mns ($M_{\rm MNS} \geq 0.9 (m_B/m_B')^2 \,M_\odot$). To accomplish this, we follow the procedure in Sec.~\ref{sec:NS-MNS-insp} and use Eq.~\eqref{eq:massthreshold}.  
For simplicity, we neglect the effects of rotation on the stellar structure. 
We also neglect possible  (SM and DM) particle number losses during the merger.
The resulting range for \mans~remnants is shown in the left panel of Fig.~\ref{fig:cutoff}. 
The right panel of this figure shows the DM mass $M_\dm$ versus inner radius $\Rin$ for values of $f/v$ of the stable remnants.   
The effect of the lower-mass threshold  $M_{\textrm{min}}$ on possible stable post-merger remnants is noticeable. 
In combination with radial stability constraints (see Sec.~\ref{sec:stability}), the minimal mass thresholds lead to a tight region in $\Mmans$--$\Rout$ inside which \mans remnants can exist and only if $f/v\gtrsim 5$.
When $f/v\lesssim 4$, on the other hand, all remnants either violate the minimum threshold condition or are radially unstable. Consequently, NS-MNS binaries in  the left panels of Fig.~\ref{fig:q_vs_mtot} and Fig.~\ref{fig:eta_vs_chirp}, corresponding to a dark sector with f/v=3, will all collapse to black holes after coalescence.

Figure~\ref{fig:remY-vs-M} shows the corresponding ranges in mirror baryon fraction $Y_\dm =N_\dm/(N_\sm+N_\dm)$ and total mass $\Mmans$ for \mans~remnants of NS/MNS mergers, when $f/v = 3, 5$ and $7$. The mass of these objects is tightly limited to the regime $(1.2 - 1.55)\,M_\odot$. The DM fraction $Y_\dm$ is also on a tight range,  from $\sim 15 \%$ to $\sim 25\%$. 
Observe that the mirror-matter fraction $Y_\dm$ 
is anti-correlated with the total mass of the star, which we can understand intuitively as follows. 
Increasing the amount of DM at the center of the star leads to stronger gravitational forces. That additional force must be balanced by larger pressure gradients, which means the pressure in the \(\mans\) will drop faster in the radial direction. Because the energy density is a monotonic function of pressure, the energy density will also drop faster, leading to a smaller total mass.  Another consequence of Figure~\ref{fig:remY-vs-M} is that all \mans produced from a \mns-\ns merger leading to a stable remnant are of the form of a DM core with a SM halo.  

%------------------------------------------------
\subsection{Black-Hole Remnants}\label{sec:post_merger_BHremnants}

If a stable \mans~cannot form, the system will eventually collapse to a black hole. Because of the lower mass thresholds on the colliding stars, this turns out to be the most likely scenario.
Black-hole remnants from MNS-NS coalescences, however, would be very light, lying in the mass range $\sim (1.3,3.7)\,M_\odot$. Black holes from MNS-MNS coalescences would be even lighter, in the interval $\sim (0.8,3.2)\,M_\odot$. 
To explain how we come to these numbers, we look at the minimum mass $M_{min}$ from $f/v=4$ that  produces no stable remnants. Then, $M_{min}^{f/v=4}\sim 0.4 M_\odot$ and the $M_{min}^\sm=0.9 M_\odot$ so our total minimum black hole mass from a coalescence of a SM NS and a MNS with $f/v=4$ is then $M_{min}^{BH}\sim1.3 M_\odot$.  Of course, for larger f/v then smaller  $M_{min}$ are possible. For instance, $M_{min}^{f/v=7}=0.3 M_\odot$.  However, the coalescence of this MNS with a  $M_{min}^\sm=0.9 M_\odot$ NS leads to a stable configuration (as shown in Fig.\ \ref{fig:cutoff}) so it would not produce a black hole. The maximum values of a black hole mass for a MNS-NS coalescence comes from taking the maximum mass for $f/v=2$  and adding that to the maximum mass of the standard model sequence such that $M_{max}^{BH}=1.6+2.1=3.7 M_\odot$. In Sec.\ \ref{sec:detect} we will focus specifically on the difference between $f/v<5$ and $f/v\geq 5$ for two reasons: i.) $f/v\geq 5$ obtain stable configurations whereas $f/v<5$ only produce black hole remnants ii.) $f/v=5$ also is the smallest $f/v$ that provides no overlap in the $\eta$ vs $\mathcal{M}$ relation shown in Fig.\ \ref{fig:eta_vs_chirp}.  Thus, for much of our following calculations, it is important to know that $M_{min}^{f/v=5}\sim 0.4$ and $M_{max}^{f/v=5}\sim 0.9$.

Such a light range of remnant black hole masses would have important implications for the GWs emitted post-merger. 
In particular, the GW ringdown, generated as the remnant settles to its final stable configuration, would occur at very high frequencies. The GW frequency of the dominant fundamental ringdown mode scales as $f_{\rm GW} \sim 0.44/(2 \pi M_{\rm R})$, where $M_{\rm R}$ is the mass of the remnant~\cite{Berti:2009kk,Yunes:2022ldq}. For a normal black-hole binary merger with component masses $m_1 = m_2 = 5 M_\odot$, one then expects $f_{\rm GW} \sim 1400$ Hz. However, for a MNS-NS coalescence (with $m_1 = m_2 = 3 M_\odot$), the black hole remnant would ring at $f_{\rm GW} \sim 2300$ Hz, while for a MNS-MNS coalescence (with $m_1 = m_2 = 1 M_\odot$), the black hole remnant would ring at $f_{\rm GW} \sim 14,000$ Hz. This means that while the ringdown of a MNS-NS coalescence may be observable, it will be incredibly hard to observe the ringdown of a MNS-MNS collision. Moreover, if somehow only the ringdown were observed, then it would be impossible to distinguish this small-mass black hole remnant from a primordial black hole.

%%%%%%%%%%%%%%%%%%%%%%%%%%%%%%%%%%%%%%%%%%%%%%%%%%%%%%%%%%%%%%
\section{Observational prospects}\label{sec:detect}

Because of the unique signatures for  binary MNSs mergers and NS-MNS mergers, we find that there are multiple ways that one could detect the presence of mirror matter within NSs. Here we will list all the unique signatures that underlie various scenarios, summarizing several discussions presented earlier in this paper.

\begin{itemize}
    \item {\bf Scenario 1 - binary MNS inspiral and merger with $f/v\geq 5$}: Both objects are dark (no electromagnetic signatures) and the symmetric mass ratio and chirp mass is entirely distinguishable from that of SM NS binaries. The tidal deformabilities are orders of magnitude smaller than that of SM NSs, and the remnant is also dark, but may produce a more massive MNS or a black hole. The black hole remnant would be very light and certainly lower than the maximum mass of a neutron star (black hole masses starting at $M\sim 0.8 M_\odot$ for $f/v=7$).  
    \item {\bf Scenario 2 - binary MNS inspiral and merger with $f/v< 5$}: Since both objects are dark (no electromagnetic signatures), we need to rely on GW measurements and need to look into the parameter space of such measurements. For the parameter space of mass ratio and $\Mbin$ (or $\eta$ and $\mathcal{M}$), there is a small overlap between SM NS binaries and MNS binaries. Therefore, for most binary systems, a MNS binary can be distinguished from a SM NS binary, and evidence of DM can be found once we detect a binary system in the MNS-MNS region.
    Tidal deformabilities are approximately one order of magnitude smaller than that of SM NSs, and the remnant is dark, but may produce a more massive MNS or a black hole. In this case, however, the black holes produced may be smaller than the maximum mass of SM NSs, i.e.~the black hole remnants can be as small as  $M\gtrsim 1.3 M_\odot$. 
    \item {\bf Scenario 3 - NS-MNS inspiral and merger with $f/v\geq 5$}: One object is dark, but the NS may produce electromagnetic signatures from tidal disruption. The  symmetric mass ratio and chirp mass are entirely distinguishable from those of SM NS binaries. The tidal deformabilities of the binary components are orders of magnitude different from each other. The remnant  may produce a \mans~or a black hole. The resulting admixed star has a significantly smaller radius than that of a SM NS. The black hole remnant may be lighter than the maximum mass of the SM NS sequence (black hole masses starting at $M\sim 1.4 M_\odot$).  
    \item {\bf Scenario 4 - NS-MNS inspiral and merger with $f/v< 5$}: One object is dark but the NS may produce electromagnetic signatures from tidal disruption. The  symmetric mass ratio and chirp mass are mostly distinguishable from those of SM NS binaries, but some small overlap region exists. The tidal deformabilities of the binary components are an order of magnitude different from each other. The remnant  will be a black hole. The black hole remnant may be lighter than the maximum mass of the SM sequence (masses starting at $M\sim 1.3 M_\odot$).
    \item {\bf Scenario 5 - \mans-\mans~ inspiral and mergers or NS-\mans~ inspiral and mergers}: Both objects may produce electromagnetic signatures from tidal disruption. The  symmetric mass ratio and chirp mass may overlap with that of SM NS binaries. Tidal deformabilities would break universal relations and may appear similar to twin stars~\cite{Tan:2021nat}. The remnant will be a black hole in most scenarios, unless the dark matter core is extremely small.
    \item {\bf Scenario 6 - Radius measurement of \(\mans\)~with X-rays}: Because a \(\mans\)~can have an outer layer (or halo) entirely composed of SM matter, it can have a hot spot on its surface, just like a SM NS. If the hot spot emits X-rays, NICER could potentially observe them and obtain the \(\Mmans\) and \(\Rout\). Notice that it is possible to observe two \(\manss\) with the same mass, but with very different radii,  because \(\manss\) exist in a 2D area in the mass-radius plane. Therefore, from an observational perspective, \mans{s} may be confused with 1st-order phase transition (1st-order PT) twin stars, which also allows two SM NSs with the same mass but very different radii. The difference between \(\manss\) and 1st-order PT twin stars (known as mass twins \cite{Gerlach:1968zz,Kampfer:1981yr}) is that mass twins are produced by a single SM EoS (one fluid only), and they lie on a single non-monotonic mass-radius curve, instead of
    a non-bijective mapping from
    $(\epsilon_c^\mathrm{SM}, \epsilon_c^\mathrm{DM})$  to the  $(R_\mathrm{out}, M)$ plane that occurs for \mans. 
    However, it may still be possible to distinguish the 2D mirror matter plane from mass twins with just  the measurement of two compact objects with the same mass, if they have drastically different radii. For instance, applying the heavy maximum mass constraint of $M_{\rm max}\sim 2M_\odot$~\cite{Cromartie:2019kug}, it is difficult to produce mass twins; only extremely large first-order phase transitions can produce a large radius difference between the different branches (e.g. see Fig. 12 and Fig. 14 from \cite{Tan:2021ahl}).  Thus, admixed mirror stars may initially mask as mass twins, but once a 2D mass-radius plane is measured, they would be clearly identified as mirror matter. 
    \item {\bf Scenario 7 - Black hole-NS Twins}: 
    Binary inspirals and mergers involving MNSs have the potential to create very light black holes. A striking possibility is that this would lead to  black hole-NS mass twins---that is, a black hole and a NS with the same gravitational mass. In order to clearly distinguish these objects from each other, one would need a sufficiently good measurement of the their tidal deformability, since black hole are expected to have none. This could prove very difficult for very massive NSs, because their deformability, $\Lambda\sim 10$, would not be sufficiently different from that of black holes. Therefore, it would be more feasible to distinguish between lighter black hole-NS twins, say with $M\approx 1.4\, M_\odot$.  
\end{itemize}
 We have outlined some of the specific scenarios for the inspiral and merger of compact objects containing mirror matter.  However, some signatures will only appear if a large population of these objects is measured.  The primary signature with large statistics is that if admixed mirror stars exist, then one could find a 2D plane in the mass-radius or the tidal deformability-mass relation.  The size and shape of the 2D mass-radius plane depend on a number of factors, such as the value of $f/v$, the minimum mass of a MNSs (i.e.~if they are subject to the Chandrasekhar limit or not),  if the admixed stars are produced due to accreted mirror matter or as a remnant of a MNS-NS merger, and details of the EOS. However, the existence of the 2D plane is robust, only possible in a two-fluid model, and would not occur from any known SM EOS. 

Thus, if it were to become clear from observations (say with GW detectors such as advanced LIGO/Virgo/KAGRA, or the X-ray telescope NICER) that the mass-radius posteriors of compact objects cannot be described through a one-dimensional sequence alone (even considering disconnected sequences), then such a measurement would be a clean-cut signature for admixed mirror stars. Such a measurement would require the confidence ellipses on the mass-radius (or mass-tidal deformability) plane of two independent observations of at least two different NSs to be non-overlapping. This could happen, for example, if one were to observe a compact object with $M=1\,M_\odot$ and $R \lesssim 6$ km, with an accuracy of $\delta R \lesssim 4$ km.  
Black holes of such a low mass could potentially be distinguishable from admixed mirror star remnants through the quasi-normal GWs they would emit as they settle down to their final stationary configuration. 

%%%%%%%%%%%%%%%%%%%%%%%%%%%%%%%%%%%%%%%%%%%%%%%%%%%%%%%%%%%%%%
\section{Conclusions and Outlook}
\label{sec:conclusion}

In this paper, we have investigated the effects of Twin Higgs mirror matter in isolated \mans, and in NS-MNS and MNS-MNS binaries. We have found that, instead of forming a single curve, \mans span over an 2D region on the mass-radius plan. 
This region ranges from the SM NS mass-radius curve on the left, past the MNS sequence,  to a curve of minimal radius, where $\Rout\sim 4$ km, and the star is filled with both DM and SM matter throughout its extension. Because the outer radius changes non-monotonically with the DM fraction $Y_\dm$ in the star, different $Y_\dm$ intervals can overlap in the mass-radius diagram, leading to ``\emph{ultimate twins},'' i.e.~stars with the same exact mass and radius, but different internal composition. While these twins have exactly the same mass and radius, their tidal deformability is different and would distinguish the two compact objects. 

We have also explored the properties of  \mans that would have formed via DM capture in NSs above a generous assumed NS formation threshold $M_\sm \gtrsim 0.9 \,M_\odot$.   Surprisingly, we find that the corresponding mass-radius region is nearly independent of the mirror Higgs scale, yielding approximately the same region for $f/v=3-7$. If we allow for the possibility of mirror matter capture, then a DM-admixed NS can support masses in the same range as a regular NS, but with a smaller radius (as low as $R\sim 8$ km). Conversely, if we begin with a mirror neutron star and allow for SM capture, then masses on the small end of NS are possible but the radii are significantly smaller $R\sim 4$ km.
A smoking-gun signature of \mans is the mass-radius plane that they inhabit, which presents a 2D structure, unlike any other object studied so far. Indeed, even objects with distinct stable branches due to first-order phase transition (mass twins) still exist in a one-dimensional sequence.

We then moved to a discussion of the effects of mirror matter in NS-MNS and MNS-MNS binary systems. We found that the amplitude of the GWs emitted by these binary systems (at fixed distance) would be smaller than, but still comparable to, the one from NS-NS binaries due to the smaller total mass of MNSs. We also emphasize that, by measuring the chirp mass $\mathcal{M}$ and the symmetric mass ratio $\eta$ alone, one can distinguish a \(\mns\)-NS or \(\mns\)-\(\mns\) system from a NS-NS one. This distinction is clear-cut only when $f/v\gtrsim 5$. We have also shown that binary Love relations cannot be na\"ively applied to break the degeneracy of a GW measurement from a NS-MNS system, because that relation is no longer EoS insensitive. These relations can, however, be applied without change to \mns-\mns binaries. 

NS-MNS mergers are of particular interest because they provide a promising way to form \(\manss\). However, stability against radial oscillations and NS and MNS mass thresholds impose very stringent constraints on the resulting remnant. For $f/v<5$, we have found no possibility of stable \(\manss\) remnants. For $f/v\geq 5$, we find a small stable mass-radius region.  In cases where a stable \mans remnant is not formed (the most likely scenario), collapse to a black hole will follow. In that case, one could potentially measure the post-merger ringdown.

One very intriguing possibility that arises from the coalescence of MNS-MNS and NS-MNS is that they can produce very light black-hole remnants, with masses as low as $0.8\, M_\odot$. Thus, depending on the masses of the MNS-MNS and NS-MNS binaries, it is possible to produce black hole remnants in the range $(0.8-3.7)\, M_\odot$, MNSs remnants in the range $(0.8-1.5)\, M_\odot$, and \(\mans\) remnants in the range $(1.2-1.6)\, M_\odot$. Thus, mirror matter provides a mechanism that allows for black holes, MNSs and \(\mans\)s remnants - all in the same mass range,  but with very different radii. 

Our work opens up a number of new possibilities and questions.  It is not clear what the precise merger and post-merger signals of such a two fluid model would be like.  Would the two-fluid nature of the compact objects affect the peak frequency of the GWs emitted at merger? What properties would we expect if spin is considered? Is there a possibility of a Kilonova if \(\mans\) merge? In order to answer these questions, a full two-fluid numerical relativity simulation is required, which we leave to future work. 

%%%%%%%%%%%%%%%%%%%%%%%%%%%%%%%%%%%%%%%%%%%%%%%%%%%%%%%%%%%%%%
\acknowledgements

H.T. and N.Y.~acknowledge financial support through NSF grants No.~PHY-1759615, PHY-1949838 and PHY-2207650.
M.H. was supported in part by the National Science Foundation (NSF) within the framework of the MUSES collaboration, under grant number OAC-2103680.
The research of DC was supported in part by a Discovery Grant from the Natural Sciences and Engineering Research Council of Canada, the Canada Research Chair program, the Alfred P. Sloan Foundation, the Ontario Early Researcher Award, and the University of Toronto McLean Award.
J.N.H. acknowledges financial support by the US-DOE Nuclear Science Grant No. DESC0020633.
E.D. acknowledges financial support from  Internship Grant Funding – Berea College Internship Program. This material is based upon work supported by the National Science Foundation and the Department of Defense under Grant PHY-1950744. Any opinions, findings, and conclusions or recommendations expressed in this material are those of the author(s) and do not necessarily reflect the views of the National Science Foundation or the Department of Defense.

\bibliography{references2, NOTinspire}

\begin{thebibliography}{90}%
\makeatletter
\providecommand \@ifxundefined [1]{%
 \@ifx{#1\undefined}
}%
\providecommand \@ifnum [1]{%
 \ifnum #1\expandafter \@firstoftwo
 \else \expandafter \@secondoftwo
 \fi
}%
\providecommand \@ifx [1]{%
 \ifx #1\expandafter \@firstoftwo
 \else \expandafter \@secondoftwo
 \fi
}%
\providecommand \natexlab [1]{#1}%
\providecommand \enquote  [1]{``#1''}%
\providecommand \bibnamefont  [1]{#1}%
\providecommand \bibfnamefont [1]{#1}%
\providecommand \citenamefont [1]{#1}%
\providecommand \href@noop [0]{\@secondoftwo}%
\providecommand \href [0]{\begingroup \@sanitize@url \@href}%
\providecommand \@href[1]{\@@startlink{#1}\@@href}%
\providecommand \@@href[1]{\endgroup#1\@@endlink}%
\providecommand \@sanitize@url [0]{\catcode `\\12\catcode `\$12\catcode
  `\&12\catcode `\#12\catcode `\^12\catcode `\_12\catcode `\%12\relax}%
\providecommand \@@startlink[1]{}%
\providecommand \@@endlink[0]{}%
\providecommand \url  [0]{\begingroup\@sanitize@url \@url }%
\providecommand \@url [1]{\endgroup\@href {#1}{\urlprefix }}%
\providecommand \urlprefix  [0]{URL }%
\providecommand \Eprint [0]{\href }%
\providecommand \doibase [0]{http://dx.doi.org/}%
\providecommand \selectlanguage [0]{\@gobble}%
\providecommand \bibinfo  [0]{\@secondoftwo}%
\providecommand \bibfield  [0]{\@secondoftwo}%
\providecommand \translation [1]{[#1]}%
\providecommand \BibitemOpen [0]{}%
\providecommand \bibitemStop [0]{}%
\providecommand \bibitemNoStop [0]{.\EOS\space}%
\providecommand \EOS [0]{\spacefactor3000\relax}%
\providecommand \BibitemShut  [1]{\csname bibitem#1\endcsname}%
\let\auto@bib@innerbib\@empty
%</preamble>
\bibitem [{\citenamefont {Lisanti}(2017)}]{Lisanti:2016jxe}%
  \BibitemOpen
  \bibfield  {author} {\bibinfo {author} {\bibfnamefont {M.}~\bibnamefont
  {Lisanti}},\ }in\ \href {\doibase 10.1142/9789813149441_0007} {\emph
  {\bibinfo {booktitle} {{Theoretical Advanced Study Institute in Elementary
  Particle Physics}: {New Frontiers in Fields and Strings}}}}\ (\bibinfo {year}
  {2017})\ pp.\ \bibinfo {pages} {399--446},\ \Eprint
  {http://arxiv.org/abs/1603.03797} {arXiv:1603.03797 [hep-ph]} \BibitemShut
  {NoStop}%
\bibitem [{\citenamefont {Lin}(2019)}]{Lin:2019uvt}%
  \BibitemOpen
  \bibfield  {author} {\bibinfo {author} {\bibfnamefont {T.}~\bibnamefont
  {Lin}},\ }\href {\doibase 10.22323/1.333.0009} {\bibfield  {journal}
  {\bibinfo  {journal} {PoS}\ }\textbf {\bibinfo {volume} {333}},\ \bibinfo
  {pages} {009} (\bibinfo {year} {2019})},\ \Eprint
  {http://arxiv.org/abs/1904.07915} {arXiv:1904.07915 [hep-ph]} \BibitemShut
  {NoStop}%
\bibitem [{\citenamefont {Li}\ \emph {et~al.}(2012)\citenamefont {Li},
  \citenamefont {Wang},\ and\ \citenamefont {Cheng}}]{Li:2012qf}%
  \BibitemOpen
  \bibfield  {author} {\bibinfo {author} {\bibfnamefont {X.}~\bibnamefont
  {Li}}, \bibinfo {author} {\bibfnamefont {F.}~\bibnamefont {Wang}}, \ and\
  \bibinfo {author} {\bibfnamefont {K.~S.}\ \bibnamefont {Cheng}},\ }\href
  {\doibase 10.1088/1475-7516/2012/10/031} {\bibfield  {journal} {\bibinfo
  {journal} {JCAP}\ }\textbf {\bibinfo {volume} {10}},\ \bibinfo {pages} {031}
  (\bibinfo {year} {2012})},\ \Eprint {http://arxiv.org/abs/1210.1748}
  {arXiv:1210.1748 [astro-ph.CO]} \BibitemShut {NoStop}%
\bibitem [{\citenamefont {Mukhopadhyay}\ and\ \citenamefont
  {Schaffner-Bielich}(2016)}]{Mukhopadhyay:2015xhs}%
  \BibitemOpen
  \bibfield  {author} {\bibinfo {author} {\bibfnamefont {P.}~\bibnamefont
  {Mukhopadhyay}}\ and\ \bibinfo {author} {\bibfnamefont {J.}~\bibnamefont
  {Schaffner-Bielich}},\ }\href {\doibase 10.1103/PhysRevD.93.083009}
  {\bibfield  {journal} {\bibinfo  {journal} {Phys. Rev. D}\ }\textbf {\bibinfo
  {volume} {93}},\ \bibinfo {pages} {083009} (\bibinfo {year} {2016})},\
  \Eprint {http://arxiv.org/abs/1511.00238} {arXiv:1511.00238 [astro-ph.HE]}
  \BibitemShut {NoStop}%
\bibitem [{\citenamefont {Baym}\ \emph {et~al.}(2018)\citenamefont {Baym},
  \citenamefont {Beck}, \citenamefont {Geltenbort},\ and\ \citenamefont
  {Shelton}}]{Baym:2018ljz}%
  \BibitemOpen
  \bibfield  {author} {\bibinfo {author} {\bibfnamefont {G.}~\bibnamefont
  {Baym}}, \bibinfo {author} {\bibfnamefont {D.~H.}\ \bibnamefont {Beck}},
  \bibinfo {author} {\bibfnamefont {P.}~\bibnamefont {Geltenbort}}, \ and\
  \bibinfo {author} {\bibfnamefont {J.}~\bibnamefont {Shelton}},\ }\href
  {\doibase 10.1103/PhysRevLett.121.061801} {\bibfield  {journal} {\bibinfo
  {journal} {Phys. Rev. Lett.}\ }\textbf {\bibinfo {volume} {121}},\ \bibinfo
  {pages} {061801} (\bibinfo {year} {2018})},\ \Eprint
  {http://arxiv.org/abs/1802.08282} {arXiv:1802.08282 [hep-ph]} \BibitemShut
  {NoStop}%
\bibitem [{\citenamefont {Ellis}\ \emph {et~al.}(2018)\citenamefont {Ellis},
  \citenamefont {H\"utsi}, \citenamefont {Kannike}, \citenamefont {Marzola},
  \citenamefont {Raidal},\ and\ \citenamefont {Vaskonen}}]{Ellis:2018bkr}%
  \BibitemOpen
  \bibfield  {author} {\bibinfo {author} {\bibfnamefont {J.}~\bibnamefont
  {Ellis}}, \bibinfo {author} {\bibfnamefont {G.}~\bibnamefont {H\"utsi}},
  \bibinfo {author} {\bibfnamefont {K.}~\bibnamefont {Kannike}}, \bibinfo
  {author} {\bibfnamefont {L.}~\bibnamefont {Marzola}}, \bibinfo {author}
  {\bibfnamefont {M.}~\bibnamefont {Raidal}}, \ and\ \bibinfo {author}
  {\bibfnamefont {V.}~\bibnamefont {Vaskonen}},\ }\href {\doibase
  10.1103/PhysRevD.97.123007} {\bibfield  {journal} {\bibinfo  {journal} {Phys.
  Rev. D}\ }\textbf {\bibinfo {volume} {97}},\ \bibinfo {pages} {123007}
  (\bibinfo {year} {2018})},\ \Eprint {http://arxiv.org/abs/1804.01418}
  {arXiv:1804.01418 [astro-ph.CO]} \BibitemShut {NoStop}%
\bibitem [{\citenamefont {Dengler}\ \emph {et~al.}(2022)\citenamefont
  {Dengler}, \citenamefont {Schaffner-Bielich},\ and\ \citenamefont
  {Tolos}}]{Dengler:2021qcq}%
  \BibitemOpen
  \bibfield  {author} {\bibinfo {author} {\bibfnamefont {Y.}~\bibnamefont
  {Dengler}}, \bibinfo {author} {\bibfnamefont {J.}~\bibnamefont
  {Schaffner-Bielich}}, \ and\ \bibinfo {author} {\bibfnamefont
  {L.}~\bibnamefont {Tolos}},\ }\href {\doibase 10.1103/PhysRevD.105.043013}
  {\bibfield  {journal} {\bibinfo  {journal} {Phys. Rev. D}\ }\textbf {\bibinfo
  {volume} {105}},\ \bibinfo {pages} {043013} (\bibinfo {year} {2022})},\
  \Eprint {http://arxiv.org/abs/2111.06197} {arXiv:2111.06197 [astro-ph.HE]}
  \BibitemShut {NoStop}%
\bibitem [{\citenamefont {Kain}(2021)}]{Kain:2021hpk}%
  \BibitemOpen
  \bibfield  {author} {\bibinfo {author} {\bibfnamefont {B.}~\bibnamefont
  {Kain}},\ }\href {\doibase 10.1103/PhysRevD.103.043009} {\bibfield  {journal}
  {\bibinfo  {journal} {Phys. Rev. D}\ }\textbf {\bibinfo {volume} {103}},\
  \bibinfo {pages} {043009} (\bibinfo {year} {2021})},\ \Eprint
  {http://arxiv.org/abs/2102.08257} {arXiv:2102.08257 [gr-qc]} \BibitemShut
  {NoStop}%
\bibitem [{\citenamefont {Sen}\ and\ \citenamefont {Guha}(2021)}]{Sen:2021wev}%
  \BibitemOpen
  \bibfield  {author} {\bibinfo {author} {\bibfnamefont {D.}~\bibnamefont
  {Sen}}\ and\ \bibinfo {author} {\bibfnamefont {A.}~\bibnamefont {Guha}},\
  }\href {\doibase 10.1093/mnras/stab1056} {\bibfield  {journal} {\bibinfo
  {journal} {Mon. Not. Roy. Astron. Soc.}\ }\textbf {\bibinfo {volume} {504}},\
  \bibinfo {pages} {3354} (\bibinfo {year} {2021})},\ \Eprint
  {http://arxiv.org/abs/2104.06141} {arXiv:2104.06141 [hep-ph]} \BibitemShut
  {NoStop}%
\bibitem [{\citenamefont {Jim\'enez}\ and\ \citenamefont
  {Fraga}(2022)}]{Jimenez:2021nmr}%
  \BibitemOpen
  \bibfield  {author} {\bibinfo {author} {\bibfnamefont {J.~C.}\ \bibnamefont
  {Jim\'enez}}\ and\ \bibinfo {author} {\bibfnamefont {E.~S.}\ \bibnamefont
  {Fraga}},\ }\href {\doibase 10.3390/universe8010034} {\bibfield  {journal}
  {\bibinfo  {journal} {Universe}\ }\textbf {\bibinfo {volume} {8}},\ \bibinfo
  {pages} {34} (\bibinfo {year} {2022})},\ \Eprint
  {http://arxiv.org/abs/2111.00091} {arXiv:2111.00091 [hep-ph]} \BibitemShut
  {NoStop}%
\bibitem [{\citenamefont {Miao}\ \emph {et~al.}(2022)\citenamefont {Miao},
  \citenamefont {Zhu}, \citenamefont {Li},\ and\ \citenamefont
  {Huang}}]{Miao:2022rqj}%
  \BibitemOpen
  \bibfield  {author} {\bibinfo {author} {\bibfnamefont {Z.}~\bibnamefont
  {Miao}}, \bibinfo {author} {\bibfnamefont {Y.}~\bibnamefont {Zhu}}, \bibinfo
  {author} {\bibfnamefont {A.}~\bibnamefont {Li}}, \ and\ \bibinfo {author}
  {\bibfnamefont {F.}~\bibnamefont {Huang}},\ }\href {\doibase
  10.3847/1538-4357/ac8544} {\bibfield  {journal} {\bibinfo  {journal}
  {Astrophys. J.}\ }\textbf {\bibinfo {volume} {936}},\ \bibinfo {pages} {69}
  (\bibinfo {year} {2022})},\ \Eprint {http://arxiv.org/abs/2204.05560}
  {arXiv:2204.05560 [astro-ph.HE]} \BibitemShut {NoStop}%
\bibitem [{\citenamefont {Horowitz}\ and\ \citenamefont
  {Reddy}(2019)}]{Horowitz:2019aim}%
  \BibitemOpen
  \bibfield  {author} {\bibinfo {author} {\bibfnamefont {C.~J.}\ \bibnamefont
  {Horowitz}}\ and\ \bibinfo {author} {\bibfnamefont {S.}~\bibnamefont
  {Reddy}},\ }\href {\doibase 10.1103/PhysRevLett.122.071102} {\bibfield
  {journal} {\bibinfo  {journal} {Phys. Rev. Lett.}\ }\textbf {\bibinfo
  {volume} {122}},\ \bibinfo {pages} {071102} (\bibinfo {year} {2019})},\
  \Eprint {http://arxiv.org/abs/1902.04597} {arXiv:1902.04597 [astro-ph.HE]}
  \BibitemShut {NoStop}%
\bibitem [{\citenamefont {Maedan}(2020)}]{Maedan:2019mgz}%
  \BibitemOpen
  \bibfield  {author} {\bibinfo {author} {\bibfnamefont {S.}~\bibnamefont
  {Maedan}},\ }\href {\doibase 10.1093/ptep/ptaa014} {\bibfield  {journal}
  {\bibinfo  {journal} {PTEP}\ }\textbf {\bibinfo {volume} {2020}},\ \bibinfo
  {pages} {033B07} (\bibinfo {year} {2020})},\ \Eprint
  {http://arxiv.org/abs/1908.00711} {arXiv:1908.00711 [hep-ph]} \BibitemShut
  {NoStop}%
\bibitem [{\citenamefont {Narain}\ \emph {et~al.}(2006)\citenamefont {Narain},
  \citenamefont {Schaffner-Bielich},\ and\ \citenamefont
  {Mishustin}}]{Narain:2006kx}%
  \BibitemOpen
  \bibfield  {author} {\bibinfo {author} {\bibfnamefont {G.}~\bibnamefont
  {Narain}}, \bibinfo {author} {\bibfnamefont {J.}~\bibnamefont
  {Schaffner-Bielich}}, \ and\ \bibinfo {author} {\bibfnamefont {I.~N.}\
  \bibnamefont {Mishustin}},\ }\href {\doibase 10.1103/PhysRevD.74.063003}
  {\bibfield  {journal} {\bibinfo  {journal} {Phys. Rev. D}\ }\textbf {\bibinfo
  {volume} {74}},\ \bibinfo {pages} {063003} (\bibinfo {year} {2006})},\
  \Eprint {http://arxiv.org/abs/astro-ph/0605724} {arXiv:astro-ph/0605724}
  \BibitemShut {NoStop}%
\bibitem [{\citenamefont {Goldman}\ \emph {et~al.}(2013)\citenamefont
  {Goldman}, \citenamefont {Mohapatra}, \citenamefont {Nussinov}, \citenamefont
  {Rosenbaum},\ and\ \citenamefont {Teplitz}}]{Goldman:2013qla}%
  \BibitemOpen
  \bibfield  {author} {\bibinfo {author} {\bibfnamefont {I.}~\bibnamefont
  {Goldman}}, \bibinfo {author} {\bibfnamefont {R.~N.}\ \bibnamefont
  {Mohapatra}}, \bibinfo {author} {\bibfnamefont {S.}~\bibnamefont {Nussinov}},
  \bibinfo {author} {\bibfnamefont {D.}~\bibnamefont {Rosenbaum}}, \ and\
  \bibinfo {author} {\bibfnamefont {V.}~\bibnamefont {Teplitz}},\ }\href
  {\doibase 10.1016/j.physletb.2013.07.017} {\bibfield  {journal} {\bibinfo
  {journal} {Phys. Lett. B}\ }\textbf {\bibinfo {volume} {725}},\ \bibinfo
  {pages} {200} (\bibinfo {year} {2013})},\ \Eprint
  {http://arxiv.org/abs/1305.6908} {arXiv:1305.6908 [astro-ph.CO]} \BibitemShut
  {NoStop}%
\bibitem [{\citenamefont {Tolos}\ \emph {et~al.}(2015)\citenamefont {Tolos},
  \citenamefont {Schaffner-Bielich},\ and\ \citenamefont
  {Dengler}}]{Tolos:2015qra}%
  \BibitemOpen
  \bibfield  {author} {\bibinfo {author} {\bibfnamefont {L.}~\bibnamefont
  {Tolos}}, \bibinfo {author} {\bibfnamefont {J.}~\bibnamefont
  {Schaffner-Bielich}}, \ and\ \bibinfo {author} {\bibfnamefont
  {Y.}~\bibnamefont {Dengler}},\ }\href {\doibase 10.1103/PhysRevD.92.123002}
  {\bibfield  {journal} {\bibinfo  {journal} {Phys. Rev. D}\ }\textbf {\bibinfo
  {volume} {92}},\ \bibinfo {pages} {123002} (\bibinfo {year} {2015})},\
  \bibinfo {note} {[Erratum: Phys.Rev.D 103, 109901 (2021)]},\ \Eprint
  {http://arxiv.org/abs/1507.08197} {arXiv:1507.08197 [astro-ph.HE]}
  \BibitemShut {NoStop}%
\bibitem [{\citenamefont {Kouvaris}\ and\ \citenamefont
  {Nielsen}(2015)}]{Kouvaris:2015rea}%
  \BibitemOpen
  \bibfield  {author} {\bibinfo {author} {\bibfnamefont {C.}~\bibnamefont
  {Kouvaris}}\ and\ \bibinfo {author} {\bibfnamefont {N.~G.}\ \bibnamefont
  {Nielsen}},\ }\href {\doibase 10.1103/PhysRevD.92.063526} {\bibfield
  {journal} {\bibinfo  {journal} {Phys. Rev. D}\ }\textbf {\bibinfo {volume}
  {92}},\ \bibinfo {pages} {063526} (\bibinfo {year} {2015})},\ \Eprint
  {http://arxiv.org/abs/1507.00959} {arXiv:1507.00959 [hep-ph]} \BibitemShut
  {NoStop}%
\bibitem [{\citenamefont {Mukhopadhyay}\ \emph {et~al.}(2017)\citenamefont
  {Mukhopadhyay}, \citenamefont {Atta}, \citenamefont {Imam}, \citenamefont
  {Basu},\ and\ \citenamefont {Samanta}}]{Mukhopadhyay:2016dsg}%
  \BibitemOpen
  \bibfield  {author} {\bibinfo {author} {\bibfnamefont {S.}~\bibnamefont
  {Mukhopadhyay}}, \bibinfo {author} {\bibfnamefont {D.}~\bibnamefont {Atta}},
  \bibinfo {author} {\bibfnamefont {K.}~\bibnamefont {Imam}}, \bibinfo {author}
  {\bibfnamefont {D.~N.}\ \bibnamefont {Basu}}, \ and\ \bibinfo {author}
  {\bibfnamefont {C.}~\bibnamefont {Samanta}},\ }\href {\doibase
  10.1140/epjc/s10052-017-5006-3} {\bibfield  {journal} {\bibinfo  {journal}
  {Eur. Phys. J. C}\ }\textbf {\bibinfo {volume} {77}},\ \bibinfo {pages} {440}
  (\bibinfo {year} {2017})},\ \bibinfo {note} {[Erratum: Eur.Phys.J.C 77, 553
  (2017)]},\ \Eprint {http://arxiv.org/abs/1612.07093} {arXiv:1612.07093
  [nucl-th]} \BibitemShut {NoStop}%
\bibitem [{\citenamefont {Gresham}\ and\ \citenamefont
  {Zurek}(2019)}]{Gresham:2018rqo}%
  \BibitemOpen
  \bibfield  {author} {\bibinfo {author} {\bibfnamefont {M.~I.}\ \bibnamefont
  {Gresham}}\ and\ \bibinfo {author} {\bibfnamefont {K.~M.}\ \bibnamefont
  {Zurek}},\ }\href {\doibase 10.1103/PhysRevD.99.083008} {\bibfield  {journal}
  {\bibinfo  {journal} {Phys. Rev. D}\ }\textbf {\bibinfo {volume} {99}},\
  \bibinfo {pages} {083008} (\bibinfo {year} {2019})},\ \Eprint
  {http://arxiv.org/abs/1809.08254} {arXiv:1809.08254 [astro-ph.CO]}
  \BibitemShut {NoStop}%
\bibitem [{\citenamefont {Collier}\ \emph {et~al.}(2022)\citenamefont
  {Collier}, \citenamefont {Croon},\ and\ \citenamefont
  {Leane}}]{Collier:2022cpr}%
  \BibitemOpen
  \bibfield  {author} {\bibinfo {author} {\bibfnamefont {M.}~\bibnamefont
  {Collier}}, \bibinfo {author} {\bibfnamefont {D.}~\bibnamefont {Croon}}, \
  and\ \bibinfo {author} {\bibfnamefont {R.~K.}\ \bibnamefont {Leane}},\
  }\href@noop {} {\  (\bibinfo {year} {2022})},\ \Eprint
  {http://arxiv.org/abs/2205.15337} {arXiv:2205.15337 [gr-qc]} \BibitemShut
  {NoStop}%
\bibitem [{\citenamefont {Das}\ \emph {et~al.}(2022)\citenamefont {Das},
  \citenamefont {Malik},\ and\ \citenamefont {Nayak}}]{Das:2020ecp}%
  \BibitemOpen
  \bibfield  {author} {\bibinfo {author} {\bibfnamefont {A.}~\bibnamefont
  {Das}}, \bibinfo {author} {\bibfnamefont {T.}~\bibnamefont {Malik}}, \ and\
  \bibinfo {author} {\bibfnamefont {A.~C.}\ \bibnamefont {Nayak}},\ }\href
  {\doibase 10.1103/PhysRevD.105.123034} {\bibfield  {journal} {\bibinfo
  {journal} {Phys. Rev. D}\ }\textbf {\bibinfo {volume} {105}},\ \bibinfo
  {pages} {123034} (\bibinfo {year} {2022})},\ \Eprint
  {http://arxiv.org/abs/2011.01318} {arXiv:2011.01318 [nucl-th]} \BibitemShut
  {NoStop}%
\bibitem [{\citenamefont {Leung}\ \emph {et~al.}(2022)\citenamefont {Leung},
  \citenamefont {Chu},\ and\ \citenamefont {Lin}}]{Leung:2022wcf}%
  \BibitemOpen
  \bibfield  {author} {\bibinfo {author} {\bibfnamefont {K.-L.}\ \bibnamefont
  {Leung}}, \bibinfo {author} {\bibfnamefont {M.-c.}\ \bibnamefont {Chu}}, \
  and\ \bibinfo {author} {\bibfnamefont {L.-M.}\ \bibnamefont {Lin}},\ }\href
  {\doibase 10.1103/PhysRevD.105.123010} {\bibfield  {journal} {\bibinfo
  {journal} {Phys. Rev. D}\ }\textbf {\bibinfo {volume} {105}},\ \bibinfo
  {pages} {123010} (\bibinfo {year} {2022})},\ \Eprint
  {http://arxiv.org/abs/2207.02433} {arXiv:2207.02433 [astro-ph.HE]}
  \BibitemShut {NoStop}%
\bibitem [{\citenamefont {Rutherford}\ \emph {et~al.}(2022)\citenamefont
  {Rutherford}, \citenamefont {Raaijmakers}, \citenamefont
  {Prescod-Weinstein},\ and\ \citenamefont {Watts}}]{Rutherford:2022xeb}%
  \BibitemOpen
  \bibfield  {author} {\bibinfo {author} {\bibfnamefont {N.}~\bibnamefont
  {Rutherford}}, \bibinfo {author} {\bibfnamefont {G.}~\bibnamefont
  {Raaijmakers}}, \bibinfo {author} {\bibfnamefont {C.}~\bibnamefont
  {Prescod-Weinstein}}, \ and\ \bibinfo {author} {\bibfnamefont
  {A.}~\bibnamefont {Watts}},\ }\href@noop {} {\  (\bibinfo {year} {2022})},\
  \Eprint {http://arxiv.org/abs/2208.03282} {arXiv:2208.03282 [astro-ph.HE]}
  \BibitemShut {NoStop}%
\bibitem [{\citenamefont {Giangrandi}\ \emph {et~al.}(2022)\citenamefont
  {Giangrandi}, \citenamefont {Sagun}, \citenamefont {Ivanytskyi},
  \citenamefont {Provid\^encia},\ and\ \citenamefont
  {Dietrich}}]{Giangrandi:2022wht}%
  \BibitemOpen
  \bibfield  {author} {\bibinfo {author} {\bibfnamefont {E.}~\bibnamefont
  {Giangrandi}}, \bibinfo {author} {\bibfnamefont {V.}~\bibnamefont {Sagun}},
  \bibinfo {author} {\bibfnamefont {O.}~\bibnamefont {Ivanytskyi}}, \bibinfo
  {author} {\bibfnamefont {C.}~\bibnamefont {Provid\^encia}}, \ and\ \bibinfo
  {author} {\bibfnamefont {T.}~\bibnamefont {Dietrich}},\ }\href@noop {} {\
  (\bibinfo {year} {2022})},\ \Eprint {http://arxiv.org/abs/2209.10905}
  {arXiv:2209.10905 [astro-ph.HE]} \BibitemShut {NoStop}%
\bibitem [{\citenamefont {Deliyergiyev}\ \emph {et~al.}(2019)\citenamefont
  {Deliyergiyev}, \citenamefont {Del~Popolo}, \citenamefont {Tolos},
  \citenamefont {Le~Delliou}, \citenamefont {Lee},\ and\ \citenamefont
  {Burgio}}]{Deliyergiyev:2019vti}%
  \BibitemOpen
  \bibfield  {author} {\bibinfo {author} {\bibfnamefont {M.}~\bibnamefont
  {Deliyergiyev}}, \bibinfo {author} {\bibfnamefont {A.}~\bibnamefont
  {Del~Popolo}}, \bibinfo {author} {\bibfnamefont {L.}~\bibnamefont {Tolos}},
  \bibinfo {author} {\bibfnamefont {M.}~\bibnamefont {Le~Delliou}}, \bibinfo
  {author} {\bibfnamefont {X.}~\bibnamefont {Lee}}, \ and\ \bibinfo {author}
  {\bibfnamefont {F.}~\bibnamefont {Burgio}},\ }\href {\doibase
  10.1103/PhysRevD.99.063015} {\bibfield  {journal} {\bibinfo  {journal} {Phys.
  Rev. D}\ }\textbf {\bibinfo {volume} {99}},\ \bibinfo {pages} {063015}
  (\bibinfo {year} {2019})},\ \Eprint {http://arxiv.org/abs/1903.01183}
  {arXiv:1903.01183 [gr-qc]} \BibitemShut {NoStop}%
\bibitem [{\citenamefont {Alexander}\ \emph {et~al.}(2016)\citenamefont
  {Alexander} \emph {et~al.}}]{Alexander:2016aln}%
  \BibitemOpen
  \bibfield  {author} {\bibinfo {author} {\bibfnamefont {J.}~\bibnamefont
  {Alexander}} \emph {et~al.}\ }(\bibinfo {year} {2016})\ \Eprint
  {http://arxiv.org/abs/1608.08632} {arXiv:1608.08632 [hep-ph]} \BibitemShut
  {NoStop}%
\bibitem [{\citenamefont {Curtin}\ and\ \citenamefont
  {Setford}(2020)}]{Curtin:2019ngc}%
  \BibitemOpen
  \bibfield  {author} {\bibinfo {author} {\bibfnamefont {D.}~\bibnamefont
  {Curtin}}\ and\ \bibinfo {author} {\bibfnamefont {J.}~\bibnamefont
  {Setford}},\ }\href {\doibase 10.1007/JHEP03(2020)041} {\bibfield  {journal}
  {\bibinfo  {journal} {JHEP}\ }\textbf {\bibinfo {volume} {03}},\ \bibinfo
  {pages} {041} (\bibinfo {year} {2020})},\ \Eprint
  {http://arxiv.org/abs/1909.04072} {arXiv:1909.04072 [hep-ph]} \BibitemShut
  {NoStop}%
\bibitem [{\citenamefont {Curtin}\ and\ \citenamefont
  {Setford}(2021)}]{Curtin:2020tkm}%
  \BibitemOpen
  \bibfield  {author} {\bibinfo {author} {\bibfnamefont {D.}~\bibnamefont
  {Curtin}}\ and\ \bibinfo {author} {\bibfnamefont {J.}~\bibnamefont
  {Setford}},\ }\href {\doibase 10.1007/JHEP03(2021)166} {\bibfield  {journal}
  {\bibinfo  {journal} {JHEP}\ }\textbf {\bibinfo {volume} {03}},\ \bibinfo
  {pages} {166} (\bibinfo {year} {2021})},\ \Eprint
  {http://arxiv.org/abs/2010.00601} {arXiv:2010.00601 [hep-ph]} \BibitemShut
  {NoStop}%
\bibitem [{\citenamefont {Hippert}\ \emph {et~al.}(2022)\citenamefont
  {Hippert}, \citenamefont {Setford}, \citenamefont {Tan}, \citenamefont
  {Curtin}, \citenamefont {Noronha-Hostler},\ and\ \citenamefont
  {Yunes}}]{Hippert:2021fch}%
  \BibitemOpen
  \bibfield  {author} {\bibinfo {author} {\bibfnamefont {M.}~\bibnamefont
  {Hippert}}, \bibinfo {author} {\bibfnamefont {J.}~\bibnamefont {Setford}},
  \bibinfo {author} {\bibfnamefont {H.}~\bibnamefont {Tan}}, \bibinfo {author}
  {\bibfnamefont {D.}~\bibnamefont {Curtin}}, \bibinfo {author} {\bibfnamefont
  {J.}~\bibnamefont {Noronha-Hostler}}, \ and\ \bibinfo {author} {\bibfnamefont
  {N.}~\bibnamefont {Yunes}},\ }\href {\doibase 10.1103/PhysRevD.106.035025}
  {\bibfield  {journal} {\bibinfo  {journal} {Phys. Rev. D}\ }\textbf {\bibinfo
  {volume} {106}},\ \bibinfo {pages} {035025} (\bibinfo {year} {2022})},\
  \Eprint {http://arxiv.org/abs/2103.01965} {arXiv:2103.01965 [astro-ph.HE]}
  \BibitemShut {NoStop}%
\bibitem [{\citenamefont {Ryan}\ and\ \citenamefont
  {Radice}(2022)}]{Ryan:2022hku}%
  \BibitemOpen
  \bibfield  {author} {\bibinfo {author} {\bibfnamefont {M.}~\bibnamefont
  {Ryan}}\ and\ \bibinfo {author} {\bibfnamefont {D.}~\bibnamefont {Radice}},\
  }\href {\doibase 10.1103/PhysRevD.105.115034} {\bibfield  {journal} {\bibinfo
   {journal} {Phys. Rev. D}\ }\textbf {\bibinfo {volume} {105}},\ \bibinfo
  {pages} {115034} (\bibinfo {year} {2022})},\ \Eprint
  {http://arxiv.org/abs/2201.05626} {arXiv:2201.05626 [astro-ph.HE]}
  \BibitemShut {NoStop}%
\bibitem [{\citenamefont {Blinnikov}\ and\ \citenamefont
  {Khlopov}(1982)}]{Blinnikov:1982eh}%
  \BibitemOpen
  \bibfield  {author} {\bibinfo {author} {\bibfnamefont {S.~I.}\ \bibnamefont
  {Blinnikov}}\ and\ \bibinfo {author} {\bibfnamefont {M.~Y.}\ \bibnamefont
  {Khlopov}},\ }\href@noop {} {\bibfield  {journal} {\bibinfo  {journal} {Sov.
  J. Nucl. Phys.}\ }\textbf {\bibinfo {volume} {36}},\ \bibinfo {pages} {472}
  (\bibinfo {year} {1982})}\BibitemShut {NoStop}%
\bibitem [{\citenamefont {Blinnikov}\ and\ \citenamefont
  {Khlopov}(1983)}]{Blinnikov:1983gh}%
  \BibitemOpen
  \bibfield  {author} {\bibinfo {author} {\bibfnamefont {S.~I.}\ \bibnamefont
  {Blinnikov}}\ and\ \bibinfo {author} {\bibfnamefont {M.}~\bibnamefont
  {Khlopov}},\ }\href@noop {} {\bibfield  {journal} {\bibinfo  {journal} {Sov.
  Astron.}\ }\textbf {\bibinfo {volume} {27}},\ \bibinfo {pages} {371}
  (\bibinfo {year} {1983})}\BibitemShut {NoStop}%
\bibitem [{\citenamefont {Khlopov}\ \emph {et~al.}(1991)\citenamefont
  {Khlopov}, \citenamefont {Beskin}, \citenamefont {Bochkarev}, \citenamefont
  {Pustylnik},\ and\ \citenamefont {Pustylnik}}]{Khlopov:1989fj}%
  \BibitemOpen
  \bibfield  {author} {\bibinfo {author} {\bibfnamefont {M.~Y.}\ \bibnamefont
  {Khlopov}}, \bibinfo {author} {\bibfnamefont {G.~M.}\ \bibnamefont {Beskin}},
  \bibinfo {author} {\bibfnamefont {N.~E.}\ \bibnamefont {Bochkarev}}, \bibinfo
  {author} {\bibfnamefont {L.~A.}\ \bibnamefont {Pustylnik}}, \ and\ \bibinfo
  {author} {\bibfnamefont {S.~A.}\ \bibnamefont {Pustylnik}},\ }\href@noop {}
  {\bibfield  {journal} {\bibinfo  {journal} {Sov. Astron.}\ }\textbf {\bibinfo
  {volume} {35}},\ \bibinfo {pages} {21} (\bibinfo {year} {1991})}\BibitemShut
  {NoStop}%
\bibitem [{\citenamefont {Berezhiani}\ \emph {et~al.}(1996)\citenamefont
  {Berezhiani}, \citenamefont {Dolgov},\ and\ \citenamefont
  {Mohapatra}}]{Berezhiani:1995am}%
  \BibitemOpen
  \bibfield  {author} {\bibinfo {author} {\bibfnamefont {Z.~G.}\ \bibnamefont
  {Berezhiani}}, \bibinfo {author} {\bibfnamefont {A.~D.}\ \bibnamefont
  {Dolgov}}, \ and\ \bibinfo {author} {\bibfnamefont {R.~N.}\ \bibnamefont
  {Mohapatra}},\ }\href {\doibase 10.1016/0370-2693(96)00219-5} {\bibfield
  {journal} {\bibinfo  {journal} {Phys. Lett. B}\ }\textbf {\bibinfo {volume}
  {375}},\ \bibinfo {pages} {26} (\bibinfo {year} {1996})},\ \Eprint
  {http://arxiv.org/abs/hep-ph/9511221} {arXiv:hep-ph/9511221} \BibitemShut
  {NoStop}%
\bibitem [{\citenamefont {Chacko}\ \emph {et~al.}(2006)\citenamefont {Chacko},
  \citenamefont {Goh},\ and\ \citenamefont {Harnik}}]{Chacko:2005pe}%
  \BibitemOpen
  \bibfield  {author} {\bibinfo {author} {\bibfnamefont {Z.}~\bibnamefont
  {Chacko}}, \bibinfo {author} {\bibfnamefont {H.-S.}\ \bibnamefont {Goh}}, \
  and\ \bibinfo {author} {\bibfnamefont {R.}~\bibnamefont {Harnik}},\ }\href
  {\doibase 10.1103/PhysRevLett.96.231802} {\bibfield  {journal} {\bibinfo
  {journal} {Phys. Rev. Lett.}\ }\textbf {\bibinfo {volume} {96}},\ \bibinfo
  {pages} {231802} (\bibinfo {year} {2006})},\ \Eprint
  {http://arxiv.org/abs/hep-ph/0506256} {arXiv:hep-ph/0506256} \BibitemShut
  {NoStop}%
\bibitem [{\citenamefont {Chacko}\ \emph {et~al.}(2017)\citenamefont {Chacko},
  \citenamefont {Craig}, \citenamefont {Fox},\ and\ \citenamefont
  {Harnik}}]{Chacko:2016hvu}%
  \BibitemOpen
  \bibfield  {author} {\bibinfo {author} {\bibfnamefont {Z.}~\bibnamefont
  {Chacko}}, \bibinfo {author} {\bibfnamefont {N.}~\bibnamefont {Craig}},
  \bibinfo {author} {\bibfnamefont {P.~J.}\ \bibnamefont {Fox}}, \ and\
  \bibinfo {author} {\bibfnamefont {R.}~\bibnamefont {Harnik}},\ }\href
  {\doibase 10.1007/JHEP07(2017)023} {\bibfield  {journal} {\bibinfo  {journal}
  {JHEP}\ }\textbf {\bibinfo {volume} {07}},\ \bibinfo {pages} {023} (\bibinfo
  {year} {2017})},\ \Eprint {http://arxiv.org/abs/1611.07975} {arXiv:1611.07975
  [hep-ph]} \BibitemShut {NoStop}%
\bibitem [{\citenamefont {Chacko}\ \emph {et~al.}(2018)\citenamefont {Chacko},
  \citenamefont {Curtin}, \citenamefont {Geller},\ and\ \citenamefont
  {Tsai}}]{Chacko:2018vss}%
  \BibitemOpen
  \bibfield  {author} {\bibinfo {author} {\bibfnamefont {Z.}~\bibnamefont
  {Chacko}}, \bibinfo {author} {\bibfnamefont {D.}~\bibnamefont {Curtin}},
  \bibinfo {author} {\bibfnamefont {M.}~\bibnamefont {Geller}}, \ and\ \bibinfo
  {author} {\bibfnamefont {Y.}~\bibnamefont {Tsai}},\ }\href {\doibase
  10.1007/JHEP09(2018)163} {\bibfield  {journal} {\bibinfo  {journal} {JHEP}\
  }\textbf {\bibinfo {volume} {09}},\ \bibinfo {pages} {163} (\bibinfo {year}
  {2018})},\ \Eprint {http://arxiv.org/abs/1803.03263} {arXiv:1803.03263
  [hep-ph]} \BibitemShut {NoStop}%
\bibitem [{\citenamefont {Burdman}\ \emph {et~al.}(2015)\citenamefont
  {Burdman}, \citenamefont {Chacko}, \citenamefont {Harnik}, \citenamefont
  {de~Lima},\ and\ \citenamefont {Verhaaren}}]{Burdman:2014zta}%
  \BibitemOpen
  \bibfield  {author} {\bibinfo {author} {\bibfnamefont {G.}~\bibnamefont
  {Burdman}}, \bibinfo {author} {\bibfnamefont {Z.}~\bibnamefont {Chacko}},
  \bibinfo {author} {\bibfnamefont {R.}~\bibnamefont {Harnik}}, \bibinfo
  {author} {\bibfnamefont {L.}~\bibnamefont {de~Lima}}, \ and\ \bibinfo
  {author} {\bibfnamefont {C.~B.}\ \bibnamefont {Verhaaren}},\ }\href {\doibase
  10.1103/PhysRevD.91.055007} {\bibfield  {journal} {\bibinfo  {journal} {Phys.
  Rev. D}\ }\textbf {\bibinfo {volume} {91}},\ \bibinfo {pages} {055007}
  (\bibinfo {year} {2015})},\ \Eprint {http://arxiv.org/abs/1411.3310}
  {arXiv:1411.3310 [hep-ph]} \BibitemShut {NoStop}%
\bibitem [{\citenamefont {Craig}\ \emph {et~al.}(2015)\citenamefont {Craig},
  \citenamefont {Katz}, \citenamefont {Strassler},\ and\ \citenamefont
  {Sundrum}}]{Craig:2015pha}%
  \BibitemOpen
  \bibfield  {author} {\bibinfo {author} {\bibfnamefont {N.}~\bibnamefont
  {Craig}}, \bibinfo {author} {\bibfnamefont {A.}~\bibnamefont {Katz}},
  \bibinfo {author} {\bibfnamefont {M.}~\bibnamefont {Strassler}}, \ and\
  \bibinfo {author} {\bibfnamefont {R.}~\bibnamefont {Sundrum}},\ }\href
  {\doibase 10.1007/JHEP07(2015)105} {\bibfield  {journal} {\bibinfo  {journal}
  {JHEP}\ }\textbf {\bibinfo {volume} {07}},\ \bibinfo {pages} {105} (\bibinfo
  {year} {2015})},\ \Eprint {http://arxiv.org/abs/1501.05310} {arXiv:1501.05310
  [hep-ph]} \BibitemShut {NoStop}%
\bibitem [{\citenamefont {Aad}\ \emph {et~al.}(2021)\citenamefont {Aad} \emph
  {et~al.}}]{ATLAS:2021hza}%
  \BibitemOpen
  \bibfield  {author} {\bibinfo {author} {\bibfnamefont {G.}~\bibnamefont
  {Aad}} \emph {et~al.} (\bibinfo {collaboration} {ATLAS}),\ }\href {\doibase
  10.1007/JHEP04(2021)165} {\bibfield  {journal} {\bibinfo  {journal} {JHEP}\
  }\textbf {\bibinfo {volume} {04}},\ \bibinfo {pages} {165} (\bibinfo {year}
  {2021})},\ \Eprint {http://arxiv.org/abs/2102.01444} {arXiv:2102.01444
  [hep-ex]} \BibitemShut {NoStop}%
\bibitem [{\citenamefont {Blanchet}(2014)}]{Blanchet:2013haa}%
  \BibitemOpen
  \bibfield  {author} {\bibinfo {author} {\bibfnamefont {L.}~\bibnamefont
  {Blanchet}},\ }\href {\doibase 10.12942/lrr-2014-2} {\bibfield  {journal}
  {\bibinfo  {journal} {Living Rev. Rel.}\ }\textbf {\bibinfo {volume} {17}},\
  \bibinfo {pages} {2} (\bibinfo {year} {2014})},\ \Eprint
  {http://arxiv.org/abs/1310.1528} {arXiv:1310.1528 [gr-qc]} \BibitemShut
  {NoStop}%
\bibitem [{\citenamefont {Kain}(2020)}]{Kain:2020zjs}%
  \BibitemOpen
  \bibfield  {author} {\bibinfo {author} {\bibfnamefont {B.}~\bibnamefont
  {Kain}},\ }\href {\doibase 10.1103/PhysRevD.102.023001} {\bibfield  {journal}
  {\bibinfo  {journal} {Phys. Rev. D}\ }\textbf {\bibinfo {volume} {102}},\
  \bibinfo {pages} {023001} (\bibinfo {year} {2020})},\ \Eprint
  {http://arxiv.org/abs/2007.04311} {arXiv:2007.04311 [gr-qc]} \BibitemShut
  {NoStop}%
\bibitem [{\citenamefont {Naidu}\ \emph {et~al.}(2021)\citenamefont {Naidu},
  \citenamefont {Carloni},\ and\ \citenamefont {Dunsby}}]{Naidu:2021nwh}%
  \BibitemOpen
  \bibfield  {author} {\bibinfo {author} {\bibfnamefont {N.~F.}\ \bibnamefont
  {Naidu}}, \bibinfo {author} {\bibfnamefont {S.}~\bibnamefont {Carloni}}, \
  and\ \bibinfo {author} {\bibfnamefont {P.}~\bibnamefont {Dunsby}},\ }\href
  {\doibase 10.1103/PhysRevD.104.044014} {\bibfield  {journal} {\bibinfo
  {journal} {Phys. Rev. D}\ }\textbf {\bibinfo {volume} {104}},\ \bibinfo
  {pages} {044014} (\bibinfo {year} {2021})},\ \Eprint
  {http://arxiv.org/abs/2102.05693} {arXiv:2102.05693 [gr-qc]} \BibitemShut
  {NoStop}%
\bibitem [{\citenamefont {Gleason}\ \emph {et~al.}(2022)\citenamefont
  {Gleason}, \citenamefont {Brown},\ and\ \citenamefont
  {Kain}}]{Gleason:2022eeg}%
  \BibitemOpen
  \bibfield  {author} {\bibinfo {author} {\bibfnamefont {T.}~\bibnamefont
  {Gleason}}, \bibinfo {author} {\bibfnamefont {B.}~\bibnamefont {Brown}}, \
  and\ \bibinfo {author} {\bibfnamefont {B.}~\bibnamefont {Kain}},\ }\href
  {\doibase 10.1103/PhysRevD.105.023010} {\bibfield  {journal} {\bibinfo
  {journal} {Phys. Rev. D}\ }\textbf {\bibinfo {volume} {105}},\ \bibinfo
  {pages} {023010} (\bibinfo {year} {2022})},\ \Eprint
  {http://arxiv.org/abs/2201.02274} {arXiv:2201.02274 [gr-qc]} \BibitemShut
  {NoStop}%
\bibitem [{\citenamefont {Strobel}\ \emph {et~al.}(1999)\citenamefont
  {Strobel}, \citenamefont {Schaab},\ and\ \citenamefont
  {Weigel}}]{Strobel:1999vn}%
  \BibitemOpen
  \bibfield  {author} {\bibinfo {author} {\bibfnamefont {K.}~\bibnamefont
  {Strobel}}, \bibinfo {author} {\bibfnamefont {C.}~\bibnamefont {Schaab}}, \
  and\ \bibinfo {author} {\bibfnamefont {M.~K.}\ \bibnamefont {Weigel}},\
  }\href@noop {} {\bibfield  {journal} {\bibinfo  {journal} {Astron.
  Astrophys.}\ }\textbf {\bibinfo {volume} {350}},\ \bibinfo {pages} {497}
  (\bibinfo {year} {1999})},\ \Eprint {http://arxiv.org/abs/astro-ph/9908132}
  {arXiv:astro-ph/9908132} \BibitemShut {NoStop}%
\bibitem [{\citenamefont {{Doroshenko}}\ \emph {et~al.}(2022)\citenamefont
  {{Doroshenko}}, \citenamefont {{Suleimanov}}, \citenamefont
  {{P{\"u}hlhofer}},\ and\ \citenamefont {{Santangelo}}}]{2022NatAs.tmp..224D}%
  \BibitemOpen
  \bibfield  {author} {\bibinfo {author} {\bibfnamefont {V.}~\bibnamefont
  {{Doroshenko}}}, \bibinfo {author} {\bibfnamefont {V.}~\bibnamefont
  {{Suleimanov}}}, \bibinfo {author} {\bibfnamefont {G.}~\bibnamefont
  {{P{\"u}hlhofer}}}, \ and\ \bibinfo {author} {\bibfnamefont {A.}~\bibnamefont
  {{Santangelo}}},\ }\href {\doibase 10.1038/s41550-022-01800-1} {\bibfield
  {journal} {\bibinfo  {journal} {Nature Astronomy}\ } (\bibinfo {year}
  {2022}),\ 10.1038/s41550-022-01800-1}\BibitemShut {NoStop}%
\bibitem [{\citenamefont {Catuneanu}\ \emph {et~al.}(2013)\citenamefont
  {Catuneanu}, \citenamefont {Heinke}, \citenamefont {Sivakoff}, \citenamefont
  {Ho},\ and\ \citenamefont {Servillat}}]{2013ApJ...764..145C}%
  \BibitemOpen
  \bibfield  {author} {\bibinfo {author} {\bibfnamefont {A.}~\bibnamefont
  {Catuneanu}}, \bibinfo {author} {\bibfnamefont {C.~O.}\ \bibnamefont
  {Heinke}}, \bibinfo {author} {\bibfnamefont {G.~R.}\ \bibnamefont
  {Sivakoff}}, \bibinfo {author} {\bibfnamefont {W.~C.~G.}\ \bibnamefont {Ho}},
  \ and\ \bibinfo {author} {\bibfnamefont {M.}~\bibnamefont {Servillat}},\
  }\href {\doibase 10.1088/0004-637X/764/2/145} {\bibfield  {journal} {\bibinfo
   {journal} {Astrophys. J.}\ }\textbf {\bibinfo {volume} {764}},\ \bibinfo
  {pages} {145} (\bibinfo {year} {2013})},\ \Eprint
  {http://arxiv.org/abs/1301.3768} {arXiv:1301.3768 [astro-ph.HE]} \BibitemShut
  {NoStop}%
\bibitem [{\citenamefont {Servillat}\ \emph {et~al.}(2012)\citenamefont
  {Servillat}, \citenamefont {Heinke}, \citenamefont {Ho}, \citenamefont
  {Grindlay}, \citenamefont {Hong}, \citenamefont {Berg},\ and\ \citenamefont
  {Bogdanov}}]{2012MNRAS.423.1556S}%
  \BibitemOpen
  \bibfield  {author} {\bibinfo {author} {\bibfnamefont {M.}~\bibnamefont
  {Servillat}}, \bibinfo {author} {\bibfnamefont {C.~O.}\ \bibnamefont
  {Heinke}}, \bibinfo {author} {\bibfnamefont {W.~C.~G.}\ \bibnamefont {Ho}},
  \bibinfo {author} {\bibfnamefont {J.~E.}\ \bibnamefont {Grindlay}}, \bibinfo
  {author} {\bibfnamefont {J.}~\bibnamefont {Hong}}, \bibinfo {author}
  {\bibfnamefont {M.~v.~d.}\ \bibnamefont {Berg}}, \ and\ \bibinfo {author}
  {\bibfnamefont {S.}~\bibnamefont {Bogdanov}},\ }\href {\doibase
  10.1111/j.1365-2966.2012.20976.x} {\bibfield  {journal} {\bibinfo  {journal}
  {Mon. Not. Roy. Astron. Soc.}\ }\textbf {\bibinfo {volume} {423}},\ \bibinfo
  {pages} {1556} (\bibinfo {year} {2012})},\ \Eprint
  {http://arxiv.org/abs/1203.5807} {arXiv:1203.5807 [astro-ph.HE]} \BibitemShut
  {NoStop}%
\bibitem [{\citenamefont {Sandin}\ and\ \citenamefont
  {Ciarcelluti}(2009)}]{Sandin:2008db}%
  \BibitemOpen
  \bibfield  {author} {\bibinfo {author} {\bibfnamefont {F.}~\bibnamefont
  {Sandin}}\ and\ \bibinfo {author} {\bibfnamefont {P.}~\bibnamefont
  {Ciarcelluti}},\ }\href {\doibase 10.1016/j.astropartphys.2009.09.005}
  {\bibfield  {journal} {\bibinfo  {journal} {Astropart. Phys.}\ }\textbf
  {\bibinfo {volume} {32}},\ \bibinfo {pages} {278} (\bibinfo {year} {2009})},\
  \Eprint {http://arxiv.org/abs/0809.2942} {arXiv:0809.2942 [astro-ph]}
  \BibitemShut {NoStop}%
\bibitem [{\citenamefont {Ciarcelluti}\ and\ \citenamefont
  {Sandin}(2011)}]{Ciarcelluti:2010ji}%
  \BibitemOpen
  \bibfield  {author} {\bibinfo {author} {\bibfnamefont {P.}~\bibnamefont
  {Ciarcelluti}}\ and\ \bibinfo {author} {\bibfnamefont {F.}~\bibnamefont
  {Sandin}},\ }\href {\doibase 10.1016/j.physletb.2010.11.021} {\bibfield
  {journal} {\bibinfo  {journal} {Phys. Lett. B}\ }\textbf {\bibinfo {volume}
  {695}},\ \bibinfo {pages} {19} (\bibinfo {year} {2011})},\ \Eprint
  {http://arxiv.org/abs/1005.0857} {arXiv:1005.0857 [astro-ph.HE]} \BibitemShut
  {NoStop}%
\bibitem [{\citenamefont {Emma}\ \emph {et~al.}(2022)\citenamefont {Emma},
  \citenamefont {Schianchi}, \citenamefont {Pannarale}, \citenamefont {Sagun},\
  and\ \citenamefont {Dietrich}}]{Emma:2022xjs}%
  \BibitemOpen
  \bibfield  {author} {\bibinfo {author} {\bibfnamefont {M.}~\bibnamefont
  {Emma}}, \bibinfo {author} {\bibfnamefont {F.}~\bibnamefont {Schianchi}},
  \bibinfo {author} {\bibfnamefont {F.}~\bibnamefont {Pannarale}}, \bibinfo
  {author} {\bibfnamefont {V.}~\bibnamefont {Sagun}}, \ and\ \bibinfo {author}
  {\bibfnamefont {T.}~\bibnamefont {Dietrich}},\ }\href {\doibase
  10.3390/particles5030024} {\bibfield  {journal} {\bibinfo  {journal}
  {Particles}\ }\textbf {\bibinfo {volume} {5}},\ \bibinfo {pages} {273}
  (\bibinfo {year} {2022})},\ \Eprint {http://arxiv.org/abs/2206.10887}
  {arXiv:2206.10887 [gr-qc]} \BibitemShut {NoStop}%
\bibitem [{\citenamefont {Hanhart}\ \emph {et~al.}(2008)\citenamefont
  {Hanhart}, \citenamefont {Pelaez},\ and\ \citenamefont
  {Rios}}]{Hanhart:2008mx}%
  \BibitemOpen
  \bibfield  {author} {\bibinfo {author} {\bibfnamefont {C.}~\bibnamefont
  {Hanhart}}, \bibinfo {author} {\bibfnamefont {J.~R.}\ \bibnamefont {Pelaez}},
  \ and\ \bibinfo {author} {\bibfnamefont {G.}~\bibnamefont {Rios}},\ }\href
  {\doibase 10.1103/PhysRevLett.100.152001} {\bibfield  {journal} {\bibinfo
  {journal} {Phys. Rev. Lett.}\ }\textbf {\bibinfo {volume} {100}},\ \bibinfo
  {pages} {152001} (\bibinfo {year} {2008})},\ \Eprint
  {http://arxiv.org/abs/0801.2871} {arXiv:0801.2871 [hep-ph]} \BibitemShut
  {NoStop}%
\bibitem [{\citenamefont {McNeile}\ \emph {et~al.}(2009)\citenamefont
  {McNeile}, \citenamefont {Michael},\ and\ \citenamefont
  {Urbach}}]{McNeile:2009mx}%
  \BibitemOpen
  \bibfield  {author} {\bibinfo {author} {\bibfnamefont {C.}~\bibnamefont
  {McNeile}}, \bibinfo {author} {\bibfnamefont {C.}~\bibnamefont {Michael}}, \
  and\ \bibinfo {author} {\bibfnamefont {C.}~\bibnamefont {Urbach}} (\bibinfo
  {collaboration} {ETM}),\ }\href {\doibase 10.1016/j.physletb.2009.03.055}
  {\bibfield  {journal} {\bibinfo  {journal} {Phys. Lett. B}\ }\textbf
  {\bibinfo {volume} {674}},\ \bibinfo {pages} {286} (\bibinfo {year}
  {2009})},\ \Eprint {http://arxiv.org/abs/0902.3897} {arXiv:0902.3897
  [hep-lat]} \BibitemShut {NoStop}%
\bibitem [{\citenamefont {Walker-Loud}\ \emph {et~al.}(2009)\citenamefont
  {Walker-Loud} \emph {et~al.}}]{Walker-Loud:2008rui}%
  \BibitemOpen
  \bibfield  {author} {\bibinfo {author} {\bibfnamefont {A.}~\bibnamefont
  {Walker-Loud}} \emph {et~al.},\ }\href {\doibase 10.1103/PhysRevD.79.054502}
  {\bibfield  {journal} {\bibinfo  {journal} {Phys. Rev. D}\ }\textbf {\bibinfo
  {volume} {79}},\ \bibinfo {pages} {054502} (\bibinfo {year} {2009})},\
  \Eprint {http://arxiv.org/abs/0806.4549} {arXiv:0806.4549 [hep-lat]}
  \BibitemShut {NoStop}%
\bibitem [{\citenamefont {Bratt}\ \emph {et~al.}(2010)\citenamefont {Bratt}
  \emph {et~al.}}]{LHPC:2010jcs}%
  \BibitemOpen
  \bibfield  {author} {\bibinfo {author} {\bibfnamefont {J.~D.}\ \bibnamefont
  {Bratt}} \emph {et~al.} (\bibinfo {collaboration} {LHPC}),\ }\href {\doibase
  10.1103/PhysRevD.82.094502} {\bibfield  {journal} {\bibinfo  {journal} {Phys.
  Rev. D}\ }\textbf {\bibinfo {volume} {82}},\ \bibinfo {pages} {094502}
  (\bibinfo {year} {2010})},\ \Eprint {http://arxiv.org/abs/1001.3620}
  {arXiv:1001.3620 [hep-lat]} \BibitemShut {NoStop}%
\bibitem [{\citenamefont {Syritsyn}\ \emph {et~al.}(2010)\citenamefont
  {Syritsyn} \emph {et~al.}}]{Syritsyn:2009mx}%
  \BibitemOpen
  \bibfield  {author} {\bibinfo {author} {\bibfnamefont {S.~N.}\ \bibnamefont
  {Syritsyn}} \emph {et~al.},\ }\href {\doibase 10.1103/PhysRevD.81.034507}
  {\bibfield  {journal} {\bibinfo  {journal} {Phys. Rev. D}\ }\textbf {\bibinfo
  {volume} {81}},\ \bibinfo {pages} {034507} (\bibinfo {year} {2010})},\
  \Eprint {http://arxiv.org/abs/0907.4194} {arXiv:0907.4194 [hep-lat]}
  \BibitemShut {NoStop}%
\bibitem [{\citenamefont {Albaladejo}\ and\ \citenamefont
  {Oller}(2012)}]{Albaladejo:2012te}%
  \BibitemOpen
  \bibfield  {author} {\bibinfo {author} {\bibfnamefont {M.}~\bibnamefont
  {Albaladejo}}\ and\ \bibinfo {author} {\bibfnamefont {J.~A.}\ \bibnamefont
  {Oller}},\ }\href {\doibase 10.1103/PhysRevD.86.034003} {\bibfield  {journal}
  {\bibinfo  {journal} {Phys. Rev. D}\ }\textbf {\bibinfo {volume} {86}},\
  \bibinfo {pages} {034003} (\bibinfo {year} {2012})},\ \Eprint
  {http://arxiv.org/abs/1205.6606} {arXiv:1205.6606 [hep-ph]} \BibitemShut
  {NoStop}%
\bibitem [{\citenamefont {Horsley}\ \emph {et~al.}(2014)\citenamefont
  {Horsley}, \citenamefont {Nakamura}, \citenamefont {Nobile}, \citenamefont
  {Rakow}, \citenamefont {Schierholz},\ and\ \citenamefont
  {Zanotti}}]{Horsley:2013ayv}%
  \BibitemOpen
  \bibfield  {author} {\bibinfo {author} {\bibfnamefont {R.}~\bibnamefont
  {Horsley}}, \bibinfo {author} {\bibfnamefont {Y.}~\bibnamefont {Nakamura}},
  \bibinfo {author} {\bibfnamefont {A.}~\bibnamefont {Nobile}}, \bibinfo
  {author} {\bibfnamefont {P.~E.~L.}\ \bibnamefont {Rakow}}, \bibinfo {author}
  {\bibfnamefont {G.}~\bibnamefont {Schierholz}}, \ and\ \bibinfo {author}
  {\bibfnamefont {J.~M.}\ \bibnamefont {Zanotti}},\ }\href {\doibase
  10.1016/j.physletb.2014.03.002} {\bibfield  {journal} {\bibinfo  {journal}
  {Phys. Lett. B}\ }\textbf {\bibinfo {volume} {732}},\ \bibinfo {pages} {41}
  (\bibinfo {year} {2014})},\ \Eprint {http://arxiv.org/abs/1302.2233}
  {arXiv:1302.2233 [hep-lat]} \BibitemShut {NoStop}%
\bibitem [{\citenamefont {Miller}\ \emph {et~al.}(2019)\citenamefont {Miller}
  \emph {et~al.}}]{Miller:2019cac}%
  \BibitemOpen
  \bibfield  {author} {\bibinfo {author} {\bibfnamefont {M.~C.}\ \bibnamefont
  {Miller}} \emph {et~al.},\ }\href {\doibase 10.3847/2041-8213/ab50c5}
  {\bibfield  {journal} {\bibinfo  {journal} {Astrophys. J. Lett.}\ }\textbf
  {\bibinfo {volume} {887}},\ \bibinfo {pages} {L24} (\bibinfo {year}
  {2019})},\ \Eprint {http://arxiv.org/abs/1912.05705} {arXiv:1912.05705
  [astro-ph.HE]} \BibitemShut {NoStop}%
\bibitem [{\citenamefont {Riley}\ \emph {et~al.}(2019)\citenamefont {Riley}
  \emph {et~al.}}]{Riley:2019yda}%
  \BibitemOpen
  \bibfield  {author} {\bibinfo {author} {\bibfnamefont {T.~E.}\ \bibnamefont
  {Riley}} \emph {et~al.},\ }\href {\doibase 10.3847/2041-8213/ab481c}
  {\bibfield  {journal} {\bibinfo  {journal} {Astrophys. J. Lett.}\ }\textbf
  {\bibinfo {volume} {887}},\ \bibinfo {pages} {L21} (\bibinfo {year}
  {2019})},\ \Eprint {http://arxiv.org/abs/1912.05702} {arXiv:1912.05702
  [astro-ph.HE]} \BibitemShut {NoStop}%
\bibitem [{\citenamefont {Miller}\ \emph {et~al.}(2021)\citenamefont {Miller}
  \emph {et~al.}}]{Miller:2021qha}%
  \BibitemOpen
  \bibfield  {author} {\bibinfo {author} {\bibfnamefont {M.~C.}\ \bibnamefont
  {Miller}} \emph {et~al.},\ }\href {\doibase 10.3847/2041-8213/ac089b}
  {\bibfield  {journal} {\bibinfo  {journal} {Astrophys. J. Lett.}\ }\textbf
  {\bibinfo {volume} {918}},\ \bibinfo {pages} {L28} (\bibinfo {year}
  {2021})},\ \Eprint {http://arxiv.org/abs/2105.06979} {arXiv:2105.06979
  [astro-ph.HE]} \BibitemShut {NoStop}%
\bibitem [{\citenamefont {Riley}\ \emph {et~al.}(2021)\citenamefont {Riley}
  \emph {et~al.}}]{Riley:2021pdl}%
  \BibitemOpen
  \bibfield  {author} {\bibinfo {author} {\bibfnamefont {T.~E.}\ \bibnamefont
  {Riley}} \emph {et~al.},\ }\href {\doibase 10.3847/2041-8213/ac0a81}
  {\bibfield  {journal} {\bibinfo  {journal} {Astrophys. J. Lett.}\ }\textbf
  {\bibinfo {volume} {918}},\ \bibinfo {pages} {L27} (\bibinfo {year}
  {2021})},\ \Eprint {http://arxiv.org/abs/2105.06980} {arXiv:2105.06980
  [astro-ph.HE]} \BibitemShut {NoStop}%
\bibitem [{\citenamefont {Abbott}\ \emph {et~al.}(2017)\citenamefont {Abbott}
  \emph {et~al.}}]{TheLIGOScientific:2017qsa}%
  \BibitemOpen
  \bibfield  {author} {\bibinfo {author} {\bibfnamefont {B.~P.}\ \bibnamefont
  {Abbott}} \emph {et~al.} (\bibinfo {collaboration} {LIGO Scientific,
  Virgo}),\ }\href {\doibase 10.1103/PhysRevLett.119.161101} {\bibfield
  {journal} {\bibinfo  {journal} {Phys. Rev. Lett.}\ }\textbf {\bibinfo
  {volume} {119}},\ \bibinfo {pages} {161101} (\bibinfo {year} {2017})},\
  \Eprint {http://arxiv.org/abs/1710.05832} {arXiv:1710.05832 [gr-qc]}
  \BibitemShut {NoStop}%
\bibitem [{\citenamefont {Abbott}\ \emph
  {et~al.}(2018{\natexlab{a}})\citenamefont {Abbott} \emph
  {et~al.}}]{Abbott:2018exr}%
  \BibitemOpen
  \bibfield  {author} {\bibinfo {author} {\bibfnamefont {B.~P.}\ \bibnamefont
  {Abbott}} \emph {et~al.} (\bibinfo {collaboration} {LIGO Scientific,
  Virgo}),\ }\href {\doibase 10.1103/PhysRevLett.121.161101} {\bibfield
  {journal} {\bibinfo  {journal} {Phys. Rev. Lett.}\ }\textbf {\bibinfo
  {volume} {121}},\ \bibinfo {pages} {161101} (\bibinfo {year}
  {2018}{\natexlab{a}})},\ \Eprint {http://arxiv.org/abs/1805.11581}
  {arXiv:1805.11581 [gr-qc]} \BibitemShut {NoStop}%
\bibitem [{\citenamefont {Abbott}\ \emph {et~al.}(2019)\citenamefont {Abbott}
  \emph {et~al.}}]{Abbott:2018wiz}%
  \BibitemOpen
  \bibfield  {author} {\bibinfo {author} {\bibfnamefont {B.~P.}\ \bibnamefont
  {Abbott}} \emph {et~al.} (\bibinfo {collaboration} {LIGO Scientific,
  Virgo}),\ }\href {\doibase 10.1103/PhysRevX.9.011001} {\bibfield  {journal}
  {\bibinfo  {journal} {Phys. Rev. X}\ }\textbf {\bibinfo {volume} {9}},\
  \bibinfo {pages} {011001} (\bibinfo {year} {2019})},\ \Eprint
  {http://arxiv.org/abs/1805.11579} {arXiv:1805.11579 [gr-qc]} \BibitemShut
  {NoStop}%
\bibitem [{\citenamefont {Baym}\ \emph {et~al.}(1971)\citenamefont {Baym},
  \citenamefont {Pethick},\ and\ \citenamefont {Sutherland}}]{Baym:1971pw}%
  \BibitemOpen
  \bibfield  {author} {\bibinfo {author} {\bibfnamefont {G.}~\bibnamefont
  {Baym}}, \bibinfo {author} {\bibfnamefont {C.}~\bibnamefont {Pethick}}, \
  and\ \bibinfo {author} {\bibfnamefont {P.}~\bibnamefont {Sutherland}},\
  }\href {\doibase 10.1086/151216} {\bibfield  {journal} {\bibinfo  {journal}
  {Astrophys. J.}\ }\textbf {\bibinfo {volume} {170}},\ \bibinfo {pages} {299}
  (\bibinfo {year} {1971})}\BibitemShut {NoStop}%
\bibitem [{\citenamefont {Aad}\ \emph {et~al.}(2020)\citenamefont {Aad} \emph
  {et~al.}}]{Aad:2019mbh}%
  \BibitemOpen
  \bibfield  {author} {\bibinfo {author} {\bibfnamefont {G.}~\bibnamefont
  {Aad}} \emph {et~al.} (\bibinfo {collaboration} {ATLAS}),\ }\href {\doibase
  10.1103/PhysRevD.101.012002} {\bibfield  {journal} {\bibinfo  {journal}
  {Phys. Rev. D}\ }\textbf {\bibinfo {volume} {101}},\ \bibinfo {pages}
  {012002} (\bibinfo {year} {2020})},\ \Eprint
  {http://arxiv.org/abs/1909.02845} {arXiv:1909.02845 [hep-ex]} \BibitemShut
  {NoStop}%
\bibitem [{\citenamefont {Yagi}\ and\ \citenamefont
  {Yunes}(2013)}]{Yagi:2013awa}%
  \BibitemOpen
  \bibfield  {author} {\bibinfo {author} {\bibfnamefont {K.}~\bibnamefont
  {Yagi}}\ and\ \bibinfo {author} {\bibfnamefont {N.}~\bibnamefont {Yunes}},\
  }\href {\doibase 10.1103/PhysRevD.88.023009} {\bibfield  {journal} {\bibinfo
  {journal} {Phys. Rev. D}\ }\textbf {\bibinfo {volume} {88}},\ \bibinfo
  {pages} {023009} (\bibinfo {year} {2013})},\ \Eprint
  {http://arxiv.org/abs/1303.1528} {arXiv:1303.1528 [gr-qc]} \BibitemShut
  {NoStop}%
\bibitem [{\citenamefont {Nelson}\ \emph {et~al.}(2019)\citenamefont {Nelson},
  \citenamefont {Reddy},\ and\ \citenamefont {Zhou}}]{Nelson:2018xtr}%
  \BibitemOpen
  \bibfield  {author} {\bibinfo {author} {\bibfnamefont {A.}~\bibnamefont
  {Nelson}}, \bibinfo {author} {\bibfnamefont {S.}~\bibnamefont {Reddy}}, \
  and\ \bibinfo {author} {\bibfnamefont {D.}~\bibnamefont {Zhou}},\ }\href
  {\doibase 10.1088/1475-7516/2019/07/012} {\bibfield  {journal} {\bibinfo
  {journal} {JCAP}\ }\textbf {\bibinfo {volume} {07}},\ \bibinfo {pages} {012}
  (\bibinfo {year} {2019})},\ \Eprint {http://arxiv.org/abs/1803.03266}
  {arXiv:1803.03266 [hep-ph]} \BibitemShut {NoStop}%
\bibitem [{\citenamefont {Rafiei~Karkevandi}\ \emph {et~al.}(2021)\citenamefont
  {Rafiei~Karkevandi}, \citenamefont {Shakeri}, \citenamefont {Sagun},\ and\
  \citenamefont {Ivanytskyi}}]{RafieiKarkevandi:2021hcc}%
  \BibitemOpen
  \bibfield  {author} {\bibinfo {author} {\bibfnamefont {D.}~\bibnamefont
  {Rafiei~Karkevandi}}, \bibinfo {author} {\bibfnamefont {S.}~\bibnamefont
  {Shakeri}}, \bibinfo {author} {\bibfnamefont {V.}~\bibnamefont {Sagun}}, \
  and\ \bibinfo {author} {\bibfnamefont {O.}~\bibnamefont {Ivanytskyi}},\ }in\
  \href@noop {} {\emph {\bibinfo {booktitle} {{16th Marcel Grossmann Meeting
  on~Recent Developments in Theoretical and Experimental General Relativity,
  Astrophysics and Relativistic Field Theories}}}}\ (\bibinfo {year} {2021})\
  \Eprint {http://arxiv.org/abs/2112.14231} {arXiv:2112.14231 [astro-ph.HE]}
  \BibitemShut {NoStop}%
\bibitem [{\citenamefont {Hinderer}(2008)}]{Hinderer:2007mb}%
  \BibitemOpen
  \bibfield  {author} {\bibinfo {author} {\bibfnamefont {T.}~\bibnamefont
  {Hinderer}},\ }\href {\doibase 10.1086/533487} {\bibfield  {journal}
  {\bibinfo  {journal} {Astrophys. J.}\ }\textbf {\bibinfo {volume} {677}},\
  \bibinfo {pages} {1216} (\bibinfo {year} {2008})},\ \Eprint
  {http://arxiv.org/abs/0711.2420} {arXiv:0711.2420 [astro-ph]} \BibitemShut
  {NoStop}%
\bibitem [{\citenamefont {Comer}\ \emph {et~al.}(1999)\citenamefont {Comer},
  \citenamefont {Langlois},\ and\ \citenamefont {Lin}}]{Comer:1999rs}%
  \BibitemOpen
  \bibfield  {author} {\bibinfo {author} {\bibfnamefont {G.~L.}\ \bibnamefont
  {Comer}}, \bibinfo {author} {\bibfnamefont {D.}~\bibnamefont {Langlois}}, \
  and\ \bibinfo {author} {\bibfnamefont {L.~M.}\ \bibnamefont {Lin}},\ }\href
  {\doibase 10.1103/PhysRevD.60.104025} {\bibfield  {journal} {\bibinfo
  {journal} {Phys. Rev. D}\ }\textbf {\bibinfo {volume} {60}},\ \bibinfo
  {pages} {104025} (\bibinfo {year} {1999})},\ \Eprint
  {http://arxiv.org/abs/gr-qc/9908040} {arXiv:gr-qc/9908040} \BibitemShut
  {NoStop}%
\bibitem [{\citenamefont {Leung}\ \emph {et~al.}(2012)\citenamefont {Leung},
  \citenamefont {Chu},\ and\ \citenamefont {Lin}}]{Leung:2012vea}%
  \BibitemOpen
  \bibfield  {author} {\bibinfo {author} {\bibfnamefont {S.~C.}\ \bibnamefont
  {Leung}}, \bibinfo {author} {\bibfnamefont {M.~C.}\ \bibnamefont {Chu}}, \
  and\ \bibinfo {author} {\bibfnamefont {L.~M.}\ \bibnamefont {Lin}},\ }\href
  {\doibase 10.1103/PhysRevD.85.103528} {\bibfield  {journal} {\bibinfo
  {journal} {Phys. Rev. D}\ }\textbf {\bibinfo {volume} {85}},\ \bibinfo
  {pages} {103528} (\bibinfo {year} {2012})},\ \Eprint
  {http://arxiv.org/abs/1205.1909} {arXiv:1205.1909 [astro-ph.CO]} \BibitemShut
  {NoStop}%
\bibitem [{\citenamefont {Chandrasekhar}(1964)}]{Chandrasekhar:1964zz}%
  \BibitemOpen
  \bibfield  {author} {\bibinfo {author} {\bibfnamefont {S.}~\bibnamefont
  {Chandrasekhar}},\ }\href {\doibase 10.1086/147938} {\bibfield  {journal}
  {\bibinfo  {journal} {Astrophys. J.}\ }\textbf {\bibinfo {volume} {140}},\
  \bibinfo {pages} {417} (\bibinfo {year} {1964})},\ \bibinfo {note} {[Erratum:
  Astrophys.J. 140, 1342 (1964)]}\BibitemShut {NoStop}%
\bibitem [{\citenamefont {Henriques}\ \emph {et~al.}(1990)\citenamefont
  {Henriques}, \citenamefont {Liddle},\ and\ \citenamefont
  {Moorhouse}}]{Henriques:1990xg}%
  \BibitemOpen
  \bibfield  {author} {\bibinfo {author} {\bibfnamefont {A.~B.}\ \bibnamefont
  {Henriques}}, \bibinfo {author} {\bibfnamefont {A.~R.}\ \bibnamefont
  {Liddle}}, \ and\ \bibinfo {author} {\bibfnamefont {R.~G.}\ \bibnamefont
  {Moorhouse}},\ }\href {\doibase 10.1016/0370-2693(90)90789-9} {\bibfield
  {journal} {\bibinfo  {journal} {Phys. Lett. B}\ }\textbf {\bibinfo {volume}
  {251}},\ \bibinfo {pages} {511} (\bibinfo {year} {1990})}\BibitemShut
  {NoStop}%
\bibitem [{\citenamefont {Nyhan}\ and\ \citenamefont
  {Kain}(2022)}]{Nyhan:2022pda}%
  \BibitemOpen
  \bibfield  {author} {\bibinfo {author} {\bibfnamefont {J.~E.}\ \bibnamefont
  {Nyhan}}\ and\ \bibinfo {author} {\bibfnamefont {B.}~\bibnamefont {Kain}},\
  }\href {\doibase 10.1103/PhysRevD.105.123016} {\bibfield  {journal} {\bibinfo
   {journal} {Phys. Rev. D}\ }\textbf {\bibinfo {volume} {105}},\ \bibinfo
  {pages} {123016} (\bibinfo {year} {2022})},\ \Eprint
  {http://arxiv.org/abs/2206.07715} {arXiv:2206.07715 [gr-qc]} \BibitemShut
  {NoStop}%
\bibitem [{\citenamefont {Weinberg}(1972)}]{Weinberg:1972kfs}%
  \BibitemOpen
  \bibfield  {author} {\bibinfo {author} {\bibfnamefont {S.}~\bibnamefont
  {Weinberg}},\ }\href@noop {} {\emph {\bibinfo {title} {{Gravitation and
  Cosmology}: {Principles and Applications of the General Theory of
  Relativity}}}}\ (\bibinfo  {publisher} {John Wiley and Sons},\ \bibinfo
  {address} {New York},\ \bibinfo {year} {1972})\BibitemShut {NoStop}%
\bibitem [{\citenamefont {Glendenning}(2000)}]{glendenningbook}%
  \BibitemOpen
  \bibfield  {author} {\bibinfo {author} {\bibfnamefont {N.~K.}\ \bibnamefont
  {Glendenning}},\ }\href@noop {} {\emph {\bibinfo {title} {Compact Stars}}}\
  (\bibinfo  {publisher} {Springer},\ \bibinfo {address} {New York},\ \bibinfo
  {year} {2000})\BibitemShut {NoStop}%
\bibitem [{\citenamefont {Flanagan}\ and\ \citenamefont
  {Hinderer}(2008)}]{Flanagan:2007ix}%
  \BibitemOpen
  \bibfield  {author} {\bibinfo {author} {\bibfnamefont {E.~E.}\ \bibnamefont
  {Flanagan}}\ and\ \bibinfo {author} {\bibfnamefont {T.}~\bibnamefont
  {Hinderer}},\ }\href {\doibase 10.1103/PhysRevD.77.021502} {\bibfield
  {journal} {\bibinfo  {journal} {Phys. Rev. D}\ }\textbf {\bibinfo {volume}
  {77}},\ \bibinfo {pages} {021502} (\bibinfo {year} {2008})},\ \Eprint
  {http://arxiv.org/abs/0709.1915} {arXiv:0709.1915 [astro-ph]} \BibitemShut
  {NoStop}%
\bibitem [{\citenamefont {Favata}(2014)}]{Favata:2013rwa}%
  \BibitemOpen
  \bibfield  {author} {\bibinfo {author} {\bibfnamefont {M.}~\bibnamefont
  {Favata}},\ }\href {\doibase 10.1103/PhysRevLett.112.101101} {\bibfield
  {journal} {\bibinfo  {journal} {Phys. Rev. Lett.}\ }\textbf {\bibinfo
  {volume} {112}},\ \bibinfo {pages} {101101} (\bibinfo {year} {2014})},\
  \Eprint {http://arxiv.org/abs/1310.8288} {arXiv:1310.8288 [gr-qc]}
  \BibitemShut {NoStop}%
\bibitem [{\citenamefont {Yagi}\ and\ \citenamefont
  {Yunes}(2014)}]{Yagi:2013baa}%
  \BibitemOpen
  \bibfield  {author} {\bibinfo {author} {\bibfnamefont {K.}~\bibnamefont
  {Yagi}}\ and\ \bibinfo {author} {\bibfnamefont {N.}~\bibnamefont {Yunes}},\
  }\href {\doibase 10.1103/PhysRevD.89.021303} {\bibfield  {journal} {\bibinfo
  {journal} {Phys. Rev. D}\ }\textbf {\bibinfo {volume} {89}},\ \bibinfo
  {pages} {021303} (\bibinfo {year} {2014})},\ \Eprint
  {http://arxiv.org/abs/1310.8358} {arXiv:1310.8358 [gr-qc]} \BibitemShut
  {NoStop}%
\bibitem [{\citenamefont {Yagi}\ and\ \citenamefont
  {Yunes}(2016)}]{Yagi:2015pkc}%
  \BibitemOpen
  \bibfield  {author} {\bibinfo {author} {\bibfnamefont {K.}~\bibnamefont
  {Yagi}}\ and\ \bibinfo {author} {\bibfnamefont {N.}~\bibnamefont {Yunes}},\
  }\href {\doibase 10.1088/0264-9381/33/13/13LT01} {\bibfield  {journal}
  {\bibinfo  {journal} {Class. Quant. Grav.}\ }\textbf {\bibinfo {volume}
  {33}},\ \bibinfo {pages} {13LT01} (\bibinfo {year} {2016})},\ \Eprint
  {http://arxiv.org/abs/1512.02639} {arXiv:1512.02639 [gr-qc]} \BibitemShut
  {NoStop}%
\bibitem [{\citenamefont {Abbott}\ \emph
  {et~al.}(2018{\natexlab{b}})\citenamefont {Abbott} \emph
  {et~al.}}]{LIGOScientific:2018cki}%
  \BibitemOpen
  \bibfield  {author} {\bibinfo {author} {\bibfnamefont {B.~P.}\ \bibnamefont
  {Abbott}} \emph {et~al.} (\bibinfo {collaboration} {LIGO Scientific,
  Virgo}),\ }\href {\doibase 10.1103/PhysRevLett.121.161101} {\bibfield
  {journal} {\bibinfo  {journal} {Phys. Rev. Lett.}\ }\textbf {\bibinfo
  {volume} {121}},\ \bibinfo {pages} {161101} (\bibinfo {year}
  {2018}{\natexlab{b}})},\ \Eprint {http://arxiv.org/abs/1805.11581}
  {arXiv:1805.11581 [gr-qc]} \BibitemShut {NoStop}%
\bibitem [{\citenamefont {Berti}\ \emph {et~al.}(2009)\citenamefont {Berti},
  \citenamefont {Cardoso},\ and\ \citenamefont {Starinets}}]{Berti:2009kk}%
  \BibitemOpen
  \bibfield  {author} {\bibinfo {author} {\bibfnamefont {E.}~\bibnamefont
  {Berti}}, \bibinfo {author} {\bibfnamefont {V.}~\bibnamefont {Cardoso}}, \
  and\ \bibinfo {author} {\bibfnamefont {A.~O.}\ \bibnamefont {Starinets}},\
  }\href {\doibase 10.1088/0264-9381/26/16/163001} {\bibfield  {journal}
  {\bibinfo  {journal} {Class. Quant. Grav.}\ }\textbf {\bibinfo {volume}
  {26}},\ \bibinfo {pages} {163001} (\bibinfo {year} {2009})},\ \Eprint
  {http://arxiv.org/abs/0905.2975} {arXiv:0905.2975 [gr-qc]} \BibitemShut
  {NoStop}%
\bibitem [{\citenamefont {Yunes}\ \emph {et~al.}(2022)\citenamefont {Yunes},
  \citenamefont {Miller},\ and\ \citenamefont {Yagi}}]{Yunes:2022ldq}%
  \BibitemOpen
  \bibfield  {author} {\bibinfo {author} {\bibfnamefont {N.}~\bibnamefont
  {Yunes}}, \bibinfo {author} {\bibfnamefont {M.~C.}\ \bibnamefont {Miller}}, \
  and\ \bibinfo {author} {\bibfnamefont {K.}~\bibnamefont {Yagi}},\ }\href
  {\doibase 10.1038/s42254-022-00420-y} {\bibfield  {journal} {\bibinfo
  {journal} {Nature Rev. Phys.}\ }\textbf {\bibinfo {volume} {4}},\ \bibinfo
  {pages} {237} (\bibinfo {year} {2022})},\ \Eprint
  {http://arxiv.org/abs/2202.04117} {arXiv:2202.04117 [gr-qc]} \BibitemShut
  {NoStop}%
\bibitem [{\citenamefont {Tan}\ \emph {et~al.}(2022{\natexlab{a}})\citenamefont
  {Tan}, \citenamefont {Dexheimer}, \citenamefont {Noronha-Hostler},\ and\
  \citenamefont {Yunes}}]{Tan:2021nat}%
  \BibitemOpen
  \bibfield  {author} {\bibinfo {author} {\bibfnamefont {H.}~\bibnamefont
  {Tan}}, \bibinfo {author} {\bibfnamefont {V.}~\bibnamefont {Dexheimer}},
  \bibinfo {author} {\bibfnamefont {J.}~\bibnamefont {Noronha-Hostler}}, \ and\
  \bibinfo {author} {\bibfnamefont {N.}~\bibnamefont {Yunes}},\ }\href
  {\doibase 10.1103/PhysRevLett.128.161101} {\bibfield  {journal} {\bibinfo
  {journal} {Phys. Rev. Lett.}\ }\textbf {\bibinfo {volume} {128}},\ \bibinfo
  {pages} {161101} (\bibinfo {year} {2022}{\natexlab{a}})},\ \Eprint
  {http://arxiv.org/abs/2111.10260} {arXiv:2111.10260 [astro-ph.HE]}
  \BibitemShut {NoStop}%
\bibitem [{\citenamefont {Gerlach}(1968)}]{Gerlach:1968zz}%
  \BibitemOpen
  \bibfield  {author} {\bibinfo {author} {\bibfnamefont {U.~H.}\ \bibnamefont
  {Gerlach}},\ }\href {\doibase 10.1103/PhysRev.172.1325} {\bibfield  {journal}
  {\bibinfo  {journal} {Phys. Rev.}\ }\textbf {\bibinfo {volume} {172}},\
  \bibinfo {pages} {1325} (\bibinfo {year} {1968})}\BibitemShut {NoStop}%
\bibitem [{\citenamefont {Kampfer}(1981)}]{Kampfer:1981yr}%
  \BibitemOpen
  \bibfield  {author} {\bibinfo {author} {\bibfnamefont {B.}~\bibnamefont
  {Kampfer}},\ }\href {\doibase 10.1088/0305-4470/14/11/009} {\bibfield
  {journal} {\bibinfo  {journal} {J. Phys. A}\ }\textbf {\bibinfo {volume}
  {14}},\ \bibinfo {pages} {L471} (\bibinfo {year} {1981})}\BibitemShut
  {NoStop}%
\bibitem [{\citenamefont {Cromartie}\ \emph {et~al.}(2019)\citenamefont
  {Cromartie} \emph {et~al.}}]{Cromartie:2019kug}%
  \BibitemOpen
  \bibfield  {author} {\bibinfo {author} {\bibfnamefont {H.~T.}\ \bibnamefont
  {Cromartie}} \emph {et~al.} (\bibinfo {collaboration} {NANOGrav}),\ }\href
  {\doibase 10.1038/s41550-019-0880-2} {\bibfield  {journal} {\bibinfo
  {journal} {Nature Astron.}\ }\textbf {\bibinfo {volume} {4}},\ \bibinfo
  {pages} {72} (\bibinfo {year} {2019})},\ \Eprint
  {http://arxiv.org/abs/1904.06759} {arXiv:1904.06759 [astro-ph.HE]}
  \BibitemShut {NoStop}%
\bibitem [{\citenamefont {Tan}\ \emph {et~al.}(2022{\natexlab{b}})\citenamefont
  {Tan}, \citenamefont {Dore}, \citenamefont {Dexheimer}, \citenamefont
  {Noronha-Hostler},\ and\ \citenamefont {Yunes}}]{Tan:2021ahl}%
  \BibitemOpen
  \bibfield  {author} {\bibinfo {author} {\bibfnamefont {H.}~\bibnamefont
  {Tan}}, \bibinfo {author} {\bibfnamefont {T.}~\bibnamefont {Dore}}, \bibinfo
  {author} {\bibfnamefont {V.}~\bibnamefont {Dexheimer}}, \bibinfo {author}
  {\bibfnamefont {J.}~\bibnamefont {Noronha-Hostler}}, \ and\ \bibinfo {author}
  {\bibfnamefont {N.}~\bibnamefont {Yunes}},\ }\href {\doibase
  10.1103/PhysRevD.105.023018} {\bibfield  {journal} {\bibinfo  {journal}
  {Phys. Rev. D}\ }\textbf {\bibinfo {volume} {105}},\ \bibinfo {pages}
  {023018} (\bibinfo {year} {2022}{\natexlab{b}})},\ \Eprint
  {http://arxiv.org/abs/2106.03890} {arXiv:2106.03890 [astro-ph.HE]}
  \BibitemShut {NoStop}%
\end{thebibliography}%
\end{document}